\pdfoutput=1
\documentclass[aps,prd,reprint,superscriptaddress,showpacs,preprintnumbers]{revtex4-1}

\usepackage{graphicx}
\usepackage{amsmath}
\usepackage{lineno}

\urldef{\docurl}\url{https://sharepoint.fnal.gov/project/lbne/LBNE%20at%20Work/SitePages/Reports%20and%20Documents.aspx}

\mathchardef\mhyphen="2D

\begin{document}


\preprint{FERMILAB-PUB-14-379-PPD}
\preprint{BNL-106192-2014-JA}

\newcommand{\Argonne}{Argonne National Lab., Argonne, IL 60439, USA}
\newcommand{\Brookhaven}{Brookhaven National Lab., Upton, NY 11973-5000, USA}
\newcommand{\ColoradoState}{Colorado State University, Fort Collins, CO 80523, USA}
\newcommand{\Fermilab}{Fermi National Accelerator Lab., Batavia, IL 60510-0500, USA}
\newcommand{\Houston}{Univ. of Houston, Houston, Texas, 77204, USA}
\newcommand{\Pennsylvania}{Univ. of Pennsylvania, Philadelphia, PA 19104-6396, USA}
\newcommand{\Pittsburgh}{Univ. of Pittsburgh, Pittsburgh, PA 15260, USA}

\title{Baseline optimization for the measurement of CP violation, mass hierarchy, and $\theta_{23}$ octant in a long-baseline neutrino oscillation experiment}

\affiliation{\Argonne}
\affiliation{\Brookhaven}
\affiliation{\ColoradoState}
\affiliation{\Fermilab}
\affiliation{\Houston}
\affiliation{\Pittsburgh}

\author{M.~Bass}
\affiliation{\ColoradoState}
\author{M.~Bishai}
\affiliation{\Brookhaven}
\author{D.~Cherdack}
\affiliation{\ColoradoState}
\author{M.~Diwan}
\affiliation{\Brookhaven}
\author{Z.~Djurcic}
\affiliation{\Argonne}
\author{J.~Hernandez}
\affiliation{\Houston}
\author{B.~Lundberg}
\affiliation{\Fermilab}
\author{V.~Paolone}
\affiliation{\Pittsburgh}
\author{X.~Qian}
\affiliation{\Brookhaven}
\author{R.~Rameika}
\affiliation{\Fermilab}
\author{L.~Whitehead}
\affiliation{\Houston}
\author{R.J.~Wilson}
\affiliation{\ColoradoState}
\author{E.~Worcester}
\affiliation{\Brookhaven}
\author{G.~Zeller}
\affiliation{\Fermilab}

\date{\today}

\begin{abstract}
Next-generation long-baseline electron neutrino appearance experiments will seek to discover CP violation, determine the mass hierarchy and resolve the $\theta_{23}$ octant.  In light of the recent precision measurements of $\theta_{13}$, we consider the sensitivity of these measurements in a study to determine the optimal baseline, including practical considerations regarding beam and detector performance.  We conclude that a detector at a baseline of at least 1000~km in a wide-band muon neutrino beam is the optimal configuration.
\end{abstract}



\maketitle

\section{Introduction}

The goals of next-generation neutrino experiments include searching for CP violation in the lepton sector and precision studies of the neutrino mixing matrix.  These measurements require an optimal combination of the experimental baseline (the distance between the neutrino source and detector) and the neutrino beam energy.  In this paper, we study the baseline optimization for a long-baseline neutrino experiment, assuming a wide-band neutrino beam originating from the Fermilab proton complex.

Experimental observations \cite{Wendell:2010md,Ahn:2006zza,Adamson:2013whj,Adamson:2013ue,Abe:2014ugx,Abe:2013hdq,Aharmim:2009gd,Abe:2008aa,An:2013zwz,Ahn:2012nd,Abe:2012tg} have shown that neutrinos have mass and undergo flavor oscillations due to mixing between the mass states and flavor states.  For three neutrino flavors, the mixing can be described by three mixing angles ($\theta_{12}$, $\theta_{13}$, $\theta_{23}$) and one CP-violating phase parameter ($\delta_{CP}$).  The probability for flavor oscillations also depends on the differences in the squared masses of the neutrinos, $\Delta m^2_{21}$ and $\Delta m^2_{31}$, where $\Delta m^2_{ij} \equiv m^2_i - m^2_j$ and $\Delta m^2_{31} = \Delta m^2_{32} + \Delta m^2_{21}$.

Five of the parameters governing neutrino oscillations have been measured: all three mixing angles and the magnitude of the two independent mass squared differences.  Because the sign of $\Delta m^2_{31}$ is not known, there are two possibilities for the ordering of the neutrino masses, called the mass hierarchy: $m_1 < m_2 < m_3$ (``normal hierarchy'') or $m_3 < m_1 < m_2$ (``inverted hierarchy'').  The value of the CP-violating phase $\delta_{CP}$ is unknown.  Another remaining question is the octant of $\theta_{23}$: measured values of $\sin^{2}(2\theta_{23})$ are close to 1 \cite{Wendell:2010md,Adamson:2013whj,Abe:2014ugx}, but the data are so far inconclusive as to whether $\theta_{23}$ is less than or greater than 45$^{\circ}$, the value for maximal mixing between $\nu_{\mu}$ and $\nu_{\tau}$. 

\begin{figure*}[htp]
\includegraphics[width=0.69\textwidth]{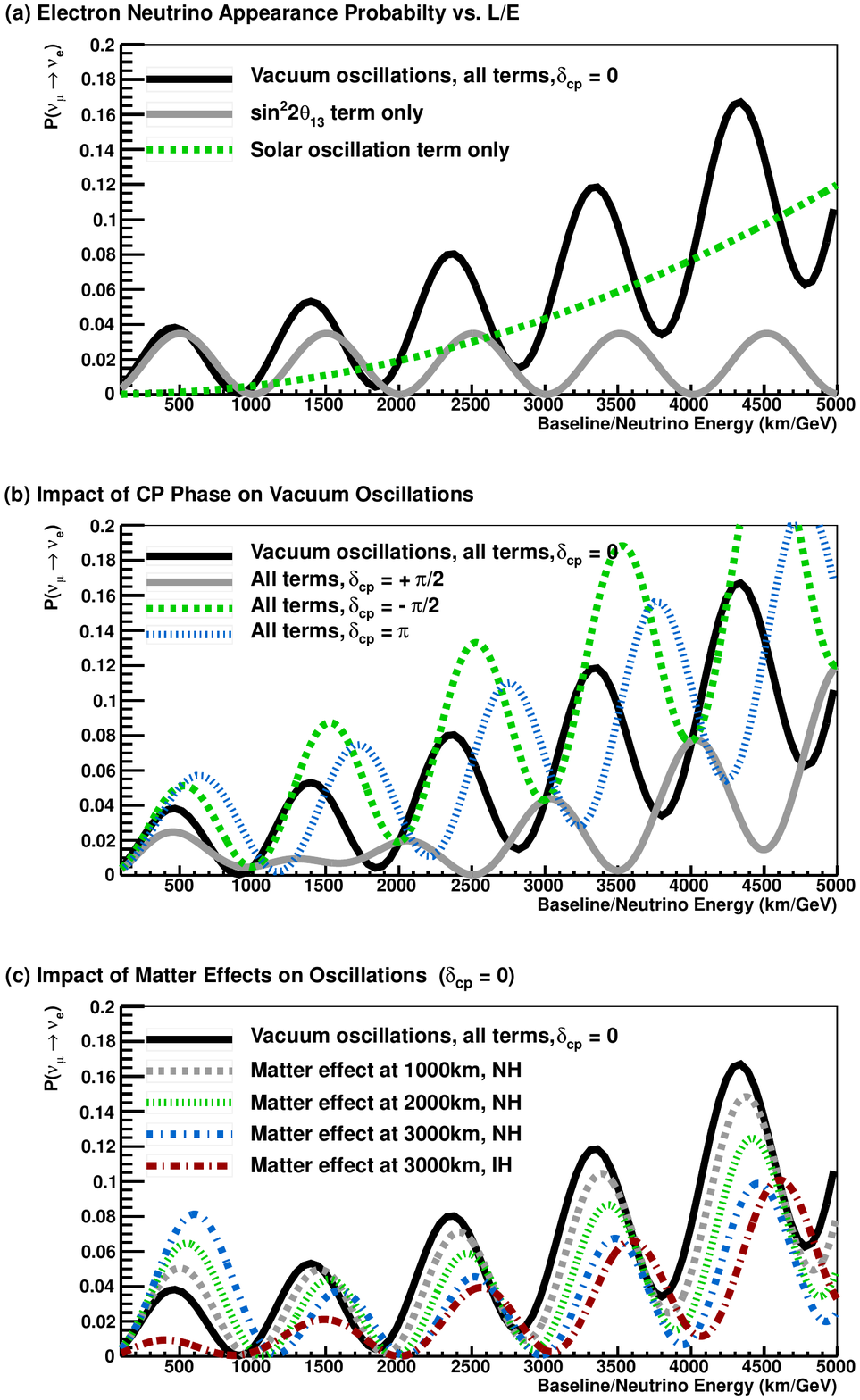}
\caption{The $\nu_\mu \rightarrow \nu_e$ oscillation probability as a function of baseline/neutrino energy ($L/E_{\nu}$) as given by Equation~\ref{eqn:prob}. The top graph (a) shows the full probability for vacuum oscillations with $\delta_{CP} = 0$ and shows the contribution from the $\sin^2(2\theta_{13})$ term and the $\sin^2(2\theta_{12})$ term (also called the solar term).  The middle graph (b) shows the full probability in vacuum for different values of $\delta_{CP}$. The bottom graph (c) shows the full probability in matter, assuming constant matter density, for different baselines compared to the vacuum probability with $\delta_{CP} = 0$.  The oscillation parameter values used to draw these curves are given in Section~\ref{sec:assumptions}.}
\label{fig:lovere}
\end{figure*}

The mass hierarchy, the value of $\delta_{CP}$, and the $\theta_{23}$ octant (value of $\sin^2\theta_{23}$) affect the muon neutrino to electron neutrino oscillation probability over a long baseline.  The oscillation probability can be approximated by \cite{Nunokawa:2007qh}
\begin{widetext}
\begin{eqnarray}
 \label{eqn:prob}
 P(\nu_{\mu} \rightarrow \nu_e) &\approx& \sin^2 \theta_{23} \sin^2 2 \theta_{13} {\sin^2(\Delta_{31} - aL) \over (\Delta_{31} - aL)^2} \Delta_{31}^2 \nonumber\\ &&
 +~\sin2\theta_{23} \sin2\theta_{13} \sin2\theta_{12}{\sin(\Delta_{31} - aL) \over (\Delta_{31} - aL)} \Delta_{31} {\sin(aL) \over (aL)} \Delta_{21} \cos(\Delta_{31} + \delta_{CP})\nonumber\\ &&
 +~\cos^2\theta_{23}\sin^22\theta_{12}{\sin^2(aL) \over (aL)^2}\Delta_{21}^2
\end{eqnarray}
\end{widetext}
where $\Delta_{ij} = \Delta m^2_{ij} L/4E_\nu$, $a = G_FN_e/\sqrt{2}$, $G_F$ is the Fermi constant, $N_e$ is the number density of electrons in the Earth, $L$ is the baseline in km, and $E_\nu$ is the neutrino energy in GeV. 
The corresponding probability for antineutrinos is the same, except that $a \rightarrow -a$ and $\delta_{CP} \rightarrow -\delta_{CP}$.  Figure~\ref{fig:lovere} shows the probability as a function of $L/E_{\nu}$ for various cases.  

The maximum oscillation probabilities in vacuum occur at
\begin{equation}
 \frac{L}{E_n} \left(\frac{\textnormal{km}}{\textnormal{GeV}}\right) \approx (2n-1)\left(\frac{\pi}{2}\right)\frac{1}{1.267 \times \Delta m_{32}^{2} (\textnormal{eV}^2)},
\end{equation}
where $E_n$ is the neutrino energy at the $n^{th}$ oscillation maximum.  For longer baselines, it is possible to observe multiple oscillation maxima in the spectra if the neutrino flux covers a wide range of energy.  At short baselines, the higher order maxima ($n>1$) are typically too low in energy to be observable with high-energy accelerator beams.

A CP-violating value of $\delta_{CP}$ ($\delta_{CP} \neq 0$ and $\delta_{CP} \neq \pi$) would cause a difference in the probabilities for $\nu_{\mu} \rightarrow \nu_e$ and
$\bar{\nu}_{\mu} \rightarrow \bar{\nu}_e$ transitions. The CP asymmetry $\mathcal{A}_{cp}$ is defined as
\begin{equation}
\mathcal{A}_{cp}(E_\nu) = \left[
\frac{{\rm P}(\nu_\mu \rightarrow \nu_e) -
  {\rm  P}(\bar{\nu}_\mu \rightarrow \bar{\nu}_e)}
{{\rm P}(\nu_\mu \rightarrow \nu_e) +
  {\rm P}(\bar{\nu}_\mu \rightarrow \bar{\nu}_e)} \right].
\end{equation}
If $\delta_{CP}=0$ or $\pi$, the transition probability for oscillations in vacuum is the same for neutrinos and antineutrinos.  For oscillations in matter, the MSW matter effect \cite{Wolfenstein:1977ue,Mikheev:1986gs} creates a difference between the neutrino and antineutrino probabilities, even for $\delta_{CP}=0$ or $\delta_{CP}=\pi$.  For oscillations in matter with $\delta_{CP} \neq 0$ and $\delta_{CP} \neq \pi$, there is an asymmetry due to both CP violation and the matter effect.  A leading-order approximation of the CP asymmetry in the three-flavor model is given by \cite{Marciano:2006uc}
\begin{align}
 \mathcal{A}_{cp}(E_\nu) \approx &\frac{\cos\theta_{23}\sin2\theta_{12}\sin\delta_{CP}}{\sin\theta_{23}\sin\theta_{13}}\left(\frac{\Delta m^2_{21}L}{4E_\nu}\right)\nonumber\\&~+~\textnormal{matter effects}.
\end{align}

In principle, a measurement of the parameter $\delta_{CP}$ could be performed based on a spectrum shape fit with only neutrino beam data.  However, long-baseline experiments seek not only to measure the parameter, but to explicitly demonstrate CP violation by observing the asymmetry between neutrinos and antineutrinos.  Additionally, a comparison of the measured value of $\delta_{CP}$ based on neutrino data alone to that based on the combined fit of neutrino and antineutrino data will be a useful cross-check.



\begin{figure*}[htp]
\includegraphics[width=\textwidth]{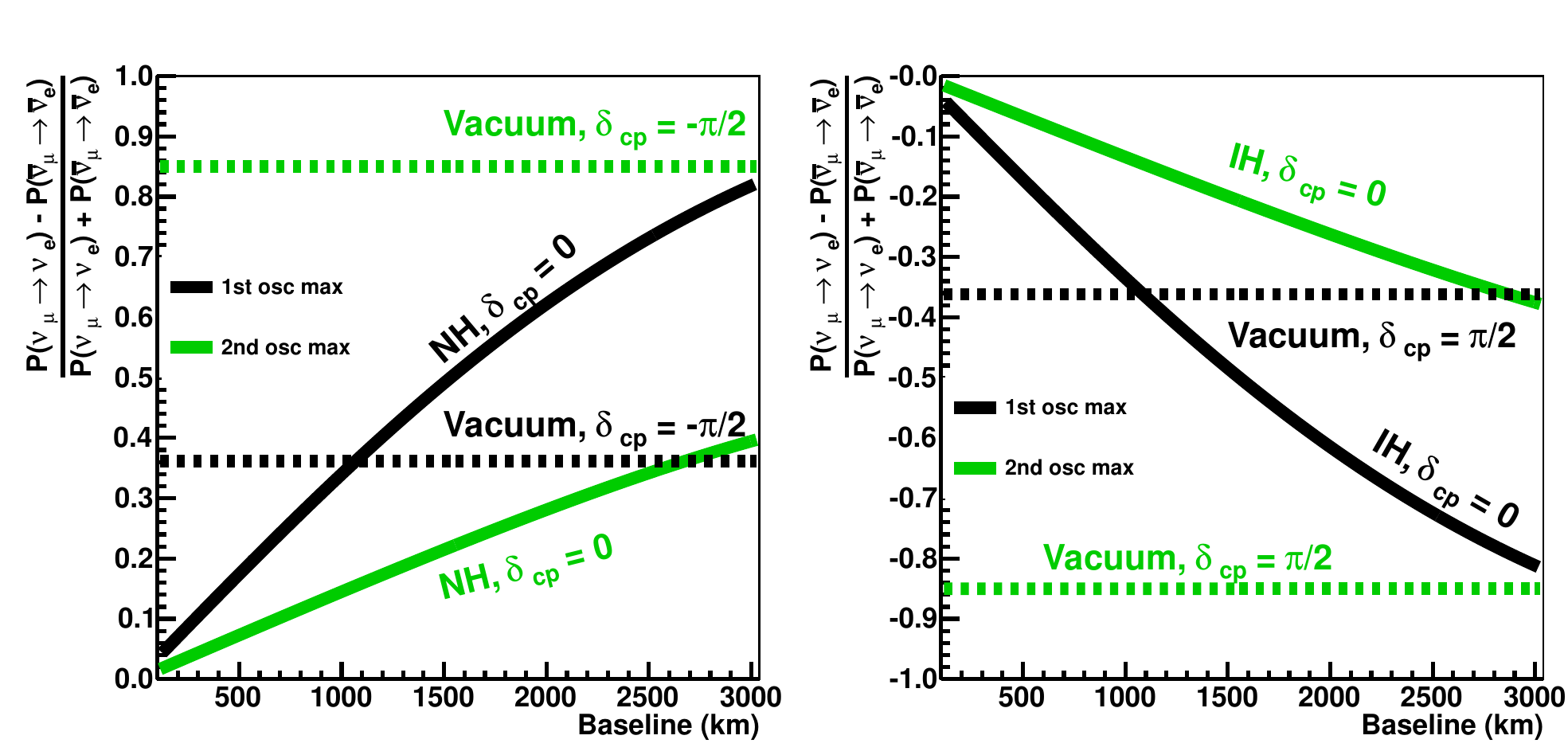}
\caption{Asymmetry vs baseline at the first (black line) and second (green line) oscillation maxima.  The solid lines show the asymmetry due to the matter effect only ($\delta_{CP}=0$) for the normal hierarchy (left plot) or inverted hierarchy (right plot).  The dashed lines show the maximum CP asymmetry in vacuum, with $\delta_{CP} = -\pi/2$ (left plot) or $\delta_{CP} = \pi/2$ (right plot).  The oscillation parameter values used to draw these curves are given in Section~\ref{sec:assumptions}.}
\label{fig:asym_1st2nd}
\end{figure*}

Figure \ref{fig:asym_1st2nd} shows the asymmetry for different baselines calculated at both the first and second oscillation maxima, since only these two maxima are accessible in practical accelerator experiments.  The asymmetry is shown assuming only matter effects ($\delta_{CP}=0$) or only maximal $\delta_{CP}$ effects (in vacuum).  The matter asymmetry grows as a function of baseline, and therefore distinguishing the normal and inverted hierarchies by measuring neutrino and antineutrino events becomes easier as the baseline increases, as long as the number of appearance events stays constant.
The asymmetry due to nonzero $\delta_{CP}$ is constant as a function of baseline at both the first and second oscillation maximum.  At the first oscillation maximum, the maximum CP asymmetry is larger than the matter asymmetry only for baselines less than $\sim$1000~km.
However, at the second oscillation maximum, the maximal CP asymmetry dominates the matter asymmetry at all baselines.  The second oscillation maximum therefore has good sensitivity to CP violation, independent of the mass hierarchy.  At short baselines, the second oscillation maximum occurs at an energy that isn't observable.  Therefore at short baselines, any observed asymmetry could be due to either the matter effect or CP violation at the first oscillation maximum; additional information is needed to determine the cause of the asymmetry.  At longer baselines with a wide-band beam, the ambiguity at the first oscillation maximum can be resolved using the information from the second oscillation maximum.  

Previous studies (for example, \cite{Diwan:2004bt,Barger:2007yw,Barger:2007jq}) have considered the optimal baseline for measurements of muon neutrino to electron neutrino oscillations using a wide-band neutrino beam from Fermilab.  However, these studies were conducted before the value of $\theta_{13}$ was measured by reactor antineutrino experiments \cite{An:2013zwz,Ahn:2012nd,Abe:2012tg}.  The measured value of $\theta_{13}$ has been incorporated into other recent long-baseline oscillation sensitivity estimates, but the study presented in this paper is unique in considering different baselines.  We reconsider the baseline optimization for an electron neutrino appearance measurement using the measured value of $\theta_{13}$ and realistic simulations of a wide-band neutrino beam facility at Fermilab.

\section{Expected Electron Neutrino Appearance Rate}
In a conventional neutrino beam, protons hit a stationary target producing secondary particles, most of which are pions.  The positively charged pions are focused in the forward direction by a toroidal magnetic field generated by magnetic horns.  The pions are then allowed to decay to produce a muon neutrino beam.  At the end of the decay region an absorber stops the remaining secondary particles from the initial proton collision, and the muons produced in the decay pipe are stopped in rock located beyond the absorber.  A muon antineutrino beam can be created by reversing the magnetic field to focus negatively charged pions.  Horn-focused beams are technologically well-matched to long-baseline experiments with neutrino energy $>$ 1~GeV since horn focusing is optimal for focusing hadrons $>$ 2~GeV and can be used effectively to charge-select the focused hadrons. In this study, we use the simulated flux from horn-focused beams to evaluate the sensitivity of long-baseline neutrino oscillation experiments at different baselines.  In this section, we discuss the expected dependence of the electron neutrino appearance rate on baseline, making ideal flux assumptions and ignoring any detector effects for simplicity.

\begin{figure*}[htp]
\includegraphics[width=0.7\textwidth]{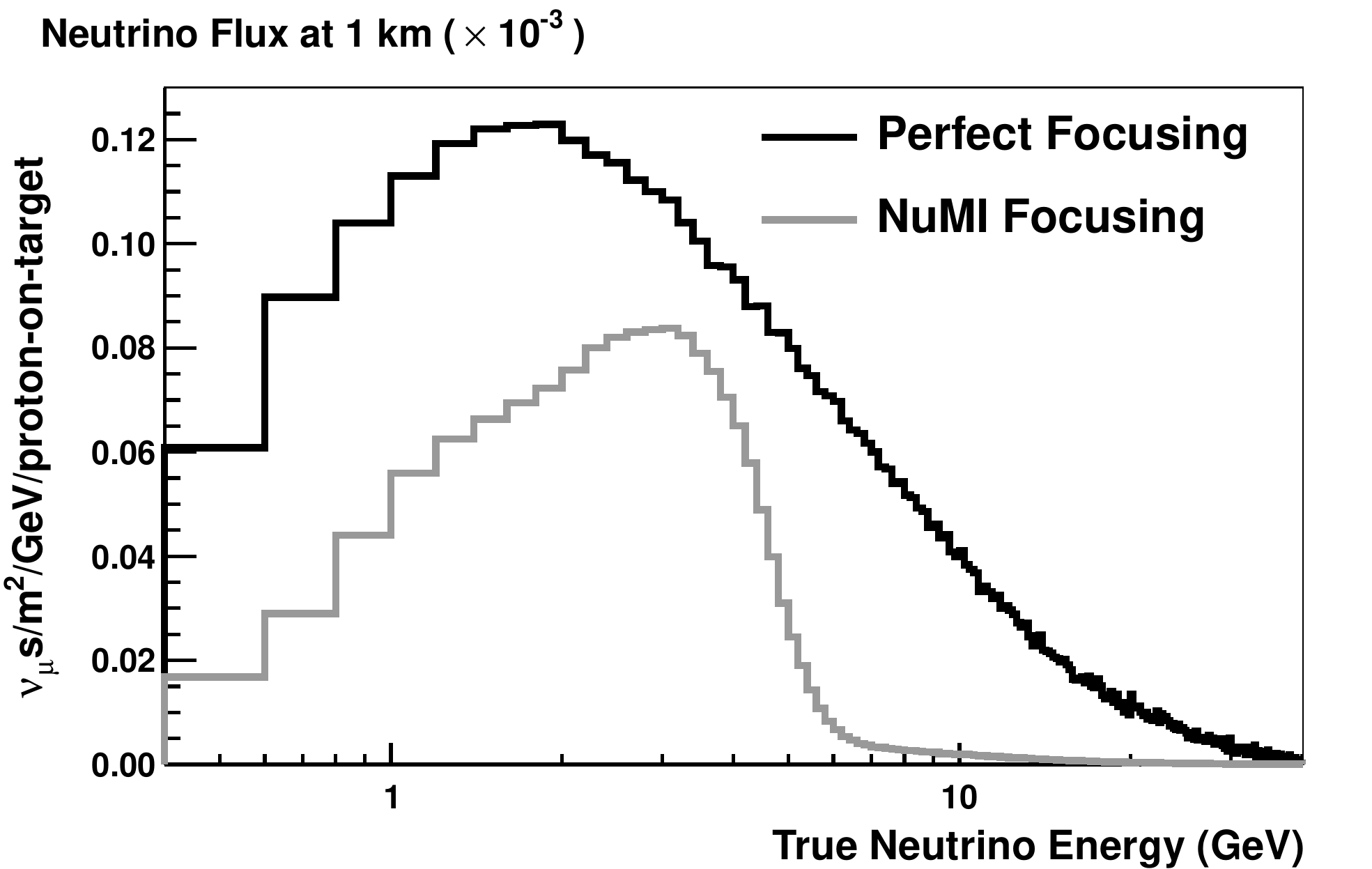}
\caption{The neutrino flux at 1~km from a 120-GeV proton beam incident on a target two interaction lengths in thickness. The black histogram is the flux from a perfectly focused hadron beam and a decay pipe 380~m in length. The gray histogram is the flux obtained using the NuMI focusing system operating at a current of 250~kA.  (The focusing system works similarly for the antineutrino spectrum.)}
\label{fig:flux1}
\end{figure*}

\begin{widetext}
The total number of electron neutrino appearance events expected for a given exposure from a muon neutrino source as a function of baseline is given as
\begin{equation}
N_{\nu_e}^{\rm appear}(L)  = N_{\rm target} \int {\Phi^{\nu_\mu}(E_{\nu},L)  \times P^{\nu_\mu \rightarrow \nu_e}(E_{\nu},L) \times \sigma^{\nu_e}(E_{\nu}) d E_{\nu} }
\label{eqn:nume}
\end{equation}
where $\Phi^{\nu_\mu}(E_{\nu},L)$ is the muon neutrino flux as a function of neutrino energy, $E_{\nu}$, and baseline, $L$, $\sigma^{\nu_e}(E_{\nu})$ is the electron neutrino inclusive charged-current cross-section per nucleon ($N$), $N_{\rm target}$ is the number of target nucleons per kt of detector fiducial volume, and $P^{\nu_\mu \rightarrow \nu_e}(E_{\nu},L)$ is the appearance probability in matter.  For this discussion, the units are always assumed to be km for $L$, GeV for $E_{\nu}$, and eV$^2$ for $\Delta m^{2}_{31}$.

For a simple estimate of the electron neutrino appearance rate as a function of baseline, we assume the neutrino beam source produces a flux that is constant in the oscillation energy region (Equation~\ref{eqn:a}), and we approximate the appearance probability with the dominant term without matter effects (Equation~\ref{eqn:b}).  The expressions for the electron neutrino charged-current cross-section and number of target nucleons per kt are given in Equations~\ref{eqn:c} and \ref{eqn:d}, respectively.
\begin{eqnarray}
\Phi^{\nu_\mu}(E_{\nu},L) & \approx & \frac{C}{L^2}, \ \ C = \textrm{number of }\nu_{\mu}/{\rm m}^2/{\rm GeV}/{\rm (MW\mhyphen yr)}  \ {\rm at \ 1~km} \label{eqn:a}\\
P^{\nu_\mu \rightarrow \nu_e}(E_{\nu},L) & \approx & \sin^2 \theta_{23} \sin^2 2\theta_{13} \sin^2 (1.27 \Delta m^2_{31} L /E_{\nu} )  \label{eqn:b}\\
\sigma^{\nu_e}(E_{\nu}) & = & 0.67 \times 10^{-42} ({\rm m}^2/{\rm GeV}/N) \times E_{\nu}, \ \ E_{\nu}>0.5 \ {\rm GeV}\label{eqn:c}\\
N_{\rm target} &= &6.022 \times 10^{32} N/{\rm kt} \label{eqn:d}
\end{eqnarray}
For a 120-GeV proton beam from the Fermilab Main Injector with a live time of $2 \times 10^7$~s/yr, 1~MW-yr corresponds to approximately $10^{21}$ protons on target.  A beam simulation with perfect hadron focusing (in which all secondary mesons are assumed to be focused towards the far detector) produces a peak flux of roughly $0.12 \times 10^{-3}~\nu_{\mu}$/m$^{2}$/GeV/proton-on-target at 1~km (see Figure \ref{fig:flux1}).  Combining these numbers and naively assuming a flat spectrum, we obtain
$C \approx 1.2 \times 10^{17} \ \nu_\mu/\rm{m}^2\rm{/GeV/(MW\mhyphen yr)}$ at 1~km. Using this assumption and the approximations in Equations \ref{eqn:a}-\ref{eqn:d}, we find that
\begin{eqnarray}
N_{\nu_e}^{\rm appear}(L) \approx (1.8 \times 10^6 {\rm events}/({\rm kt\mhyphen MW\mhyphen yr})) ({\rm km}/{\rm GeV})^2 \times \int_{x_0}^{x_1} \frac{\sin^2(ax)}{x^3} dx,\label{eqn:nume1}\\
x \equiv L/E_{\nu},~a \equiv 1.27 \Delta m^2_{31}.\nonumber
\end{eqnarray}
Integrating Equation~\ref{eqn:nume1} over the region of the first two oscillation maxima such that $x_0 = 100$~km/GeV and $x_1 = 2000$~km/GeV (see Figure~\ref{fig:lovere}), yields 
\begin{equation}
N_{\nu_e}^{\rm appear}(L) \sim \mathcal{O}(20) \ \rm{events}/\rm{(kt\mhyphen MW\mhyphen yr)} 
\label{eqn:const}
\end{equation}

\end{widetext}

As seen from the simplified discussion presented above, the $\nu_e$ appearance rate for vacuum oscillations is a constant that is largely independent of baseline for baselines $>$300~km.  The event rates at experiments with baselines $<$300~km are lower because the neutrino cross-sections at energies $<0.5$~GeV are not linear with energy.  For oscillations in matter, the electron neutrino appearance probability at the first oscillation maximum increases with baseline for the case of normal hierarchy (as shown in Figure~\ref{fig:lovere}c) and decreases for inverted hierarchy. 

\begin{figure*}[htp]
\includegraphics[width=0.7\textwidth]{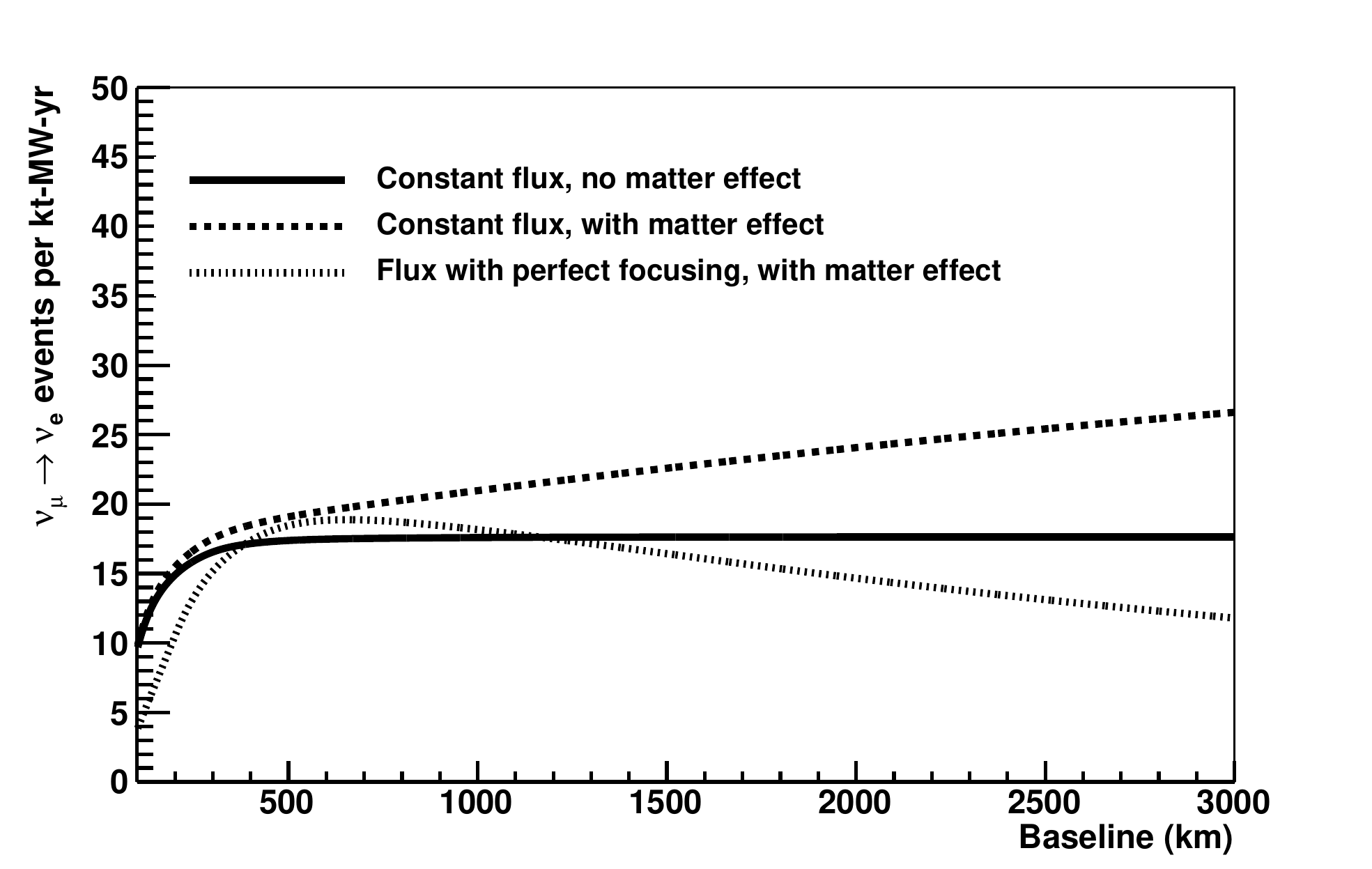}
\caption{Estimated $\nu_e$ appearance rates (with no detector effects) integrated over the first two oscillation maxima as a function of baseline for three conditions. The solid and dashed lines are the event rates obtained using a neutrino flux that is constant as a function of energy. The solid line assumes oscillations in vacuum and the dashed line assumes oscillations in a constant matter density with normal hierarchy. The dotted-dashed line is the event rate calculation using a neutrino-beam flux obtained from a perfect-focusing system with a 120-GeV primary beam and a fixed decay pipe length of 380~m assuming normal hierarchy. }
\label{fig:evtrate}
\end{figure*}

For real neutrino beams generated from pion decays in flight, it is not possible to produce a neutrino flux that is constant over a large range of energies. Hadron production from the proton target and decay kinematics due to the finite decay volume will produce a reduced neutrino flux at higher energies when a fixed proton beam energy is used. The lower flux at higher energies, and hence longer baselines, will counteract the event rate increase from the matter effect (assuming a neutrino beam and normal hierarchy).  This effect is illustrated in Figure~\ref{fig:evtrate}. We calculate the $\nu_e$ appearance event rate integrated over the region of the first two oscillation maxima using the appearance probability given by Equation~\ref{eqn:prob} assuming a constant flux or a horn-focused beam with a perfect-focusing system. 
For the constant flux assumption, we use Equation~\ref{eqn:a} with $C \approx 1.2 \times 10^{17} \ \nu_\mu/\rm{m}^2\rm{/GeV/(MW\mhyphen yr)}$ as before.
The results from this more detailed calculation are comparable to Equation~\ref{eqn:const} and illustrate the slowly varying dependence on baseline.

\begin{figure*}[htp]
\includegraphics[width=0.49\textwidth]{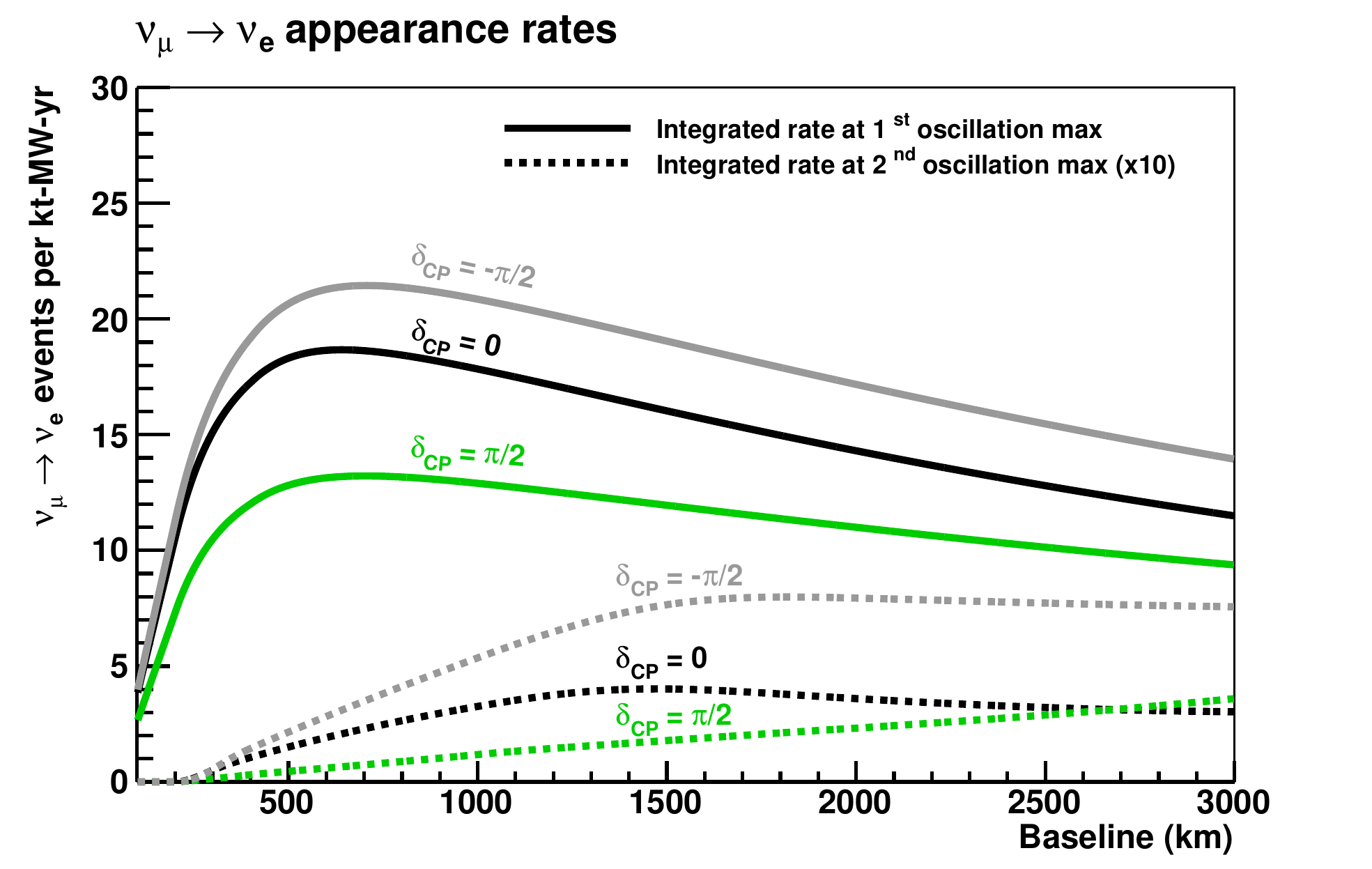}
\includegraphics[width=0.49\textwidth]{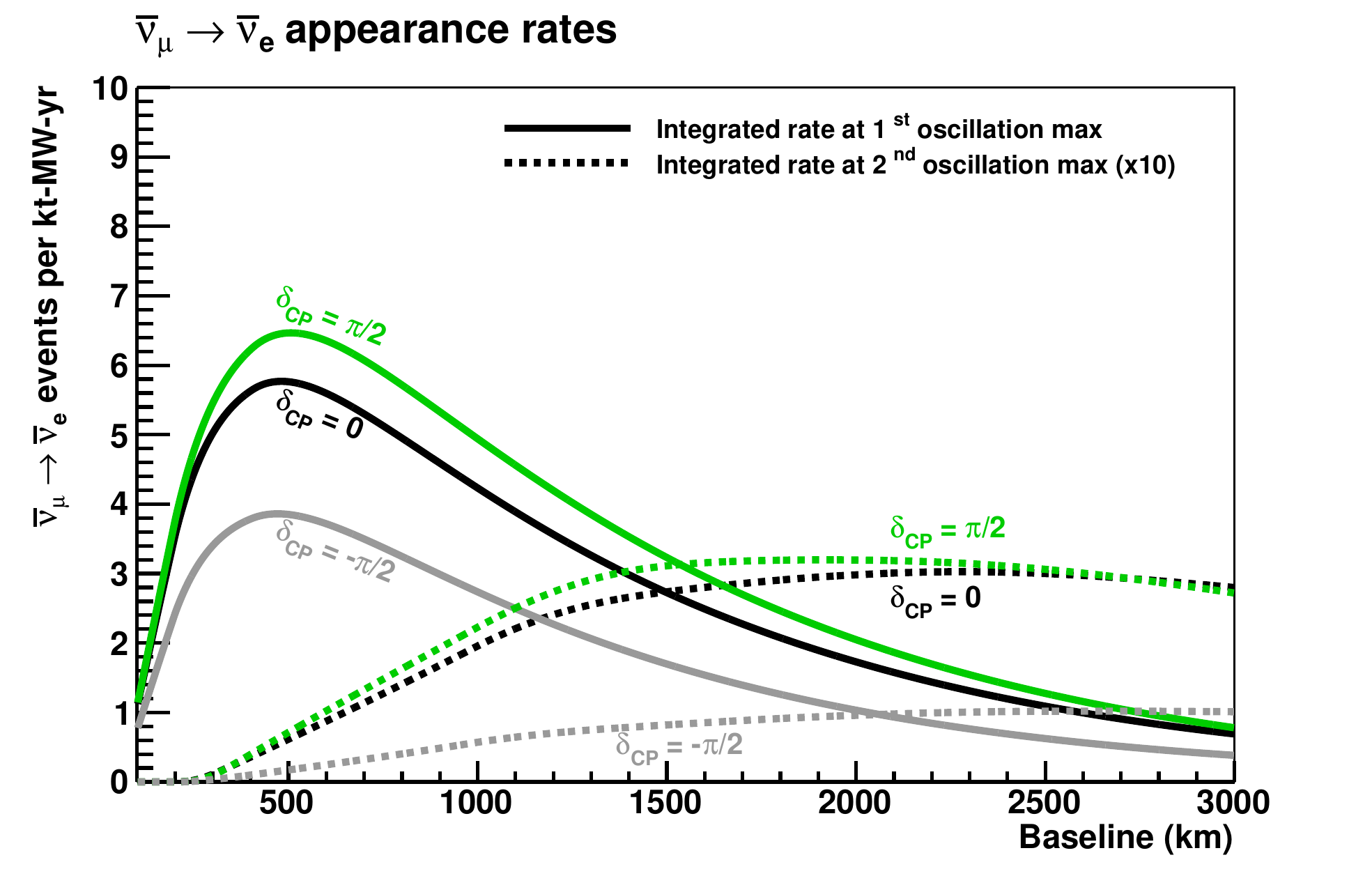}
\caption{Estimated $\nu_e$ (left) and $\bar{\nu}_e$ (right) appearance rates (with no detector effects) integrated over an energy region around the first oscillation maximum (solid line) or the second oscillation maximum (dashed line) assuming  the flux obtained from a perfect-focusing system with a 120-GeV primary beam and a fixed decay pipe length of 380~m. The curves are shown for different values of $\delta_{CP}$.  Matter effects are included assuming normal hierarchy. The rates for the second oscillation maximum have been increased by a factor of 10 for visibility.}
\label{fig:evtrate_1st2nd}
\end{figure*}

Figure \ref{fig:evtrate_1st2nd} shows separately the event rate in each of the first two oscillation maxima assuming the flux obtained from a perfect-focusing system with a fixed decay pipe length. With perfect focusing, the event rate in the region of the second oscillation maximum is relatively constant for baselines greater than 1200~km.  Therefore, based on these considerations, we don't expect the sensitivity to CP violation to increase as a function of baseline, but remain roughly the same.  The event rate in the region of the first maximum decreases with baseline due to the decreasing flux from the beam and increasing impact from the matter asymmetry. For longer baselines, the decrease in flux in the region of the first maximum can be ameliorated by using longer decay pipes to increase the number of pion decays at higher energy.  

The perfect-focusing system assumed in Figure \ref{fig:evtrate_1st2nd} uses a 120-GeV primary proton beam which can be produced at the Fermilab Main Injector.  An alternate strategy of focusing on the second oscillation maximum at long baselines by using a lower primary proton beam energy will not be considered in this study. With a lower proton energy, the integrated rates in the first and second oscillation maxima are more similar, and the second maximum contributes greatly to the CP violation sensitivity. At 120~GeV however, the rates at the first maximum dominate at all baselines in neutrino mode (shown in Figure \ref{fig:evtrate_1st2nd}), and accessing the second maximum by going to longer baselines is unlikely to yield significant enhancements to the sensitivity.  The optimal baseline therefore depends on the energy of the primary proton beam.  The highest power from the Fermilab proton complex is currently available at an energy of 120~GeV, and that is the only proton beam energy considered in this study.

The sensitivity studies in this paper assume a horn-focused neutrino beam with realistic focusing and include detector effects. The NuMI~\cite{numi} design of double-parabolic horns was chosen as the basis for these simulations because the NuMI horns were designed to be used as tunable, focusing magnetic lenses over a wide range of hadron energy.  The comparison between perfect focusing and focusing using the NuMI system is given in Figure~\ref{fig:flux1}, which shows the neutrino flux at 1~km from a 120-GeV proton beam incident on a target two interaction lengths in thickness.
An evacuated hadron decay pipe 380~m in length and 4~m in diameter is assumed for this comparison.  The next section discusses the details of the simulated fluxes with realistic focusing used for the sensitivity calculations at each baseline.

\section{Beam Simulations}

This study uses neutrino and antineutrino fluxes derived from GEANT3 \cite{geant3} beamline simulations optimized to cover the energy region of the first oscillation maximum (and the second maximum if possible) at each baseline considered. The beam simulation at each baseline assumes a 1.2-MW 120-GeV primary proton beam that delivers $1 \times 10^{21}$ protons-on-target per year. A graphite target with a diameter of 1.2 cm and a length equivalent to two interaction lengths is assumed. The double-parabolic NuMI focusing horn design is used \cite{numi}, with a horn current of 
250 kA.  The separation between the two horns is assumed to be 6~m.  The decay pipe is 4~m in diameter and is assumed to be evacuated. The beamline parameters that are varied for different baselines (distance between target and horn, decay pipe length, and off-axis angle) are summarized in Table \ref{tab:BLbeams}.  We change the decay pipe length in increments of 100 m to match the decay length of a pion whose energy corresponds to the neutrino energy of the first oscillation maximum. For baselines $>$ 1000~km, when considering different configurations that cover the oscillation region appropriately, we choose the configuration at each baseline that maximizes CP sensitivity.

\begin{table*}[htp]
\caption{The beam configuration used at each baseline to
  determine the optimal baseline for the next generation long-baseline
  experiment. The beam parameters are chosen to cover the first oscillation maximum at each baseline.\\}
\begin{tabular}{|c|c|c|c|}
\hline
~~~Baseline~~~ & Target-Horn 1 distance & Decay pipe length & Off-axis
angle \\ \hline
300~km & 30~cm & 280~m & 2$^\circ$ \\
500~km & 30~cm & 280~m & 1.5$^\circ$\\
750~km & 30~cm & 280~m & 1.0$^\circ$ \\
1000~km & 0~cm & 280~m & 0$^\circ$ \\
1300~km & 30~cm & 380~m & 0$^\circ$ \\
1700~km & 30~cm & 480~m & 0$^\circ$ \\
2000~km & 70~cm & 580~m & 0$^\circ$ \\
2500~km & 70~cm & 680~m & 0$^\circ$ \\
3000~km & 100~cm & 780~m & 0$^\circ$ \\ \hline
\end{tabular}
\label{tab:BLbeams}
\end{table*}

\begin{figure*}[htp]
\includegraphics[width=0.7\textwidth]{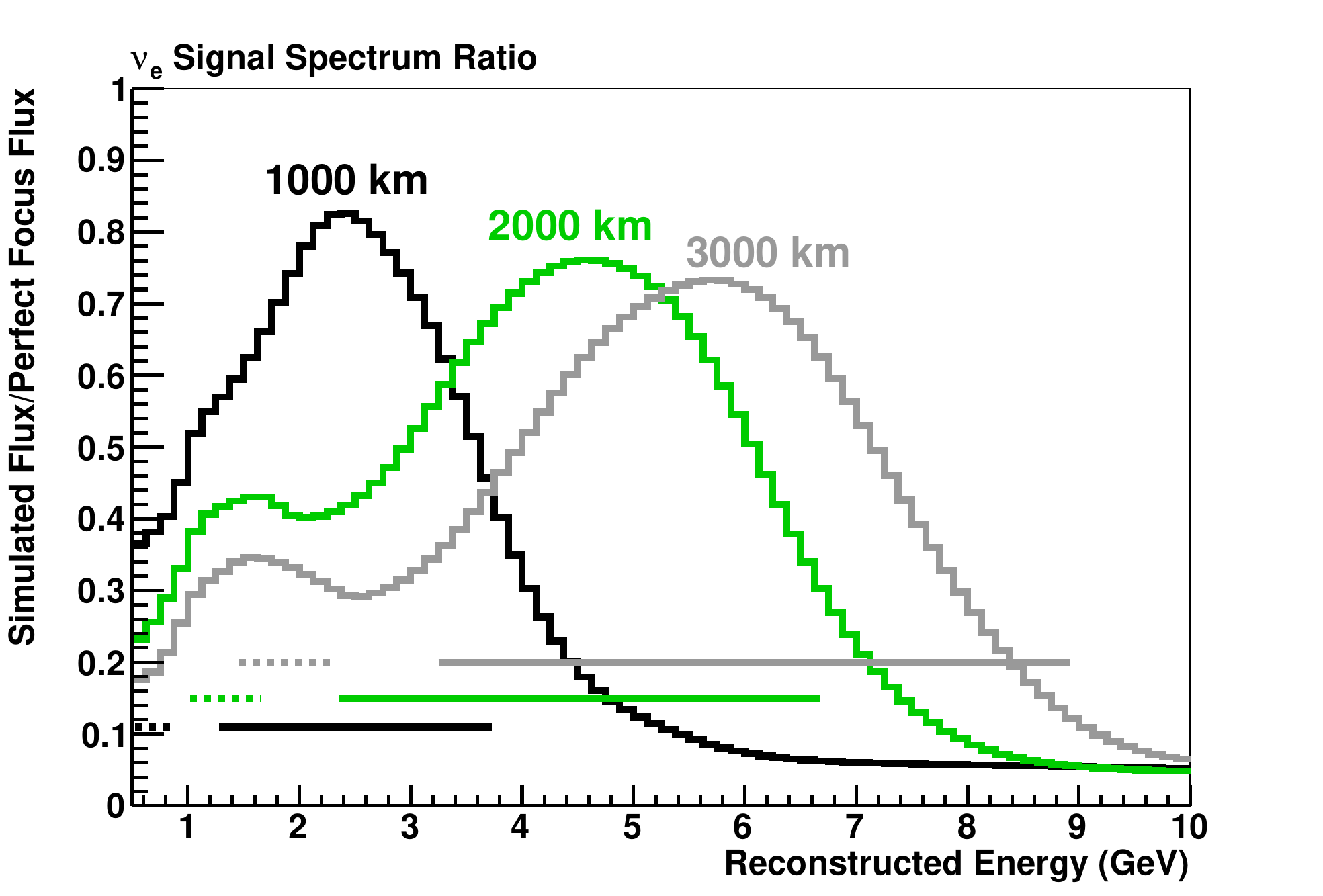}
\caption{Ratio of the $\nu_{e}$ appearance signal spectra used in this study to the appearance spectra obtained assuming perfect hadron focusing for baselines of 1000, 2000, and 3000~km.  The solid horizontal lines show the energy region around the first maximum, where the limits are the energies above and below the first maximum at which the appearance probability is 50\% of the maximum.  The dashed horizontal lines show the energy region around the second maximum, where the limits are the energies above and below the second maximum at which the appearance probability is 10\% of the maximum.}
\label{fig:perfectratio}
\end{figure*}

Figure \ref{fig:perfectratio} shows the ratio of the $\nu_{e}$ appearance signal spectra used in this study (assuming $\delta_{CP}=0$ and normal hierarchy) to the appearance spectra obtained assuming perfect hadron focusing for baselines of 1000, 2000, and 3000~km.  The perfect-focusing fluxes use the same decay pipe lengths for each baseline as given in Table \ref{tab:BLbeams}.  The simulated fluxes achieve up to $\sim$80\% of the perfect-focusing flux in the region of interest, indicated by the horizontal lines for each spectrum; our calculations are based on the achievable and excellent performance characteristics of such beams.

Because conventional neutrino beams (with currently understood technological limits on magnetic field strengths) are not efficient at focusing hadrons with energy less than 1~GeV, we use off-axis beams to generate the flux required for baselines shorter than 1000~km. The off-axis beams are tuned to match the energy range of the first oscillation maximum. 
Studies have indicated that in addition to varying the focusing geometry and off-axis angle, further optimization could be obtained by varying the proton beam energy for the shorter baselines, but the highest power from the Fermilab proton complex is available at an energy of 120~GeV.  As a point of comparison, we considered an on-axis beam with an 8-GeV primary proton beam at 300~km with the same exposure and found that the event rate wasn't better than the off-axis 120-GeV beam (580 signal events and 290 total background events integrated in reconstructed energy over the first maximum, to be compared with numbers in Table \ref{tab:rates_nu}).

While further refinements could be made to the flux optimization, the fluxes used in this study are nearly optimal for each baseline and are realistic representations of what could be achieved with a neutrino beam facility at Fermilab.

\section{Experimental Assumptions}
\label{sec:assumptions}

The signal of muon (anti)neutrino to electron (anti)neutrino oscillations is an excess of $\nu_{e}$ or $\bar{\nu}_e$ charged-current (CC) interactions over background.  $\nu_e$ ($\bar{\nu}_e$) CC events are identified by the $e^{-}$ ($e^{+}$) in the final state.  An irreducible background is caused by $\nu_e$ and $\bar{\nu}_e$ intrinsic to the beam, most of which are created by decays of kaons and muons in the decay region.  Neutral-current (NC) interactions create background when the hadronic shower has an electromagnetic component, often caused by the decay of $\pi^{0}$s.  $\nu_{\mu}$ CC interactions create background when the final state muon is not identified.  Due to $\nu_{\mu} \rightarrow \nu_{\tau}$ oscillations, there is background contribution from $\nu_{\tau}$ CC interactions in which the decay products of the $\tau$ mimic a signal event.  We expect that additional kinematic cuts can be applied to the selected sample to reduce the background from $\nu_{\tau}$ CC interactions without a significant loss of signal events.  In this study we consider two 
cases: the maximum $\nu_{\tau}$ CC background assuming no background reduction is possible and zero $\nu_{\tau}$ CC background assuming it can be completely eliminated with no reduction in signal.  The $\nu_{\tau}$  CC background is most important for the longer baseline configurations ($>$1500~km), in which a significant portion of the neutrino flux has energy above the $\tau$ production threshold.

In the neutrino-beam mode, there is a small background from wrong-sign (antineutrino) contamination in the beam, which we consider negligible.  However, in the antineutrino-beam mode, the wrong-sign (neutrino) contamination is much more substantial, and is therefore included in this study.

\begin{table*}[htp]
\caption{Detector performance parameters related to the identification of $\nu_e$ CC events.}
\begin{center}
\begin{tabular}{|l|c|} \hline
{\bf Parameter} &   {\bf Value}   \\ \hline\hline
$\nu_e$ CC efficiency           & 80\%   \\ \hline
NC mis-identification rate     & 1\%   \\ \hline
$\nu_\mu$ CC mis-identification rate       & 1\%   \\ \hline
$\nu_\tau$ CC mis-identification rate & $\sim$20\% \\ \hline
Other background                                   & 0\% \\ \hline
$\nu_e$ CC energy resolution &  $15\%/\sqrt{E(GeV)}$ \\ \hline
\end{tabular}
\label{tab:det}
\end{center}
\end{table*}

As a reference, we use a liquid argon (LAr) TPC with an exposure of 350~\mbox{kt-MW-yr} (which roughly corresponds to a 6-year exposure of a 50-kt detector in a 1.2-MW beam). Our results, however, can be easily extrapolated to other combinations of detector size and beam intensity.  Parameters describing the selection efficiency and detector energy response were input into the GLoBES software package \cite{Huber:2004ka,Huber:2007ji} to calculate and analyze the predicted spectra. The detector performance parameters used for the study are shown in Table \ref{tab:det}.  Most of these parameters are derived from studies of LAr TPC simulations and studies of the ICARUS detector performance \cite{Ankowski:2008aa,Amoruso:2003sw,Ankowski:2006ts,t2k2km}.  The NC true-to-visible energy conversion is based on a fast MC simulation developed for LBNE \cite{Adams:2013qkq,fastMC}, a planned experiment which will use a muon neutrino beam to study electron neutrino appearance.  The MC uses flux simulations (derived from GEANT3 beamline simulations as previously described) and the GENIE event generator \cite{Andreopoulos:2009rq} to generate neutrino interactions on argon.  Events can be classified as $\nu_e$ CC-like based on event-level reconstructed quantities.  The $\nu_{\tau}$ CC background energy-dependent mis-identification rate and true-to-visible energy conversion is also calculated based on the fast MC. 

The oscillation parameter values and uncertainties assumed in this study are: $\theta_{12} = 34 \pm 1^{\circ}$, $\theta_{13} = 9 \pm 1^{\circ}$, $\theta_{23} = 38 \pm 3^{\circ}$ ($\theta_{23} = 52 \pm 3^{\circ}$ when the second octant solution is considered), $\Delta m^2_{21} = (7.54 \pm 0.24)\times 10^{-5}\textnormal{~eV}^2$ and $|\Delta m^2_{31}| = (2.47 \pm 0.08)\times 10^{-3}\textnormal{~eV}^2$, mostly based on the global fit in \cite{Fogli:2012ua}. The uncertainty on $\theta_{13}$ comes from the systematic uncertainty given in \cite{An:2012bu}, on the assumption that the statistical uncertainty on $\theta_{13}$ will be negligible within a few years.  The oscillation probability calculation in GLoBES is exact, and matter effects are incorporated in GLoBES assuming a constant matter density equal to the average matter density from the PREM \cite{matterdensity1,matterdensity2} onion shell model of the earth. Using the PREM matter profile built into GLoBES rather than the average matter density produces a negligible change in the oscillation signal rates (at most 1\%).

\begin{figure*}[htp]
\includegraphics[width=0.32\textwidth]{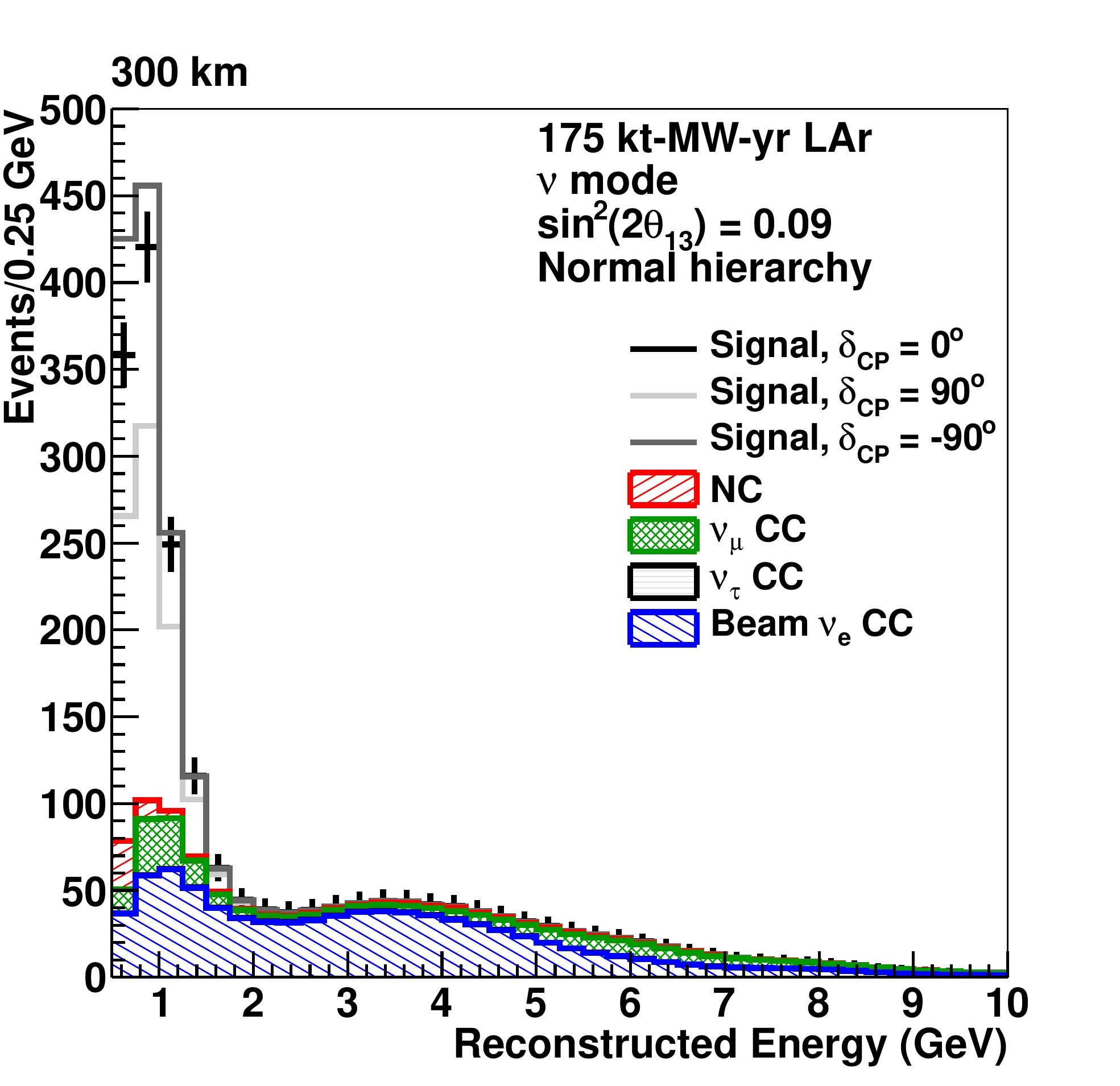}
\includegraphics[width=0.32\textwidth]{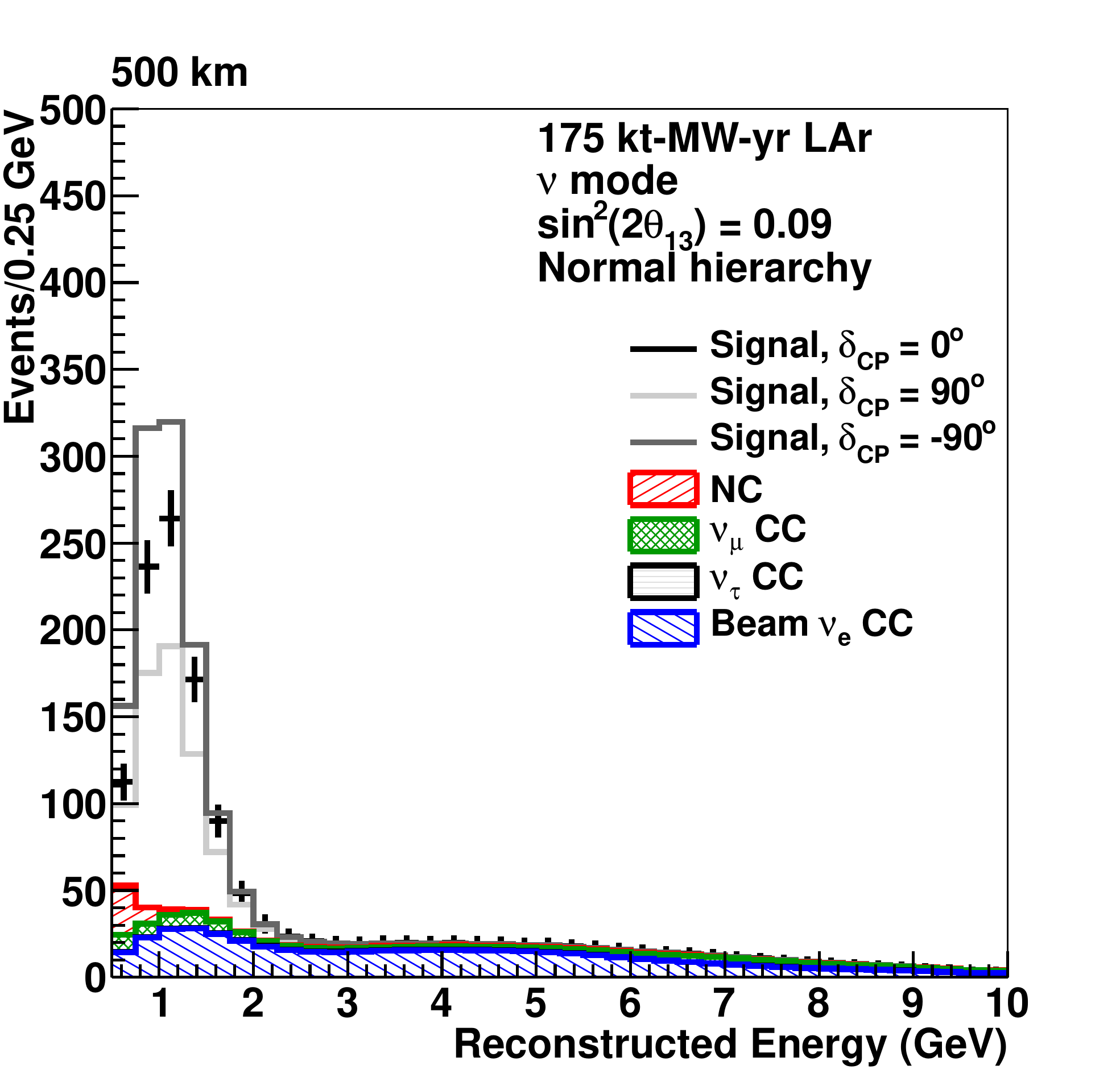}
\includegraphics[width=0.32\textwidth]{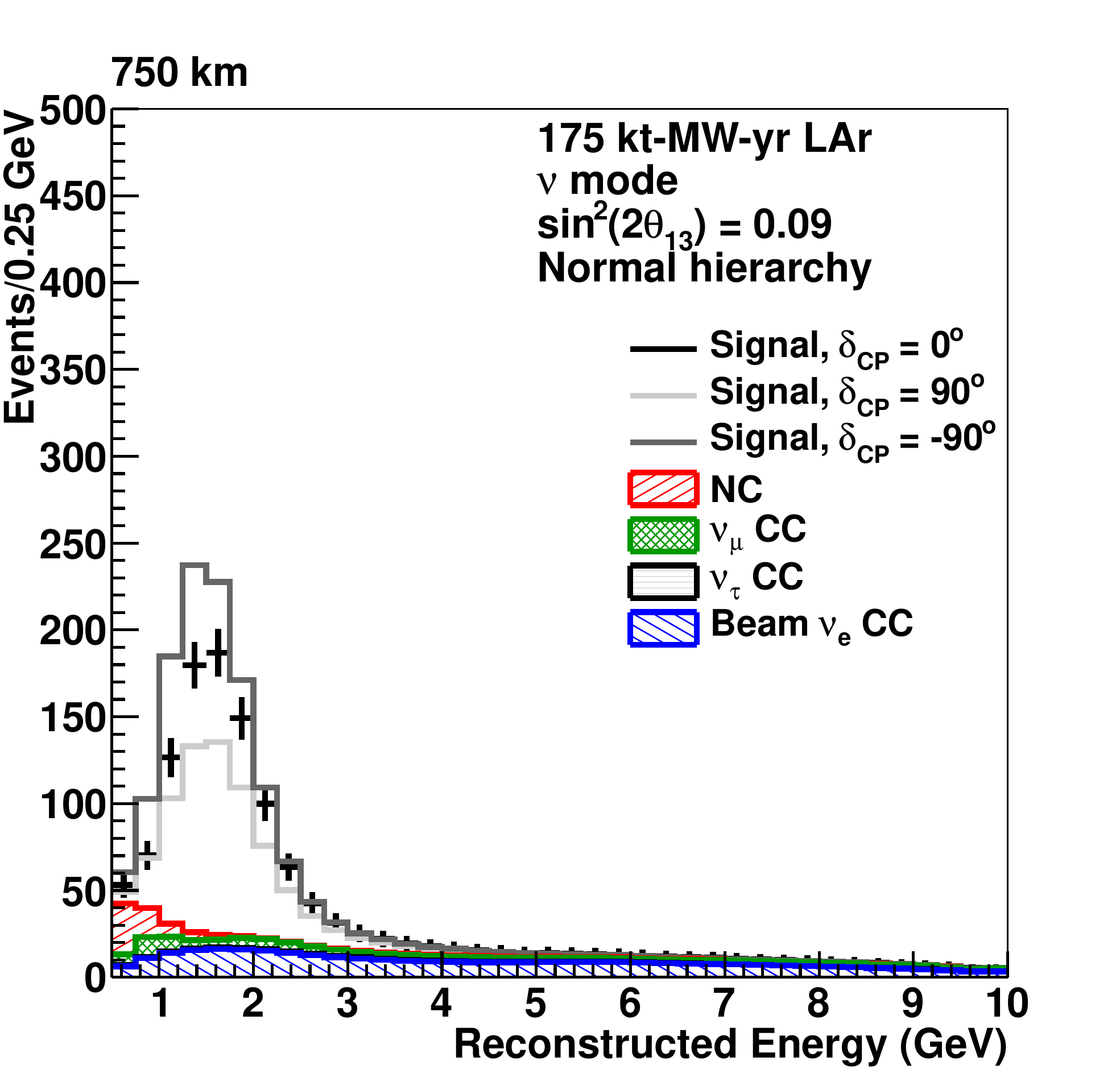}
\includegraphics[width=0.32\textwidth]{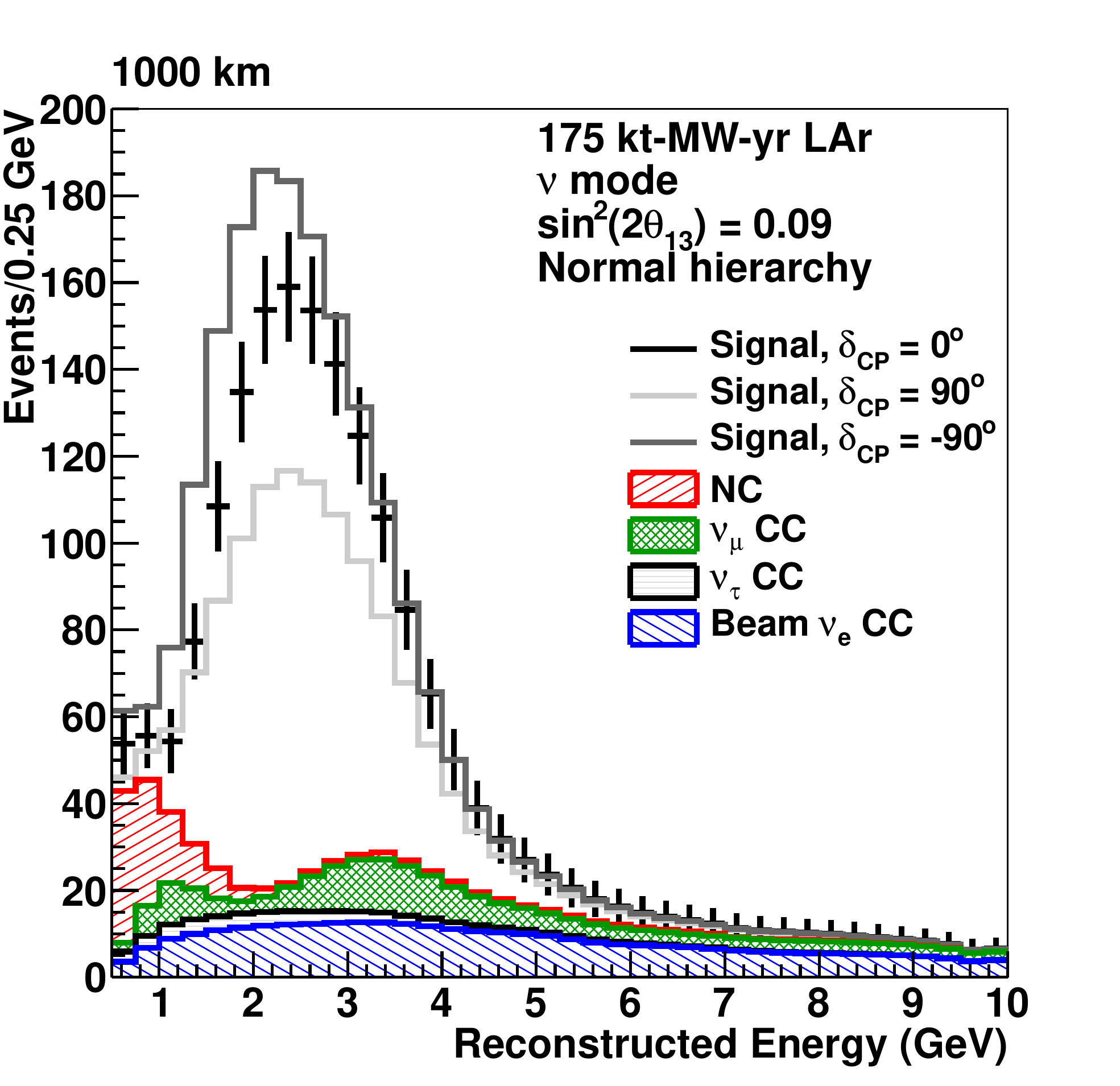}
\includegraphics[width=0.32\textwidth]{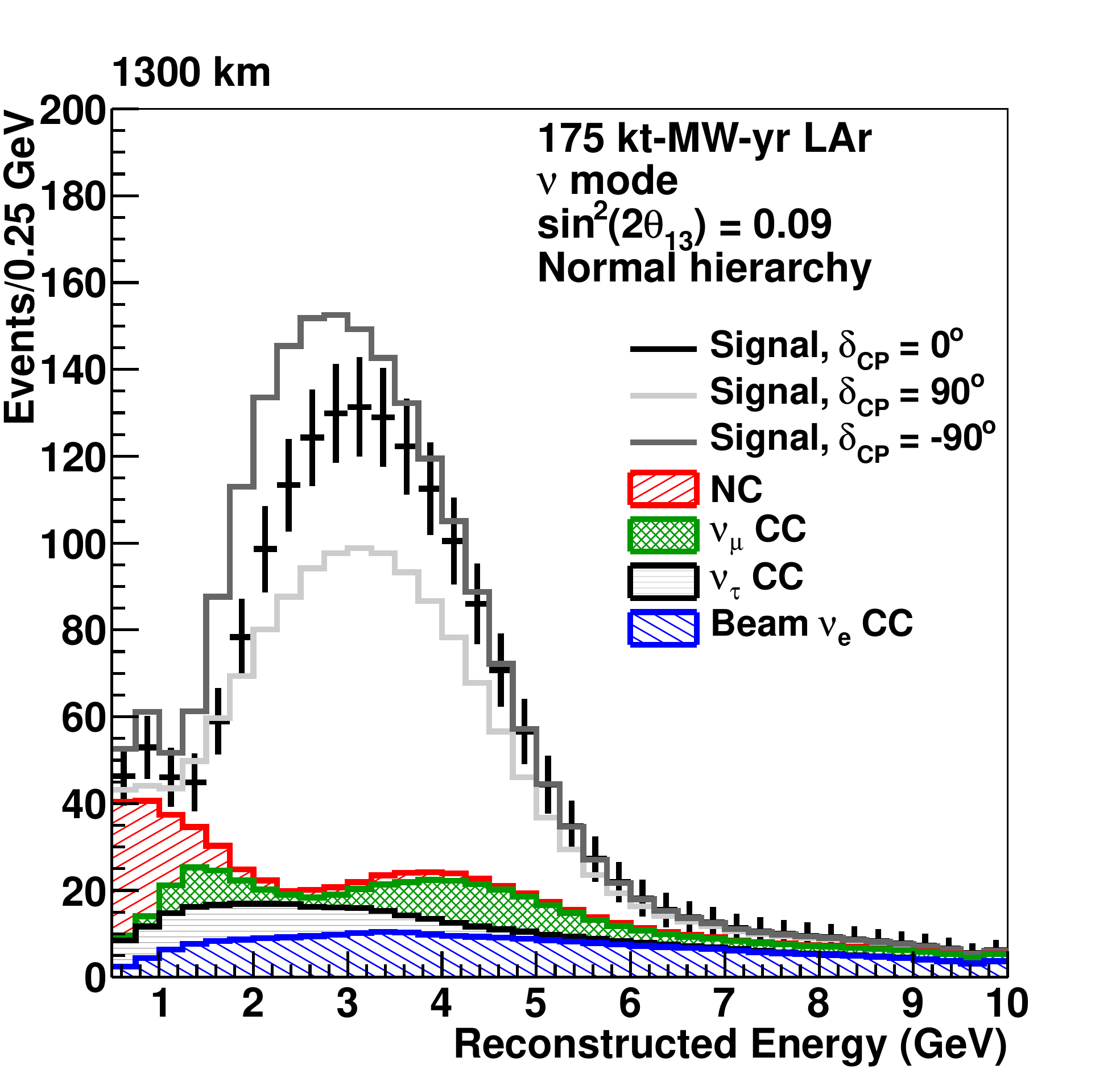}
\includegraphics[width=0.32\textwidth]{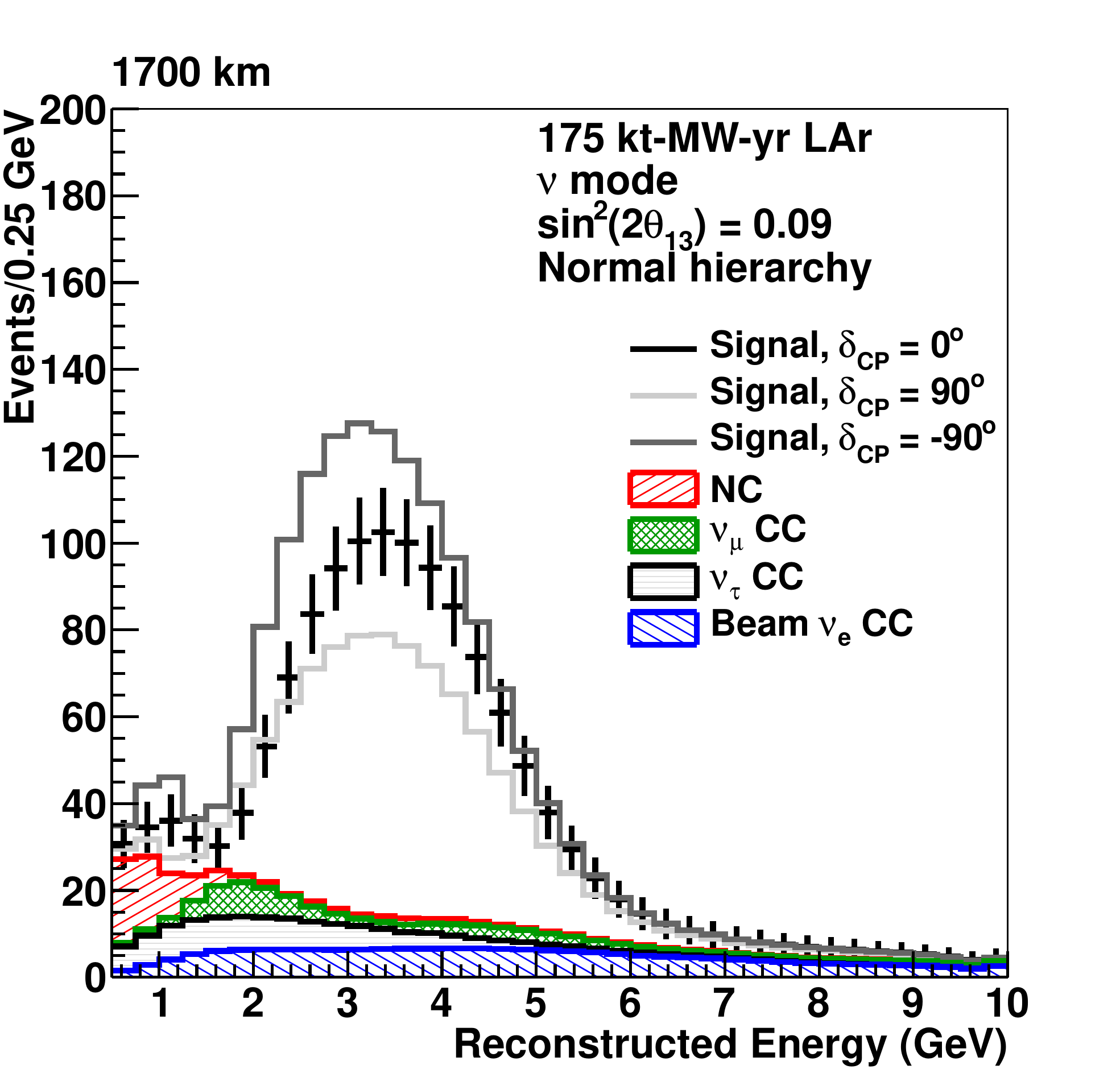}
\includegraphics[width=0.32\textwidth]{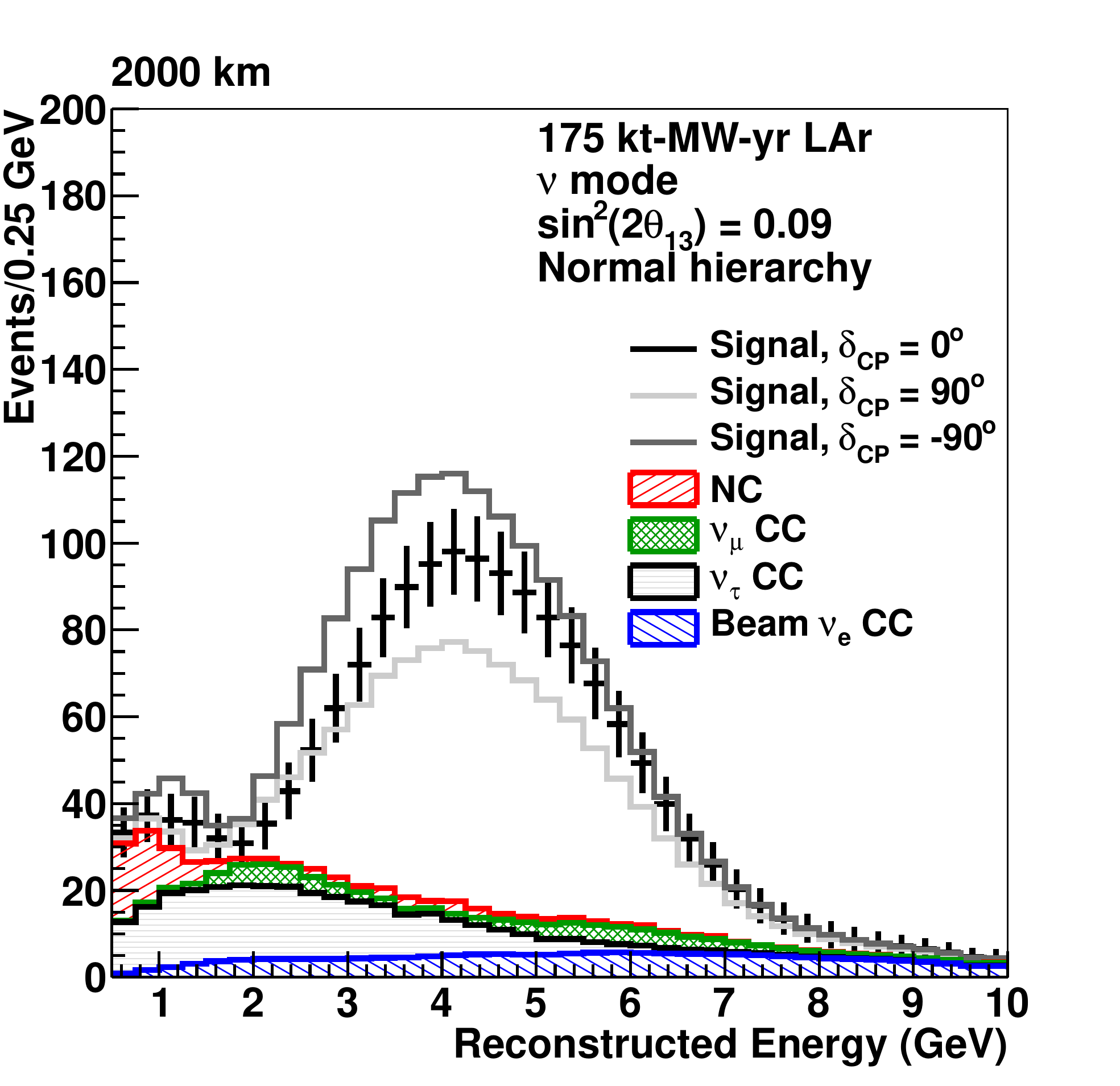}
\includegraphics[width=0.32\textwidth]{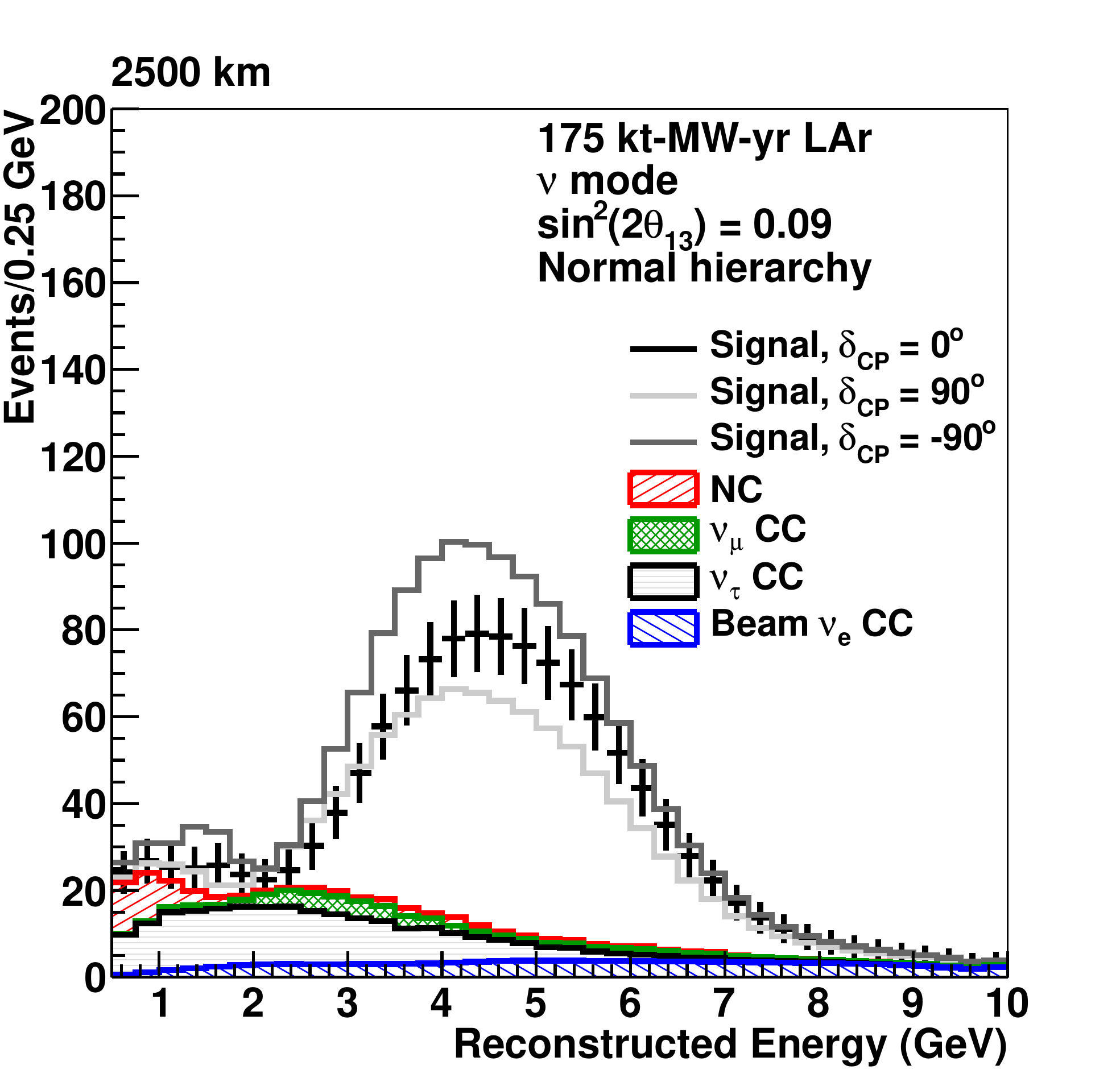}
\includegraphics[width=0.32\textwidth]{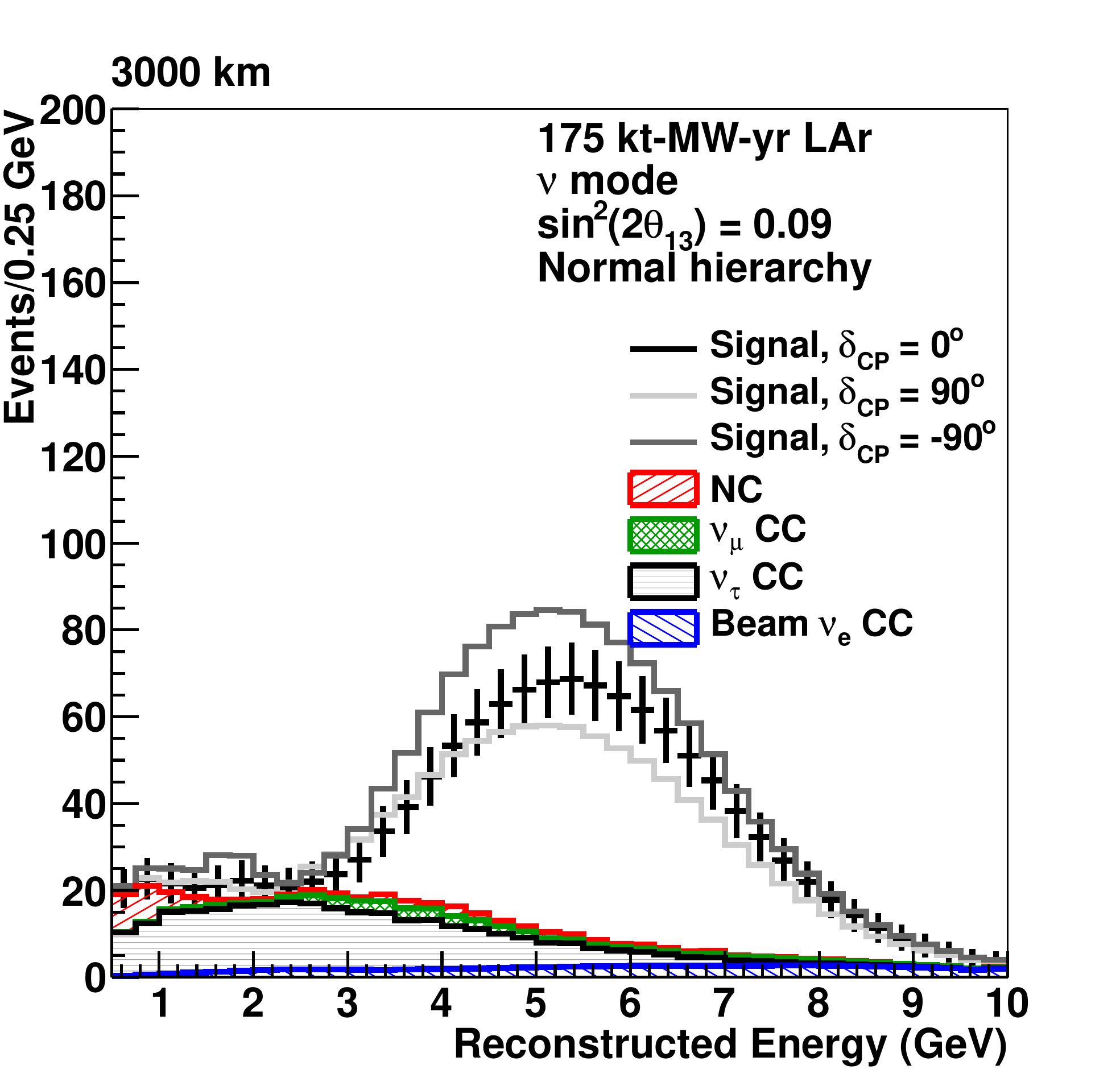}
\caption{Neutrino spectra for normal hierarchy: Reconstructed energy distribution of selected $\nu_e$ CC-like events assuming a 175~\mbox{kt-MW-yr} exposure in the neutrino-beam mode at each baseline.  The plots assume normal mass hierarchy.  The signal contribution is shown for various values of $\delta_{CP}$.}
\label{fig:nuspectra_nh}
\end{figure*}

\begin{figure*}[htp]
\includegraphics[width=0.32\textwidth]{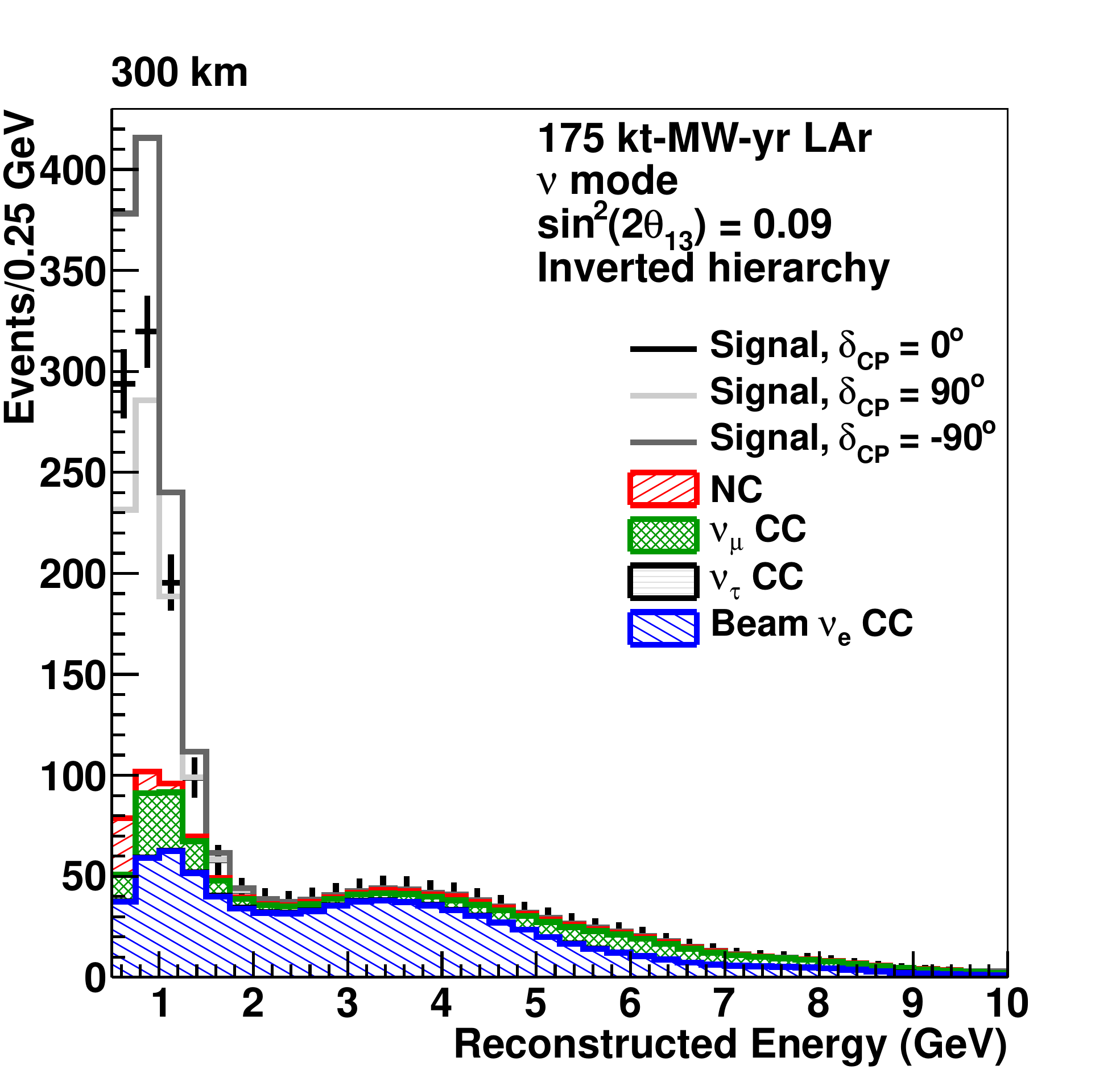}
\includegraphics[width=0.32\textwidth]{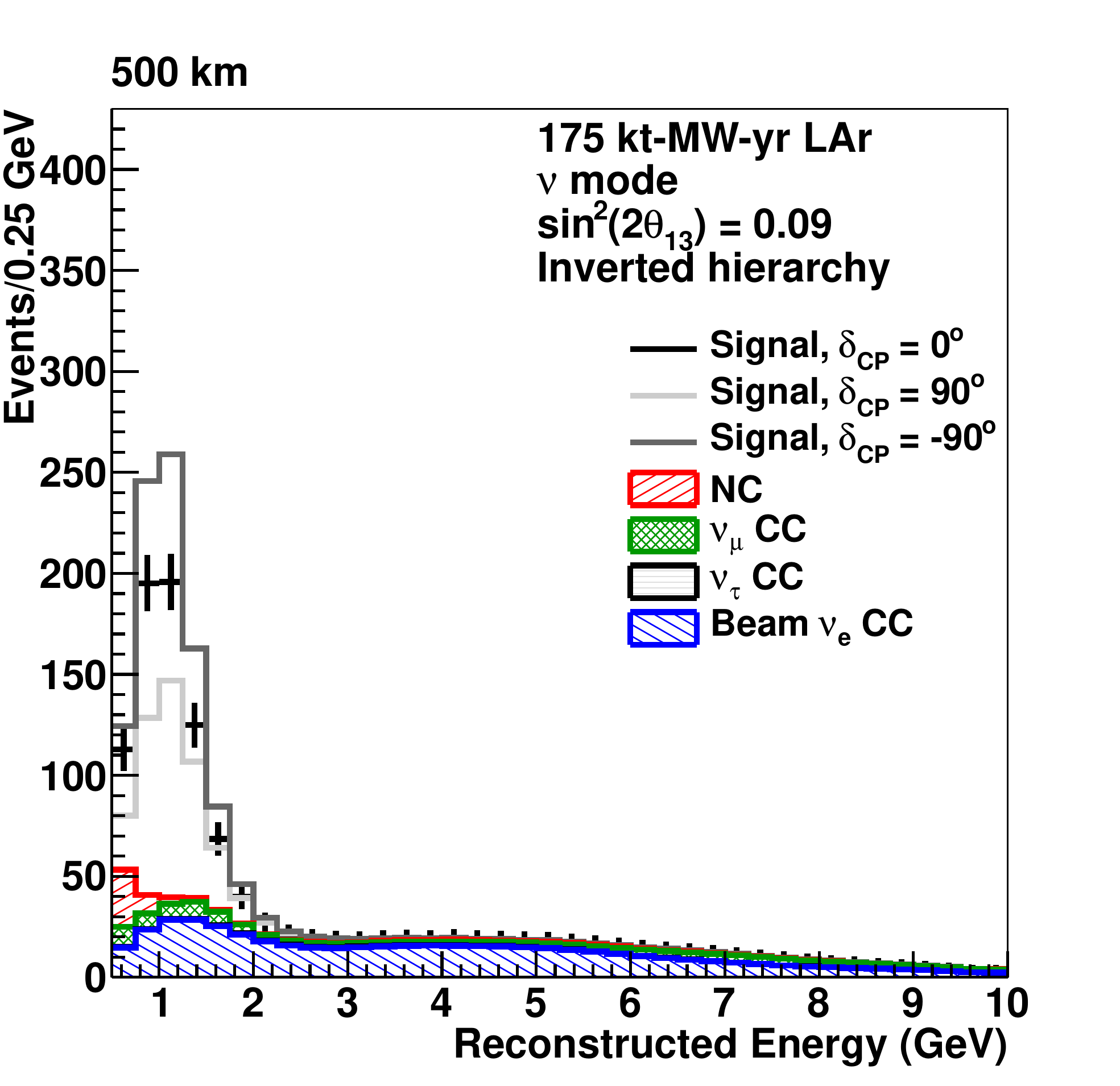}
\includegraphics[width=0.32\textwidth]{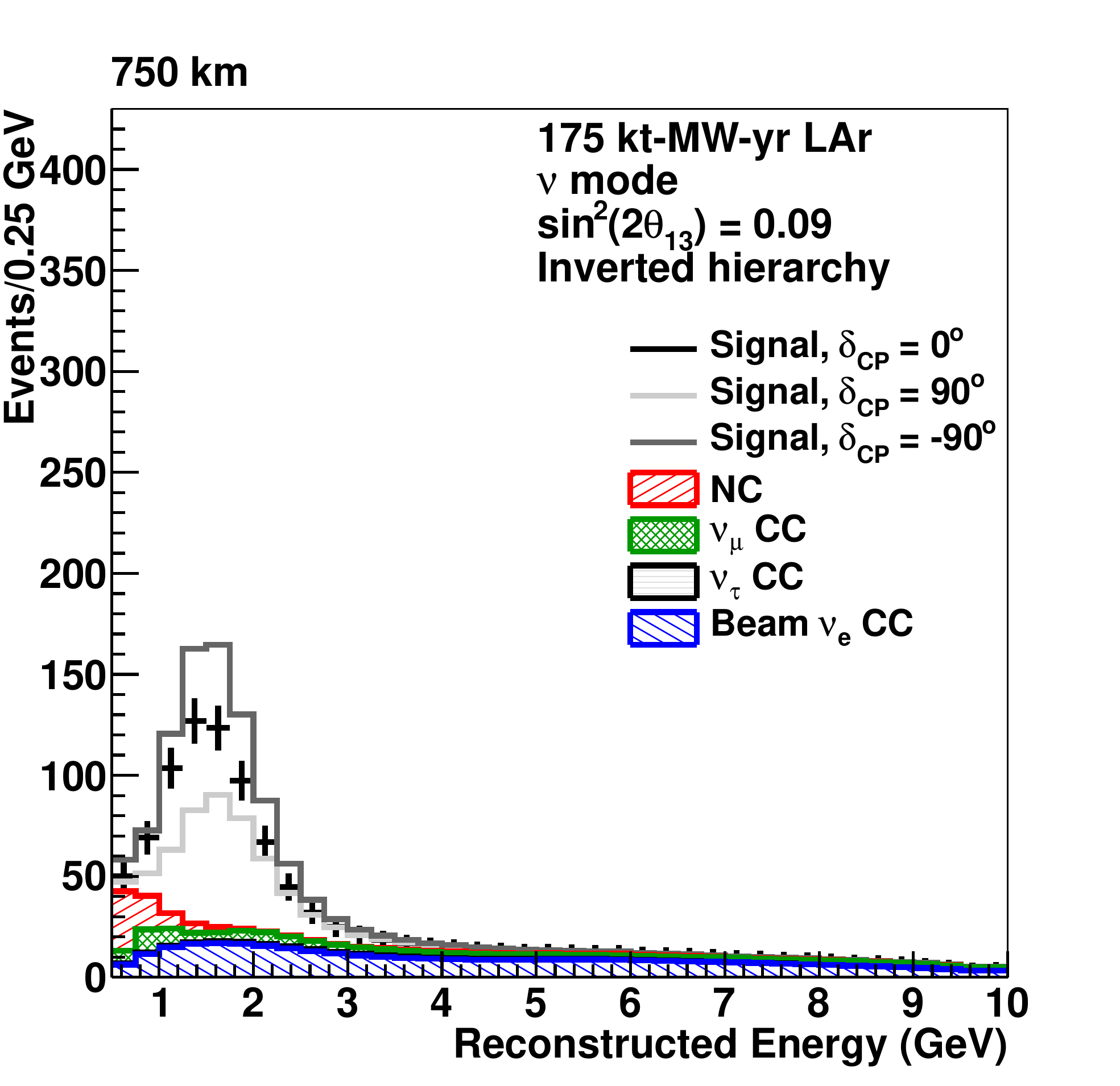}
\includegraphics[width=0.32\textwidth]{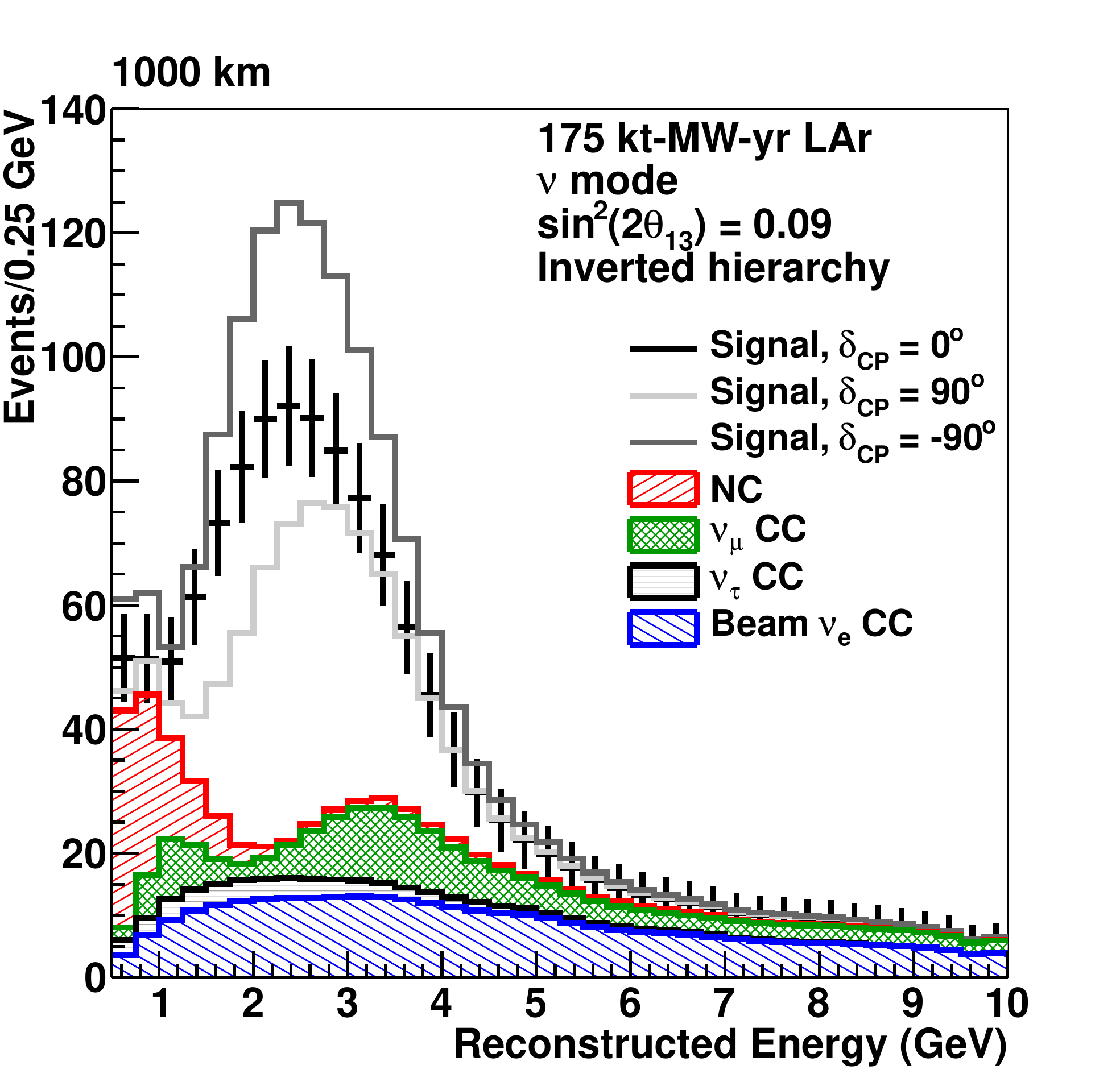}
\includegraphics[width=0.32\textwidth]{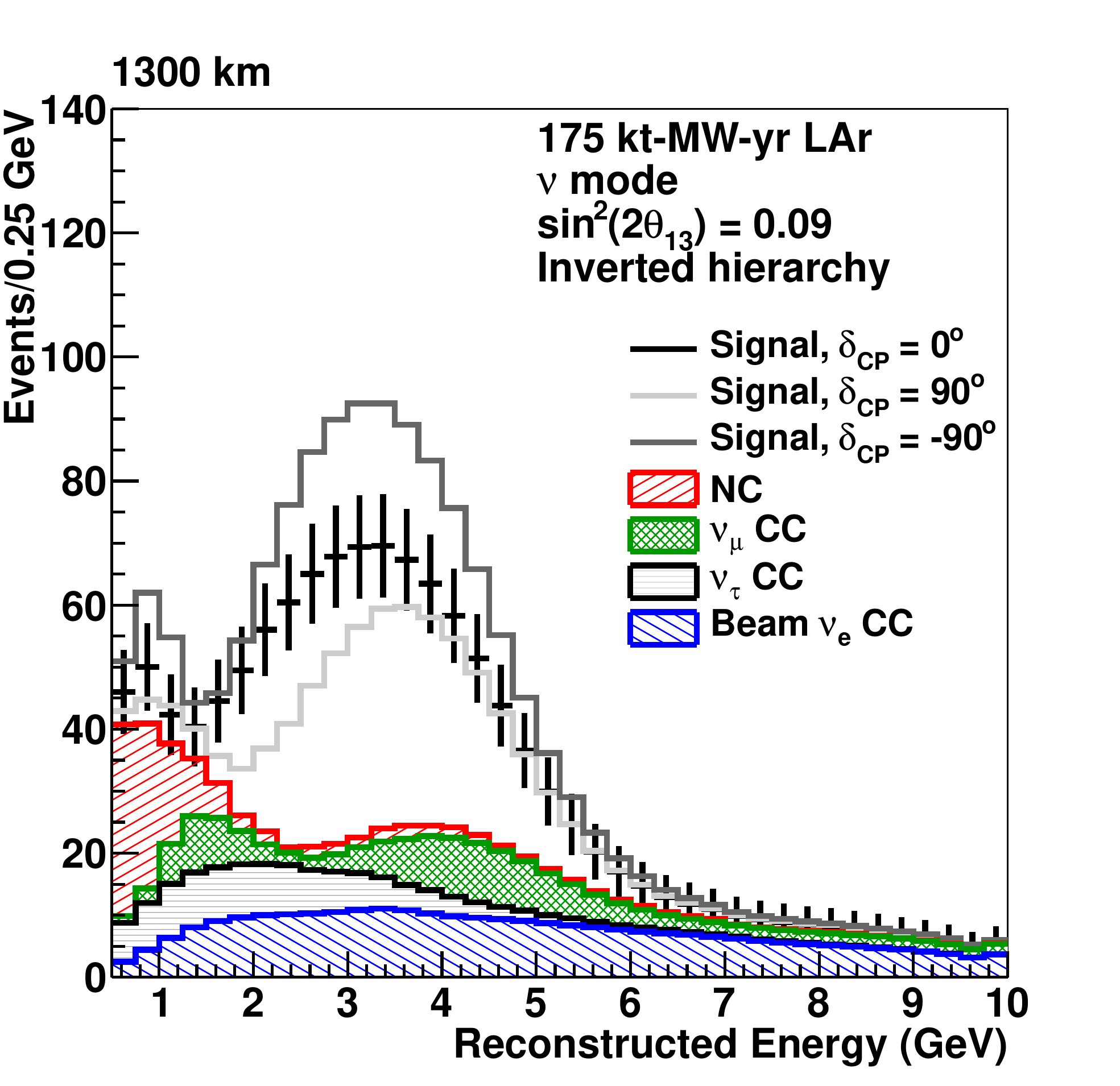}
\includegraphics[width=0.32\textwidth]{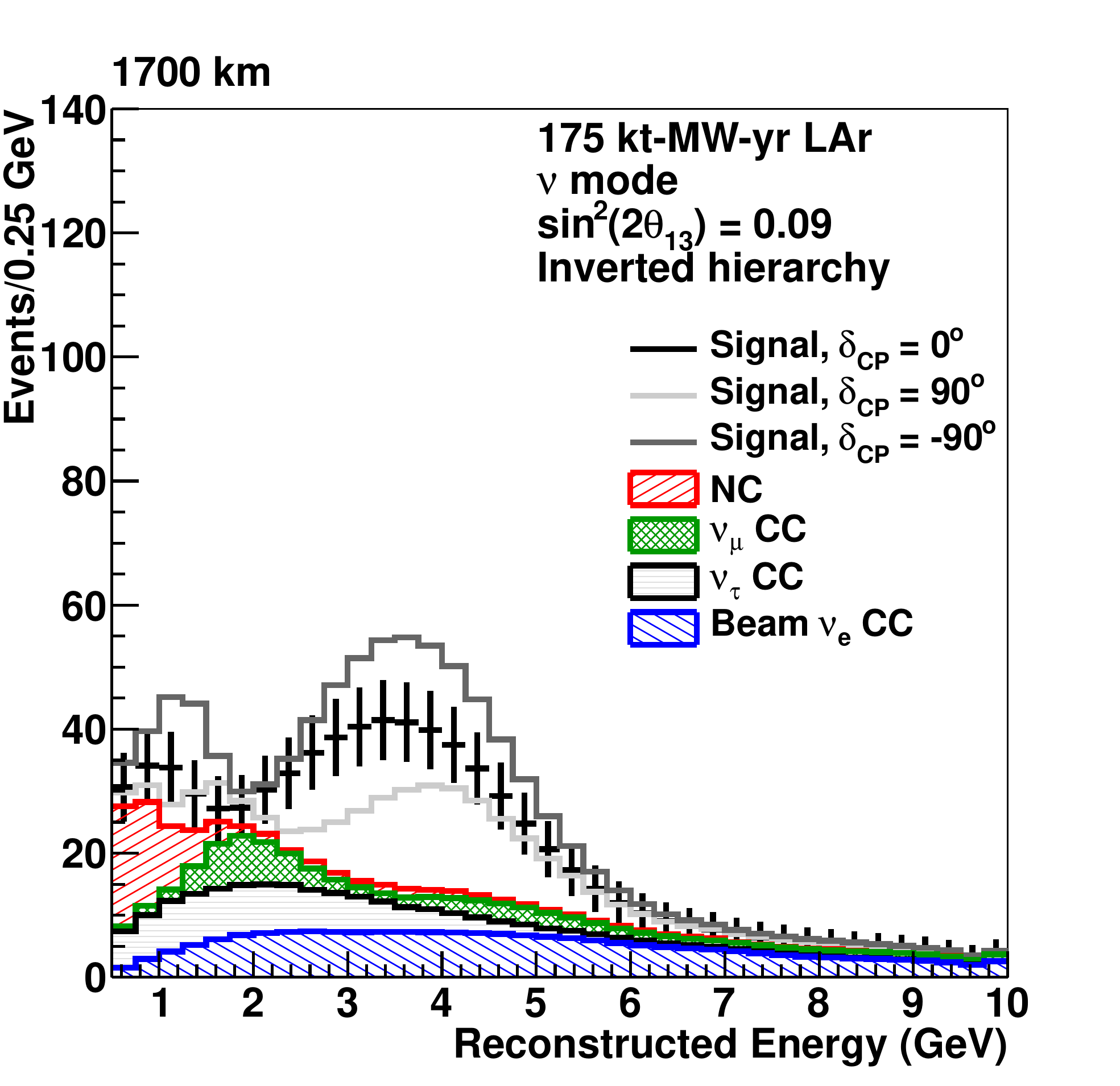}
\includegraphics[width=0.32\textwidth]{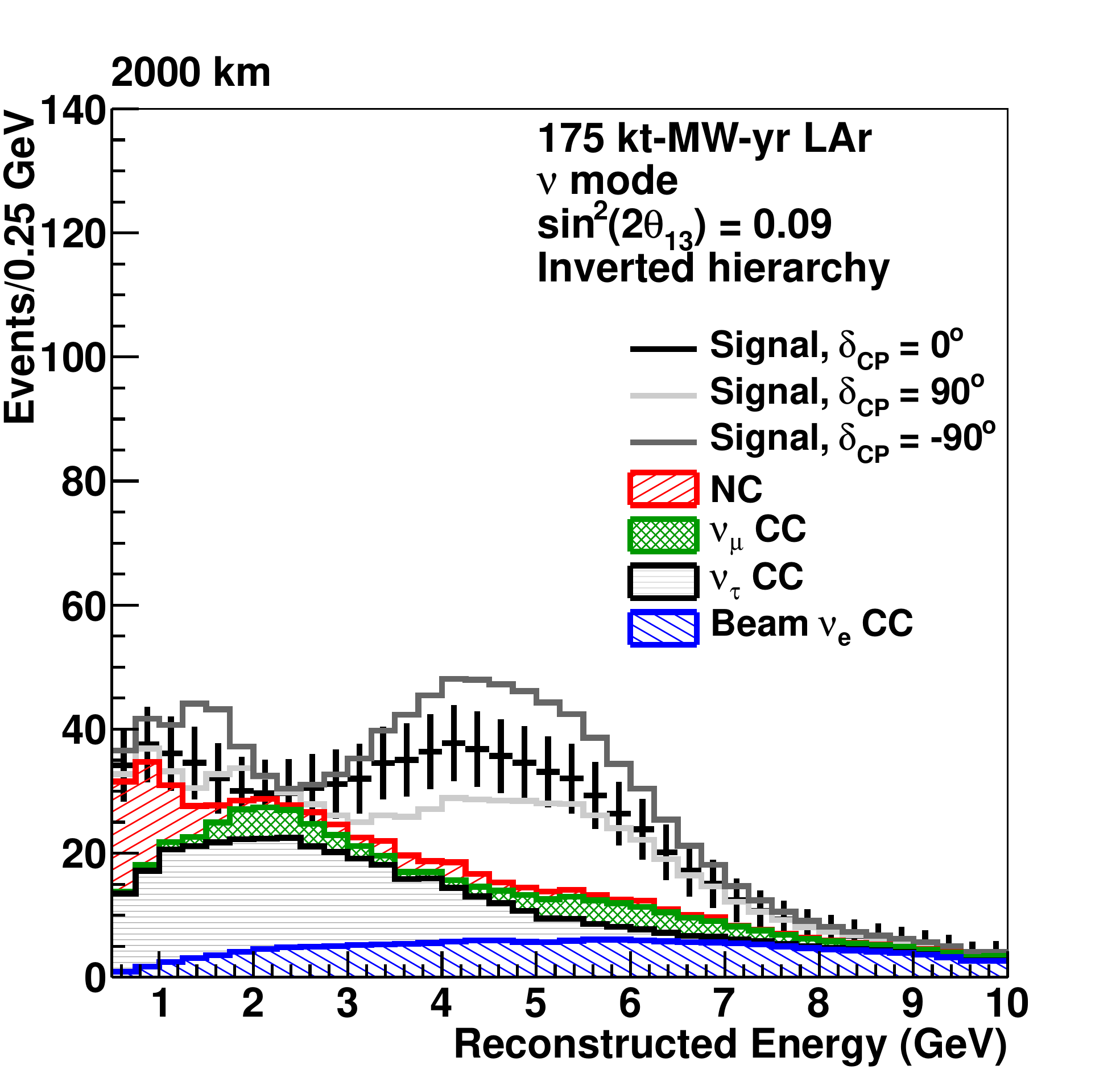}
\includegraphics[width=0.32\textwidth]{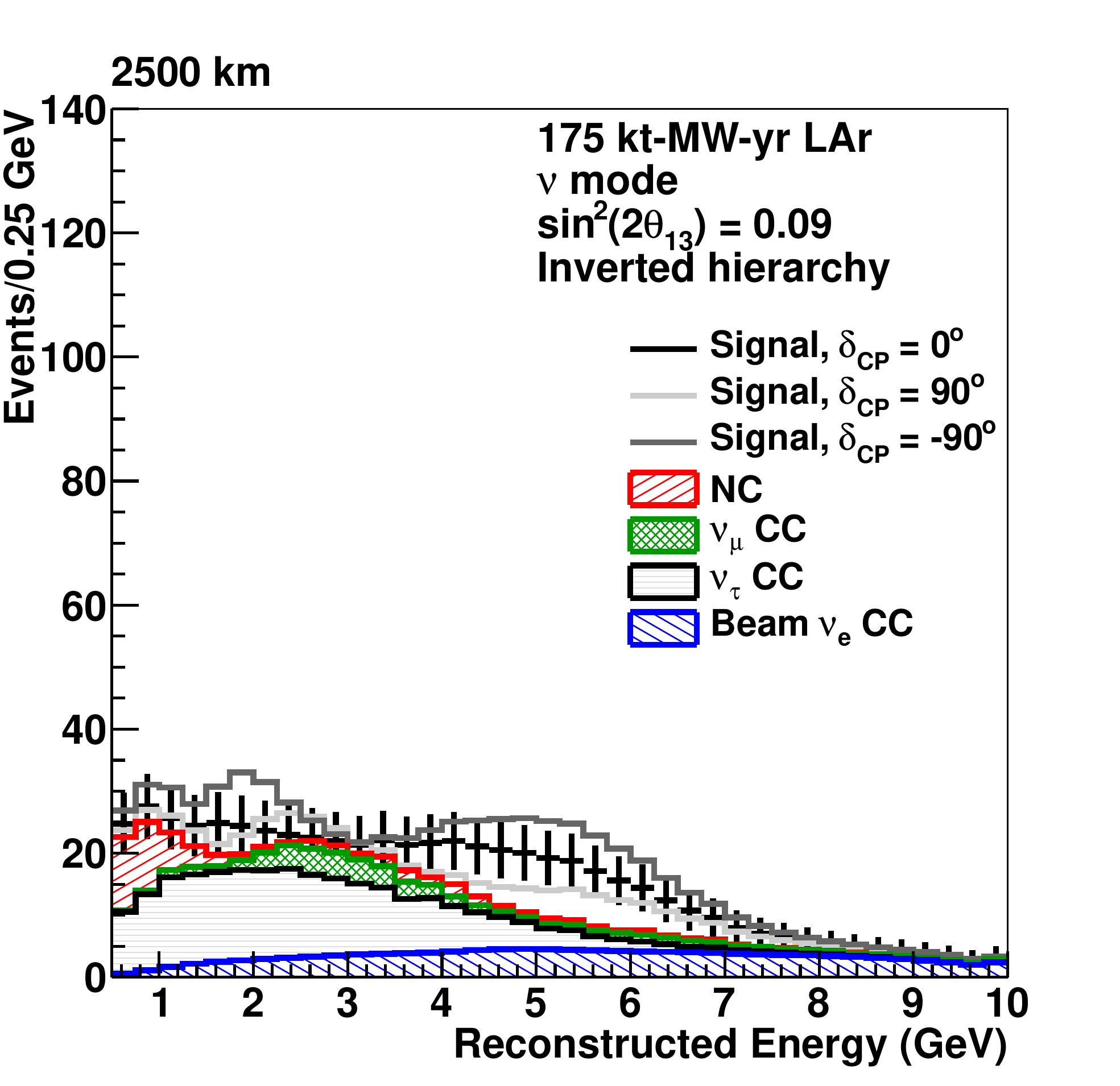}
\includegraphics[width=0.32\textwidth]{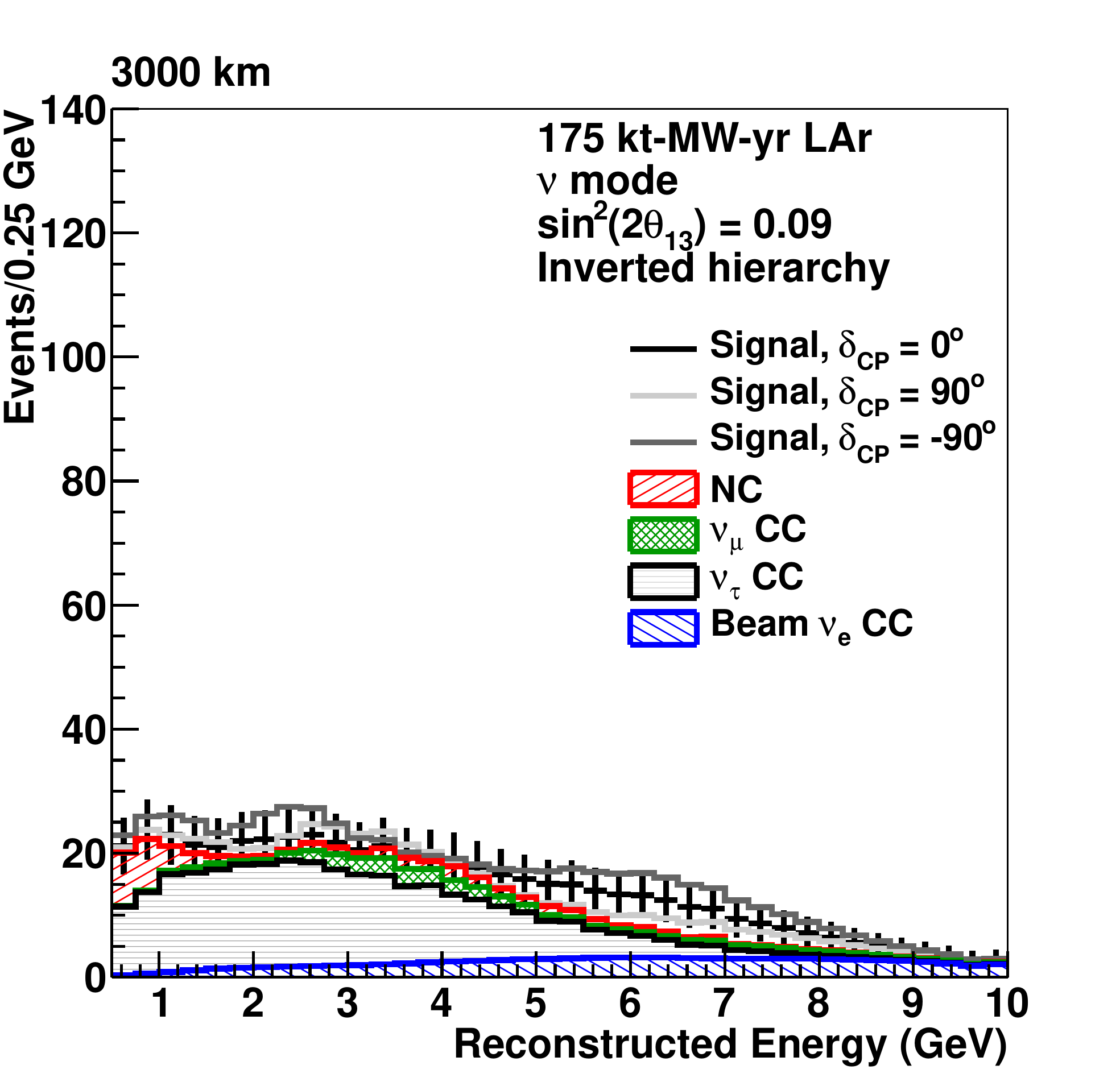}
\caption{Neutrino spectra for inverted hierarchy: Reconstructed energy distribution of selected $\nu_e$ CC-like events assuming a 175~\mbox{kt-MW-yr} exposure in the neutrino-beam mode at each baseline.  The plots assume inverted mass hierarchy.  The signal contribution is shown for various values of $\delta_{CP}$.}
\label{fig:nuspectra_ih}
\end{figure*}

\begin{figure*}[htp]
\includegraphics[width=0.32\textwidth]{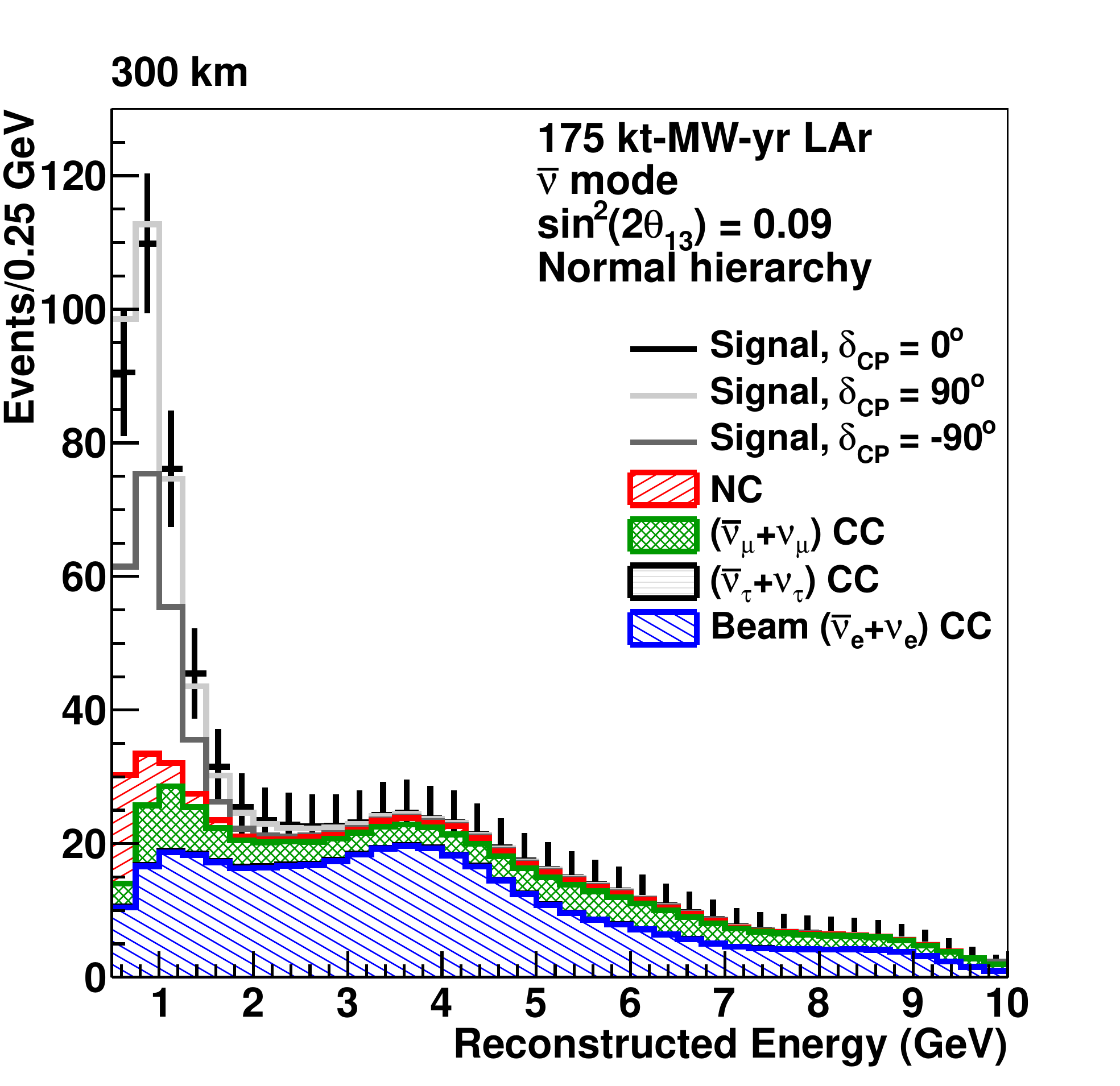}
\includegraphics[width=0.32\textwidth]{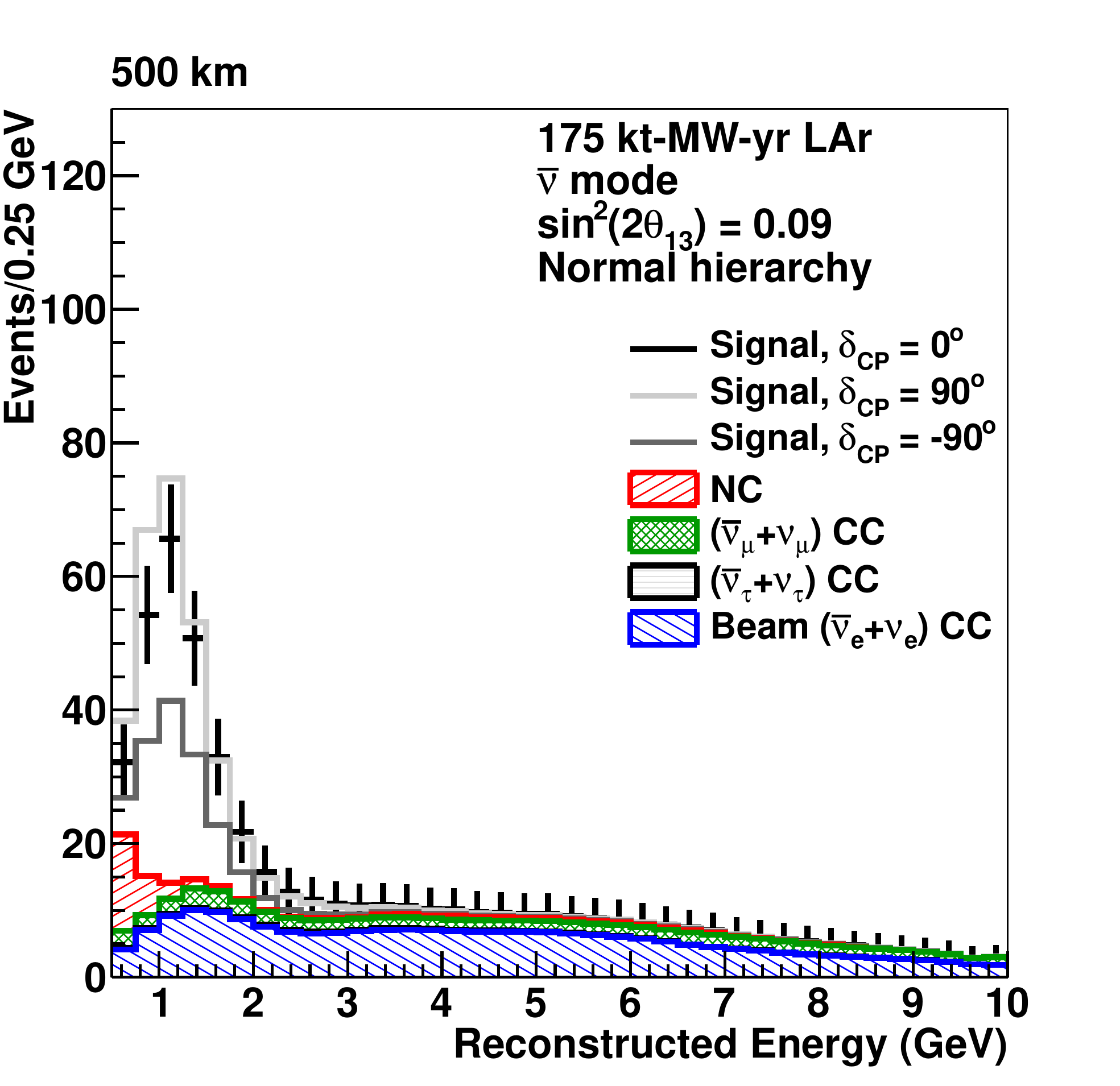}
\includegraphics[width=0.32\textwidth]{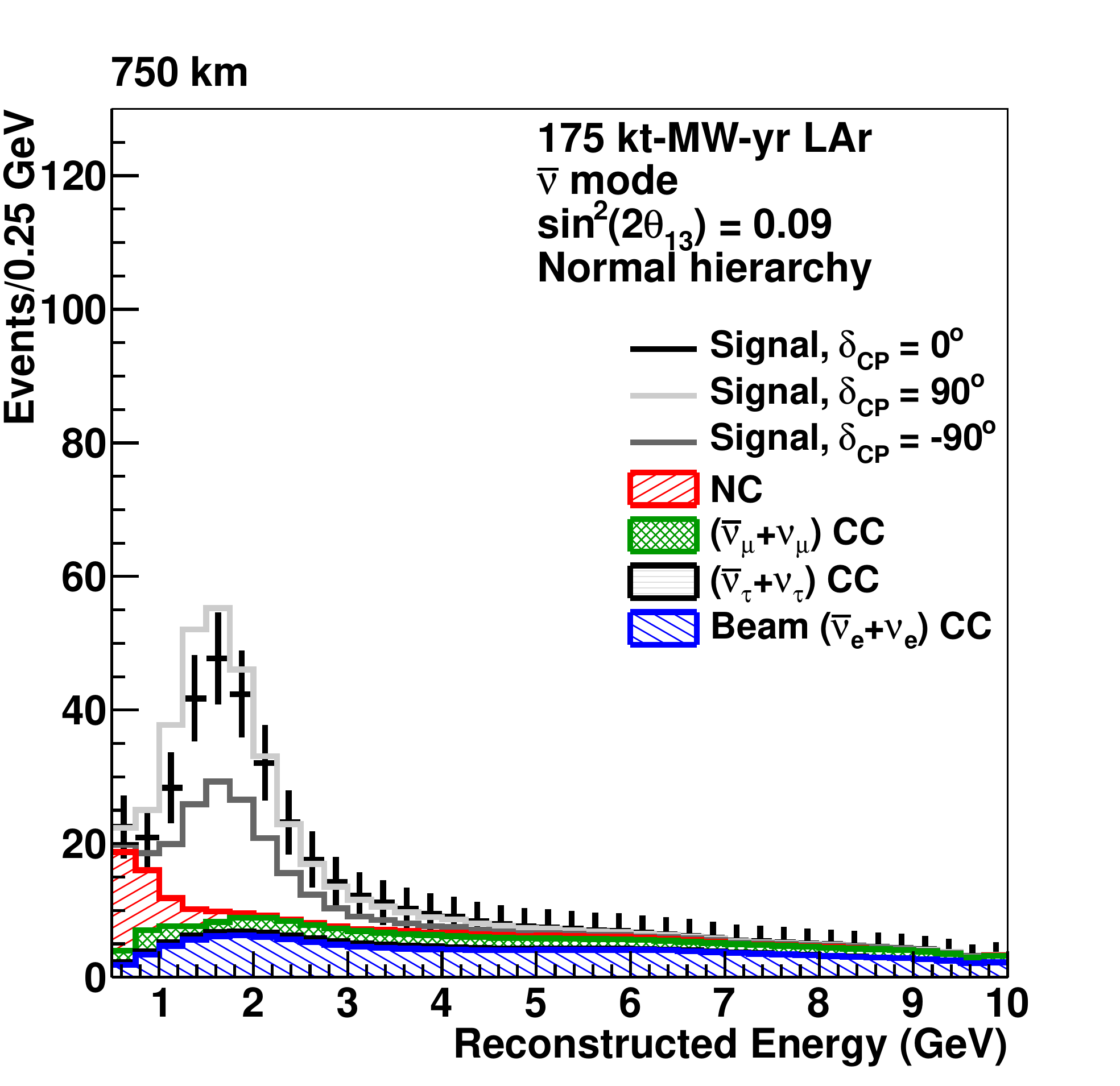}
\includegraphics[width=0.32\textwidth]{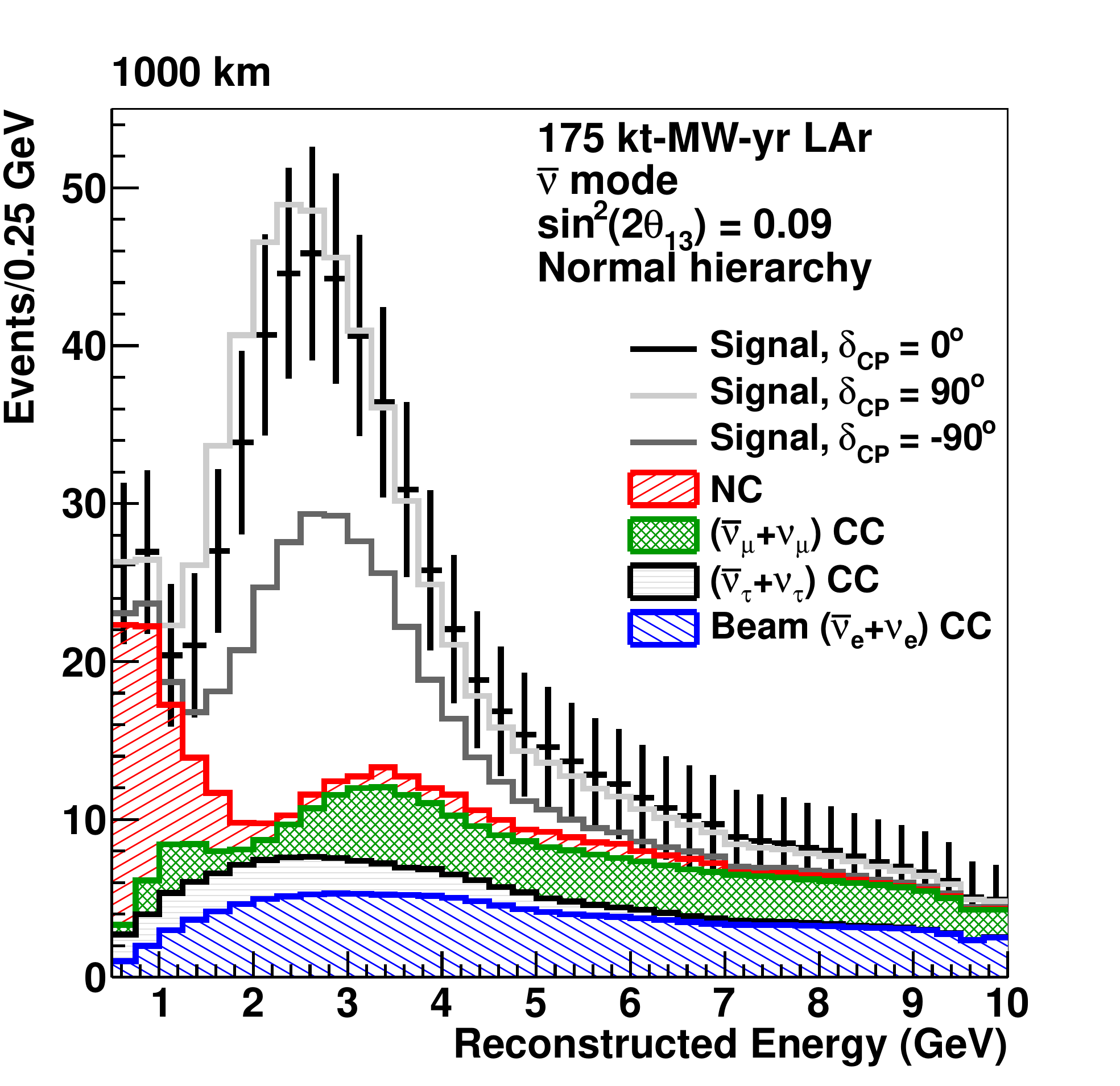}
\includegraphics[width=0.32\textwidth]{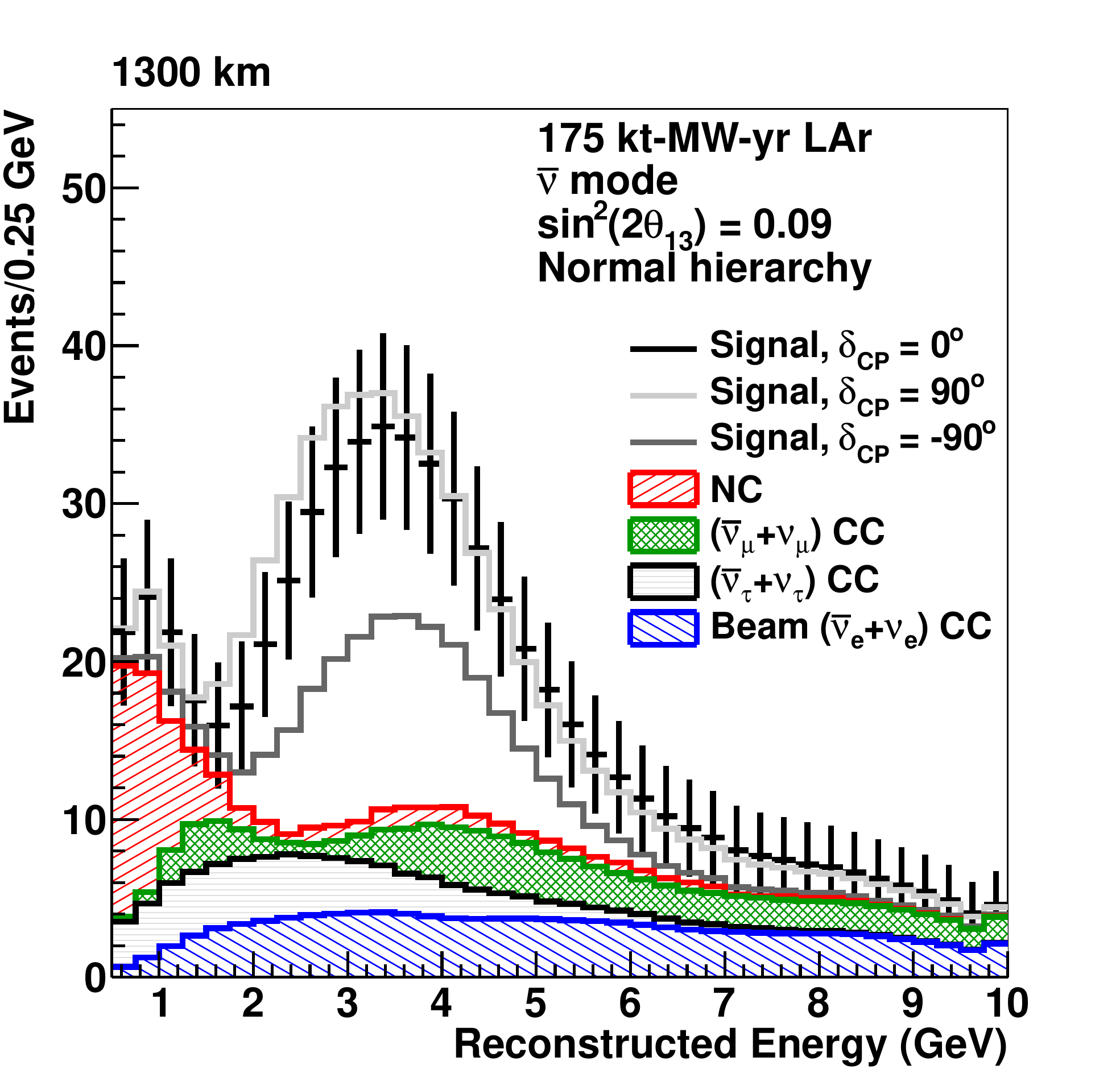}
\includegraphics[width=0.32\textwidth]{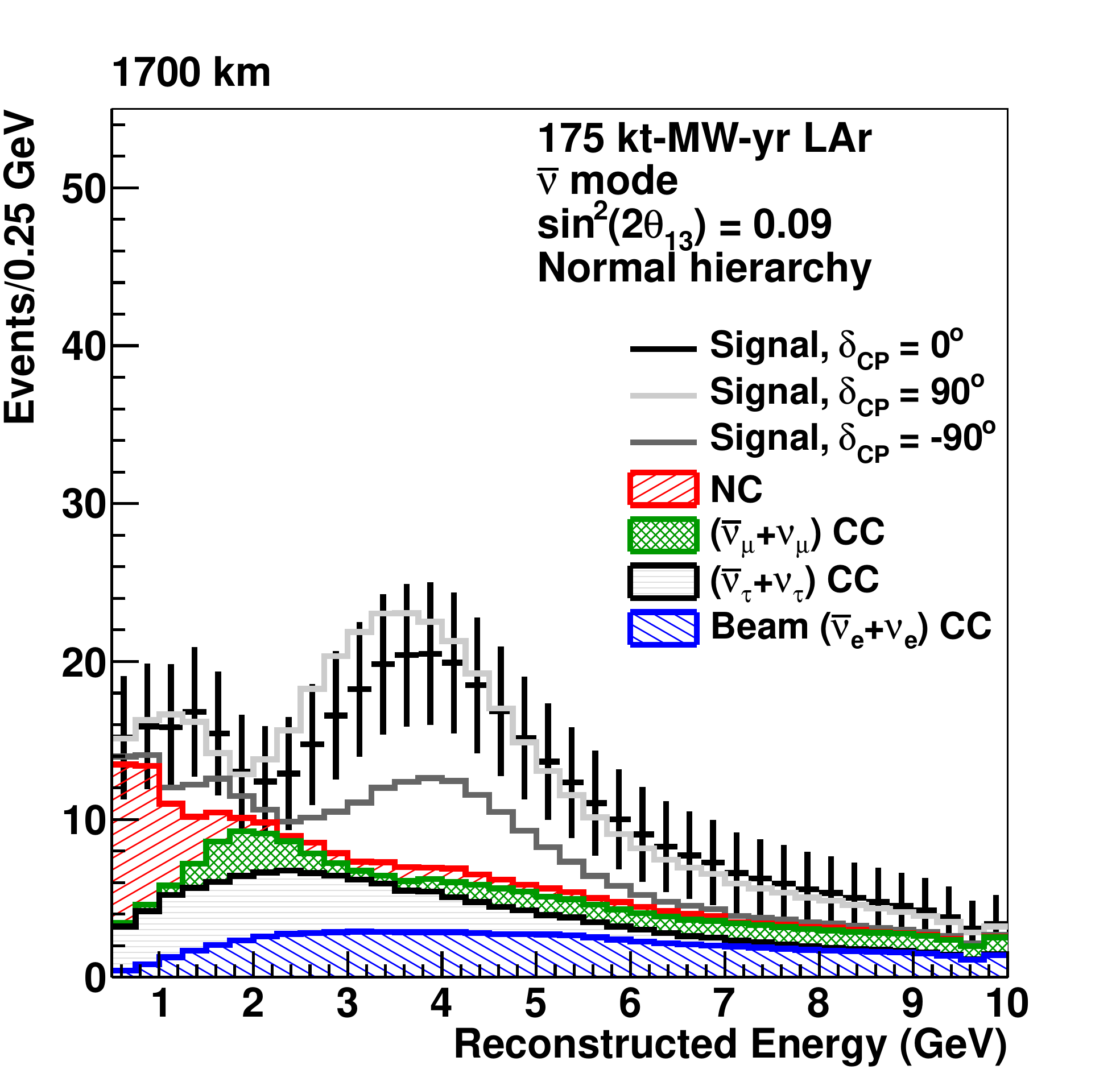}
\includegraphics[width=0.32\textwidth]{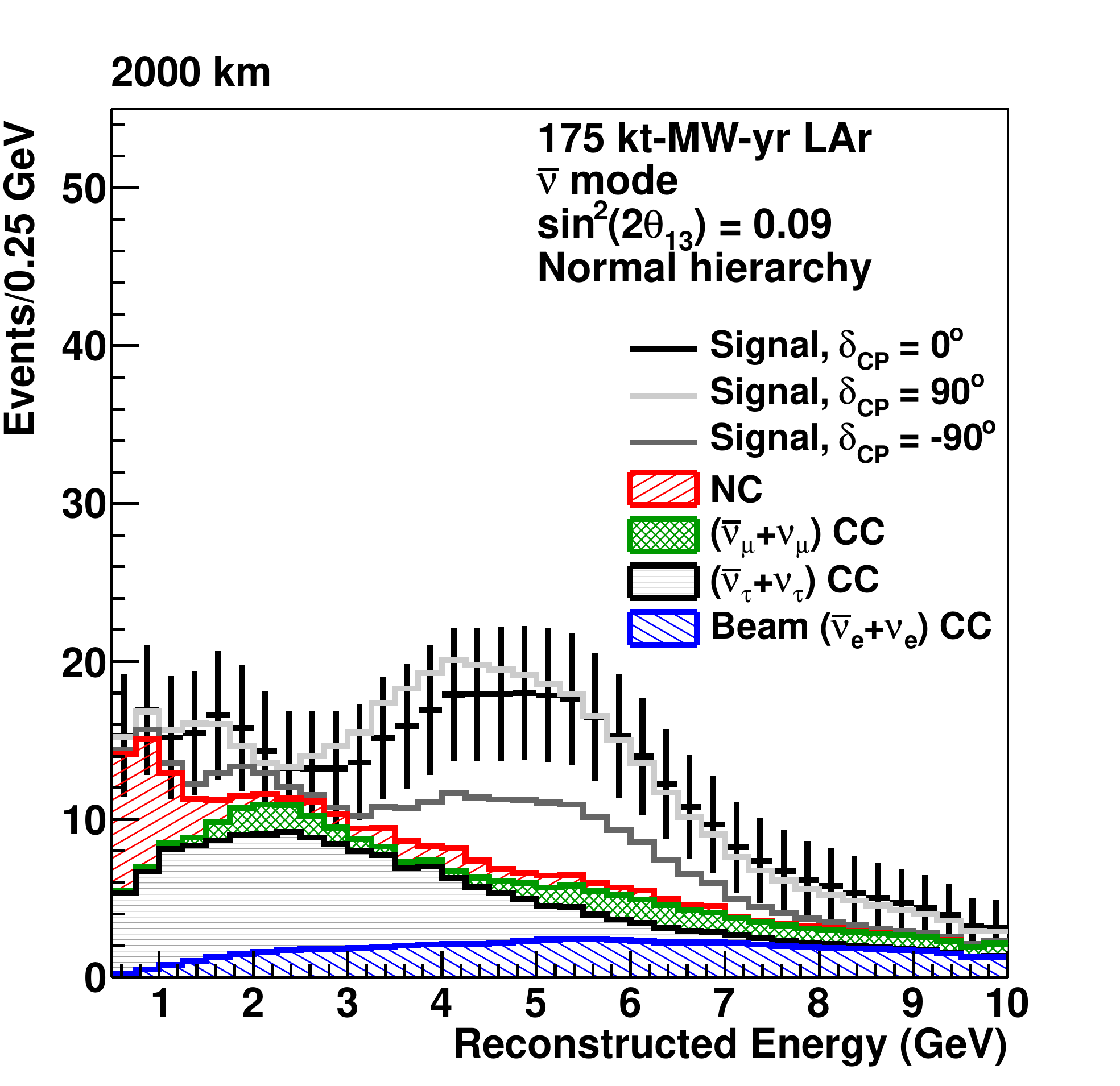}
\includegraphics[width=0.32\textwidth]{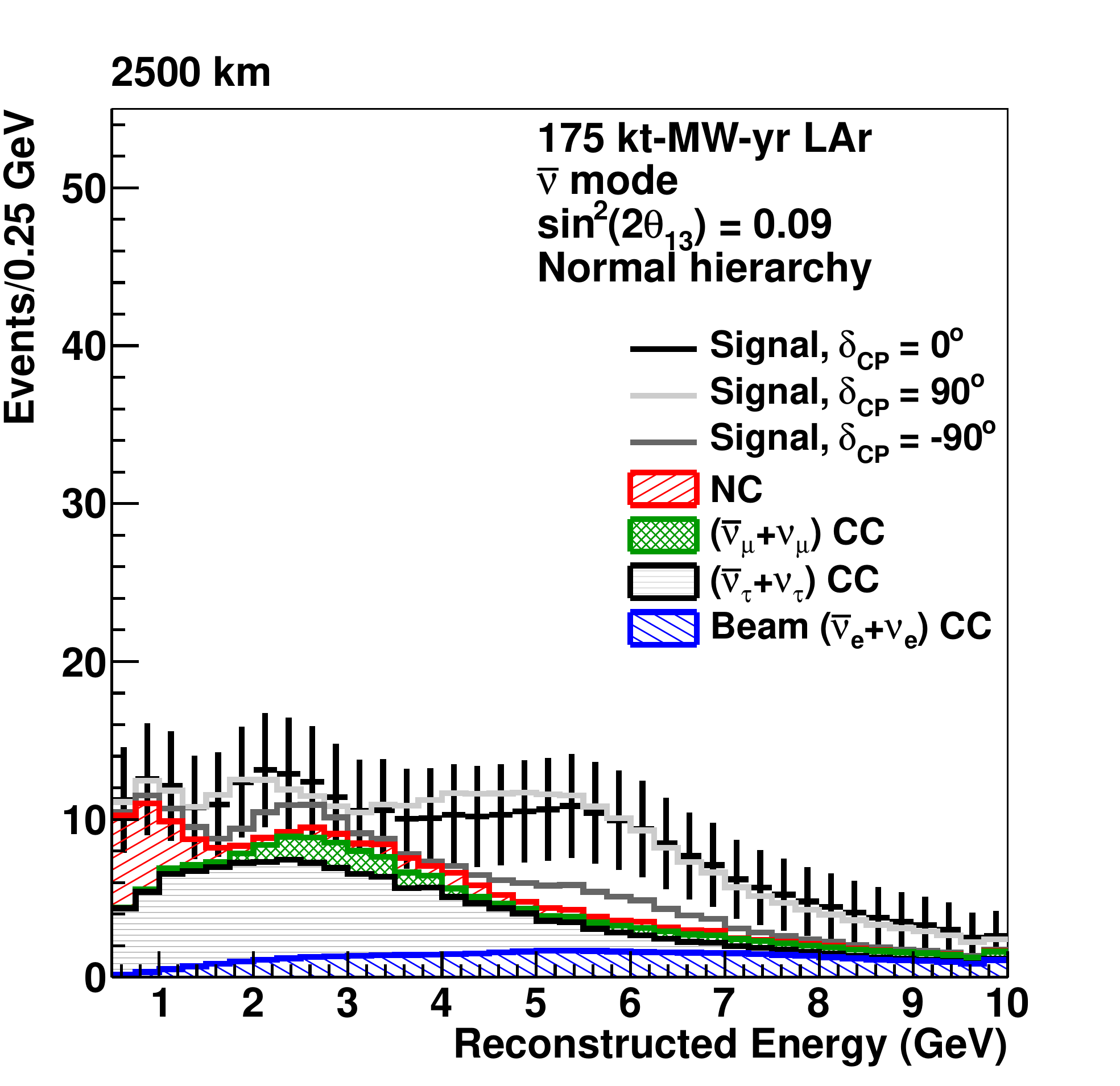}
\includegraphics[width=0.32\textwidth]{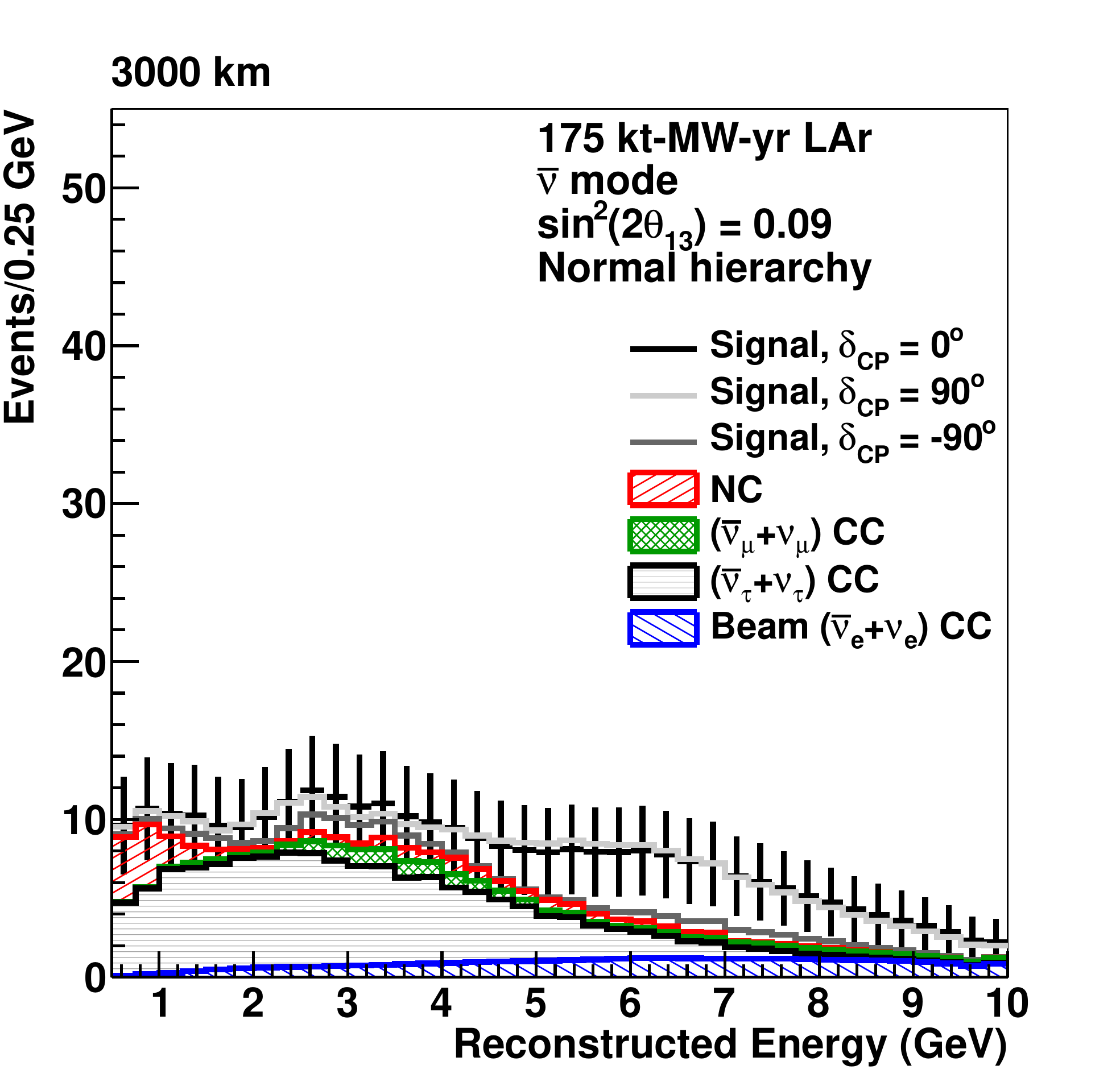}
\caption{Antineutrino spectra for normal hierarchy: Reconstructed energy distribution of selected $\nu_e$ and $\bar{\nu}_e$ CC-like events assuming a 175~\mbox{kt-MW-yr} exposure in the antineutrino-beam mode at each baseline.  The plots assume normal mass hierarchy.  The signal contribution is shown for various values of $\delta_{CP}$.}
\label{fig:anuspectra_nh}
\end{figure*}

\begin{figure*}[htp]
\includegraphics[width=0.32\textwidth]{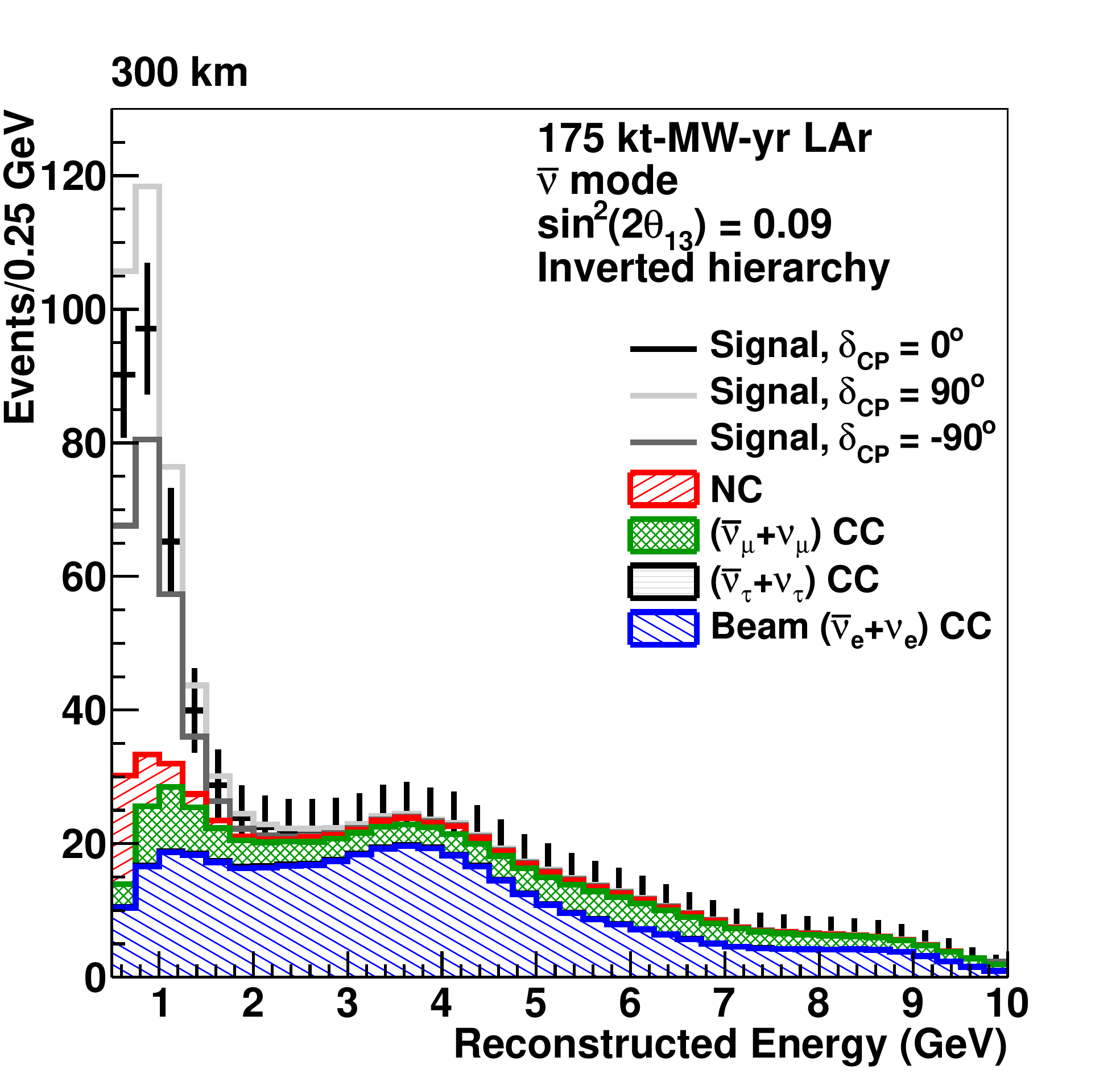}
\includegraphics[width=0.32\textwidth]{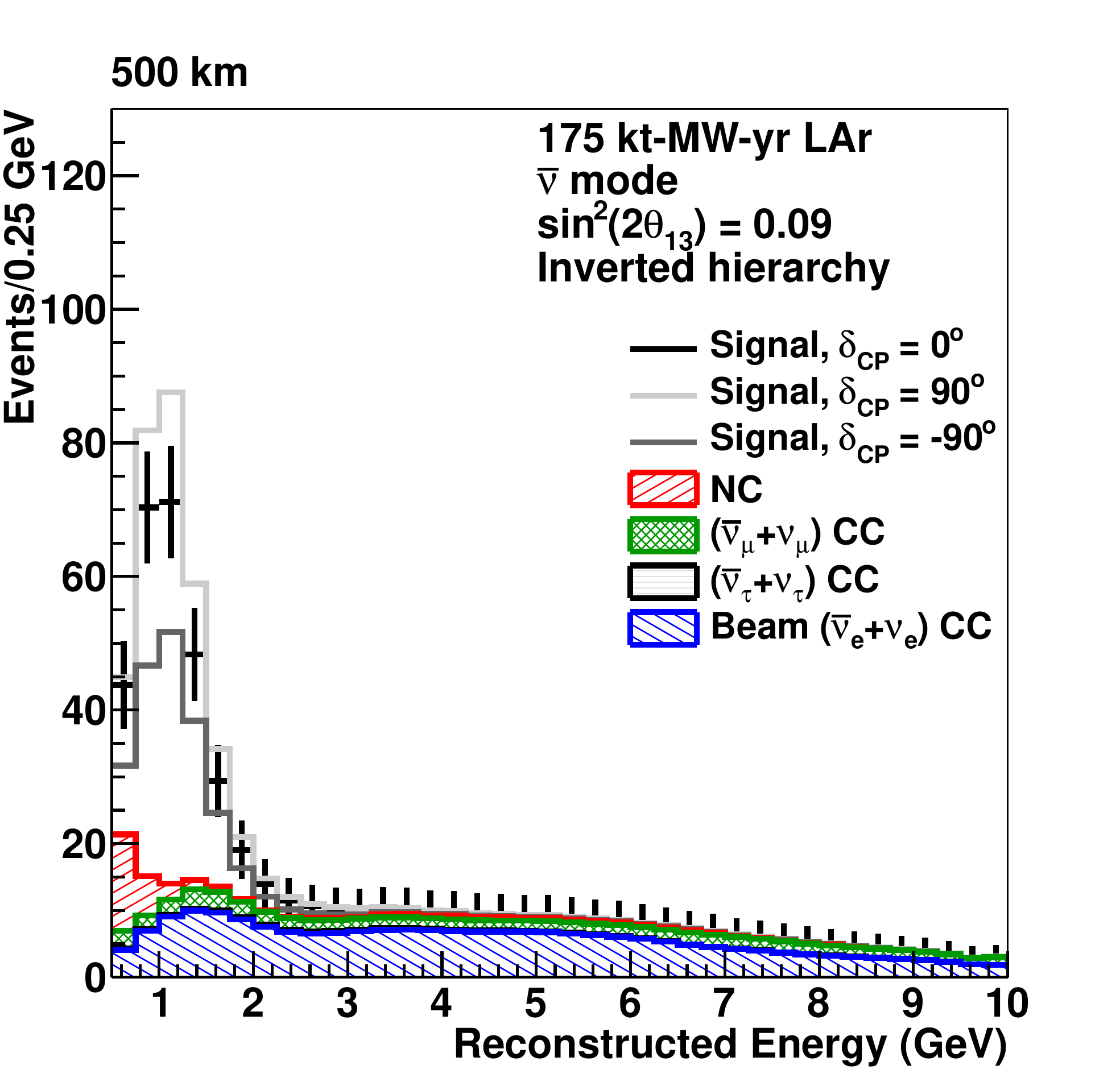}
\includegraphics[width=0.32\textwidth]{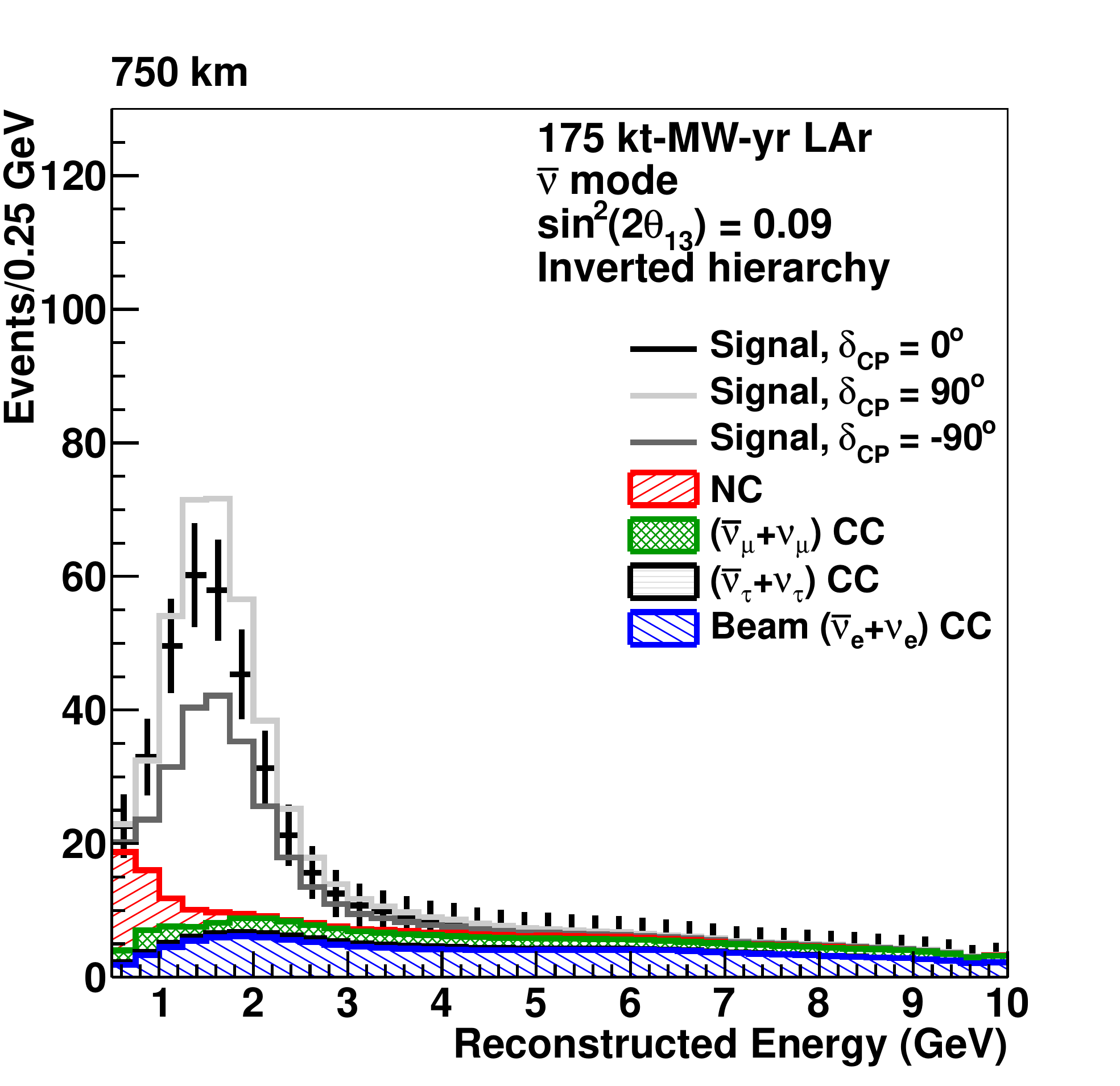}
\includegraphics[width=0.32\textwidth]{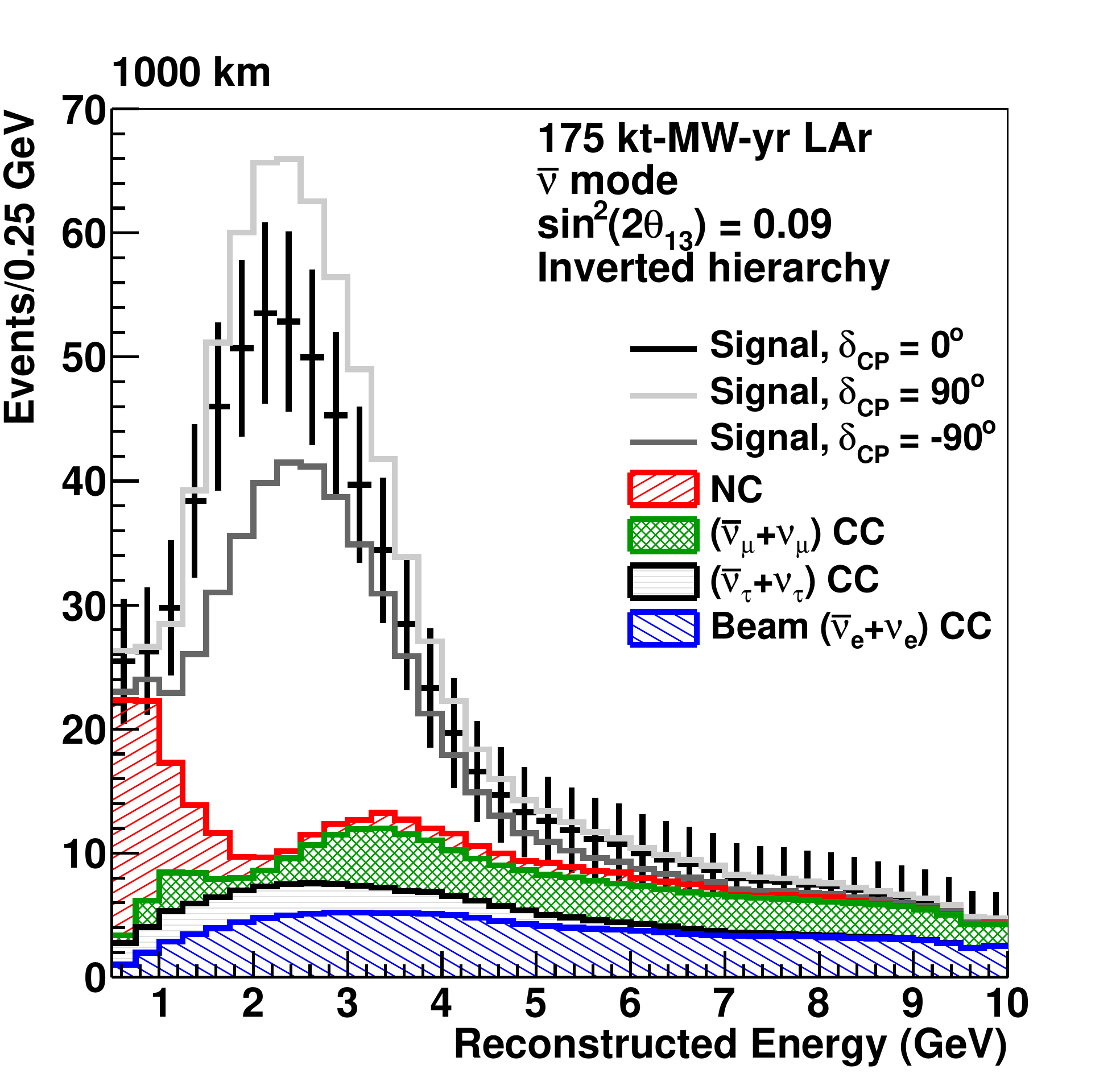}
\includegraphics[width=0.32\textwidth]{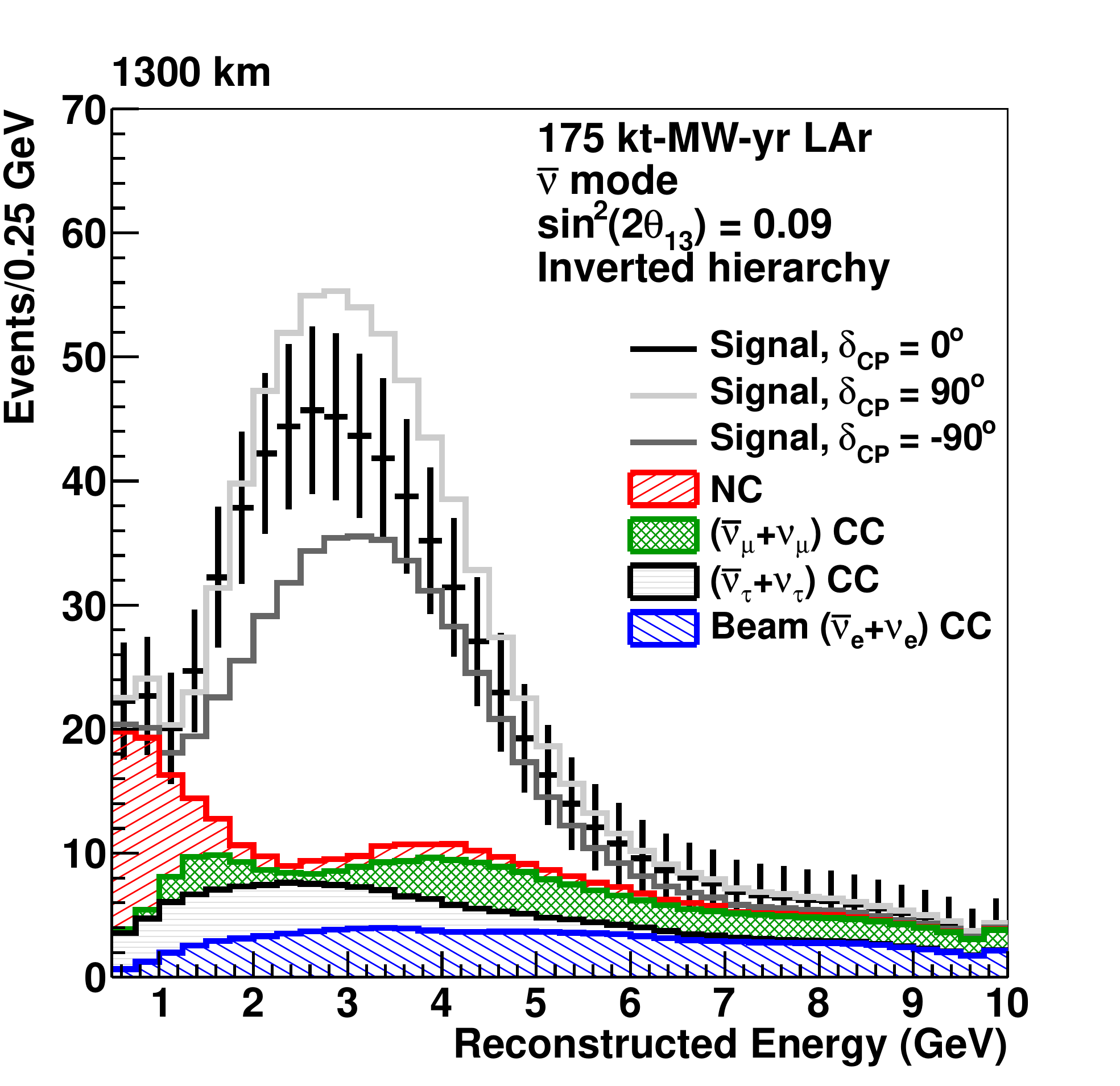}
\includegraphics[width=0.32\textwidth]{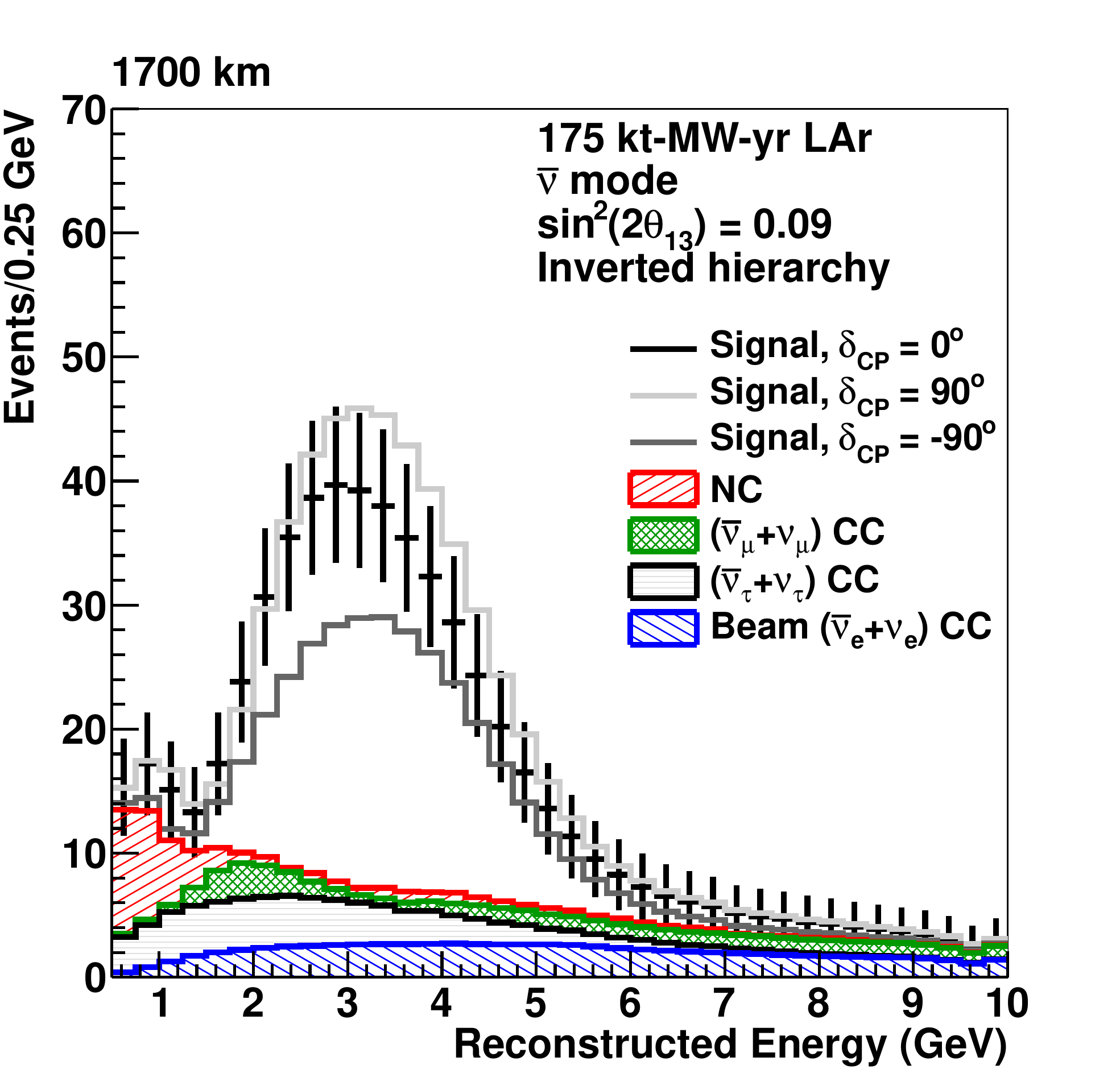}
\includegraphics[width=0.32\textwidth]{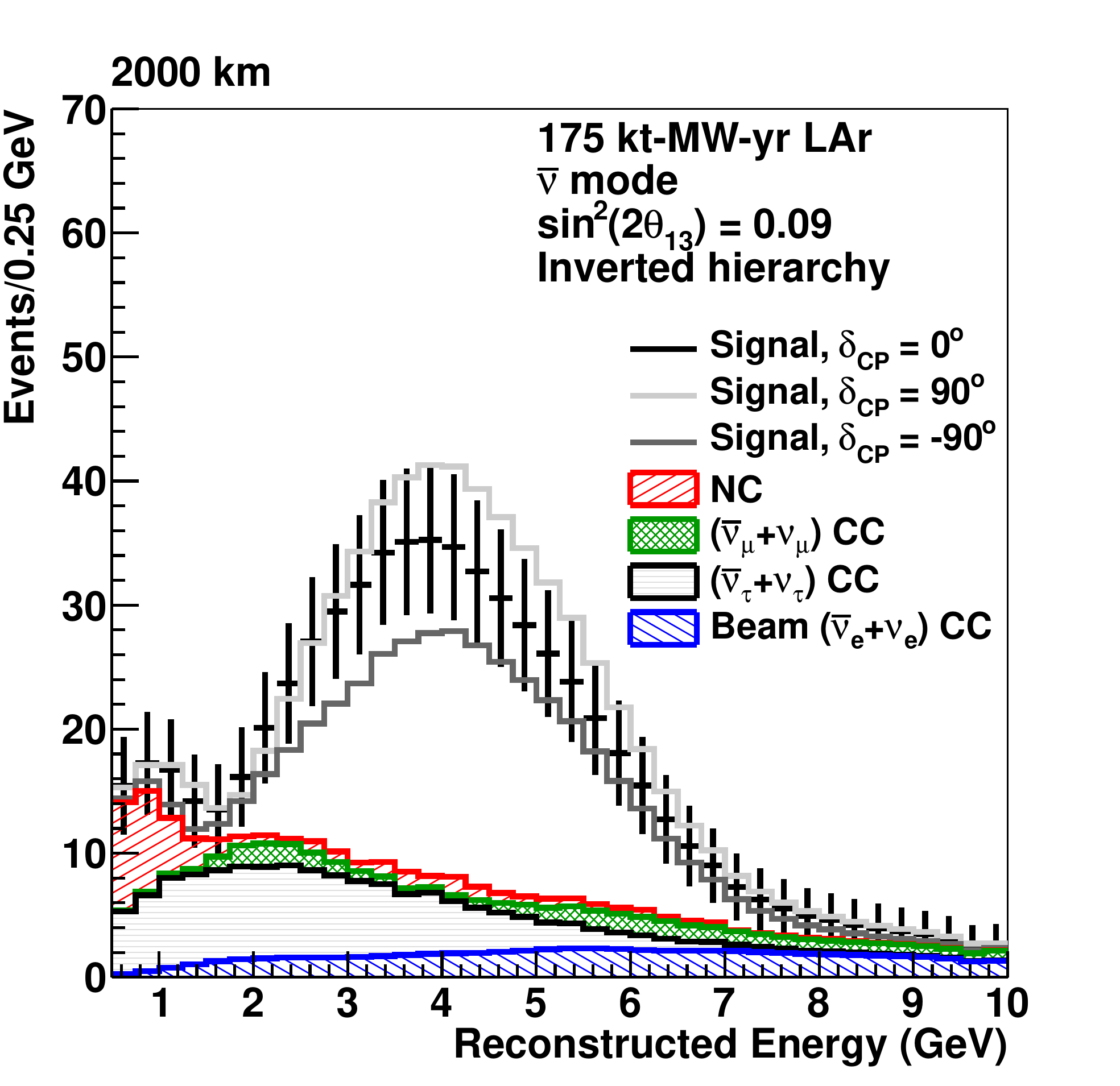}
\includegraphics[width=0.32\textwidth]{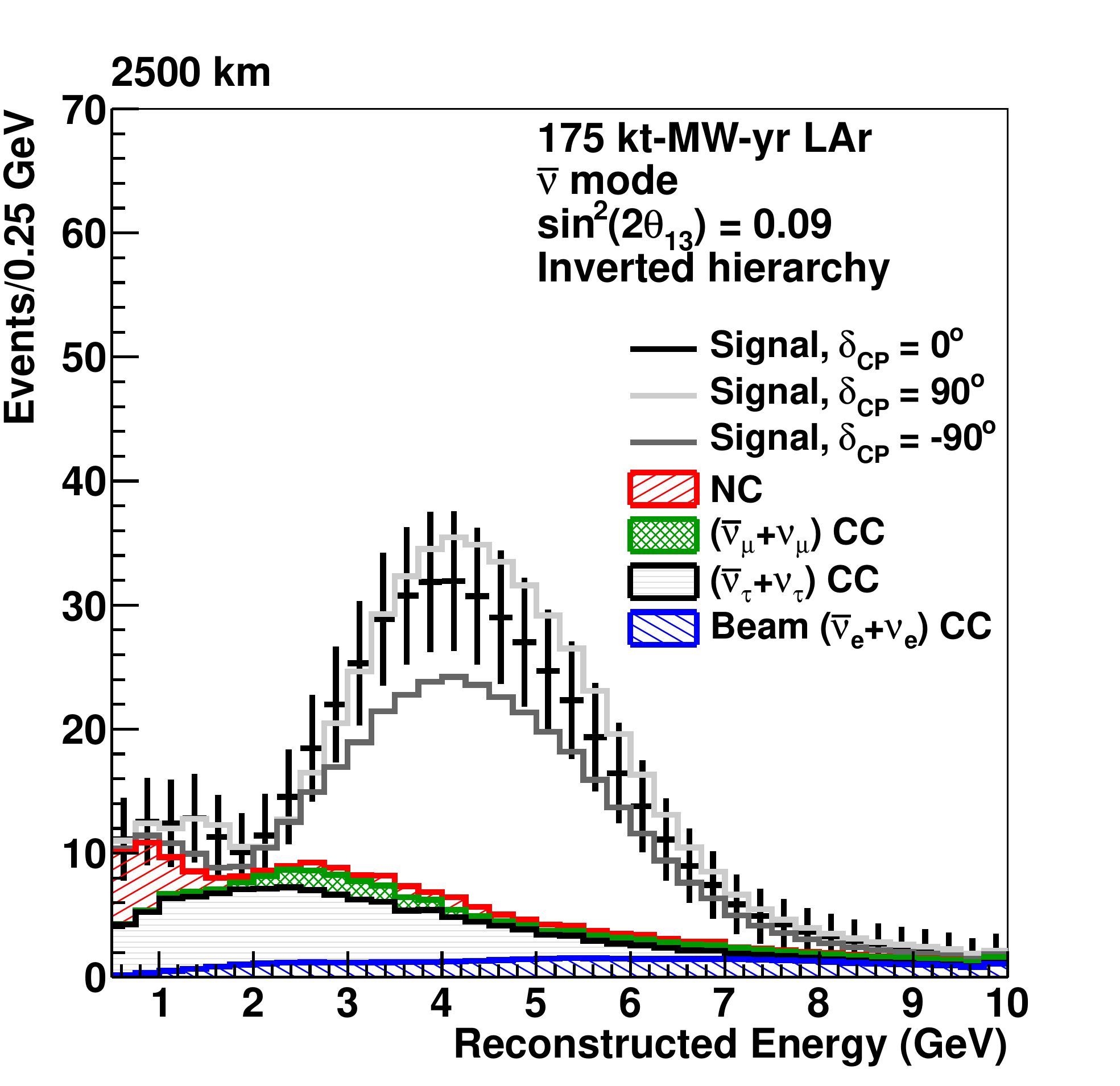}
\includegraphics[width=0.32\textwidth]{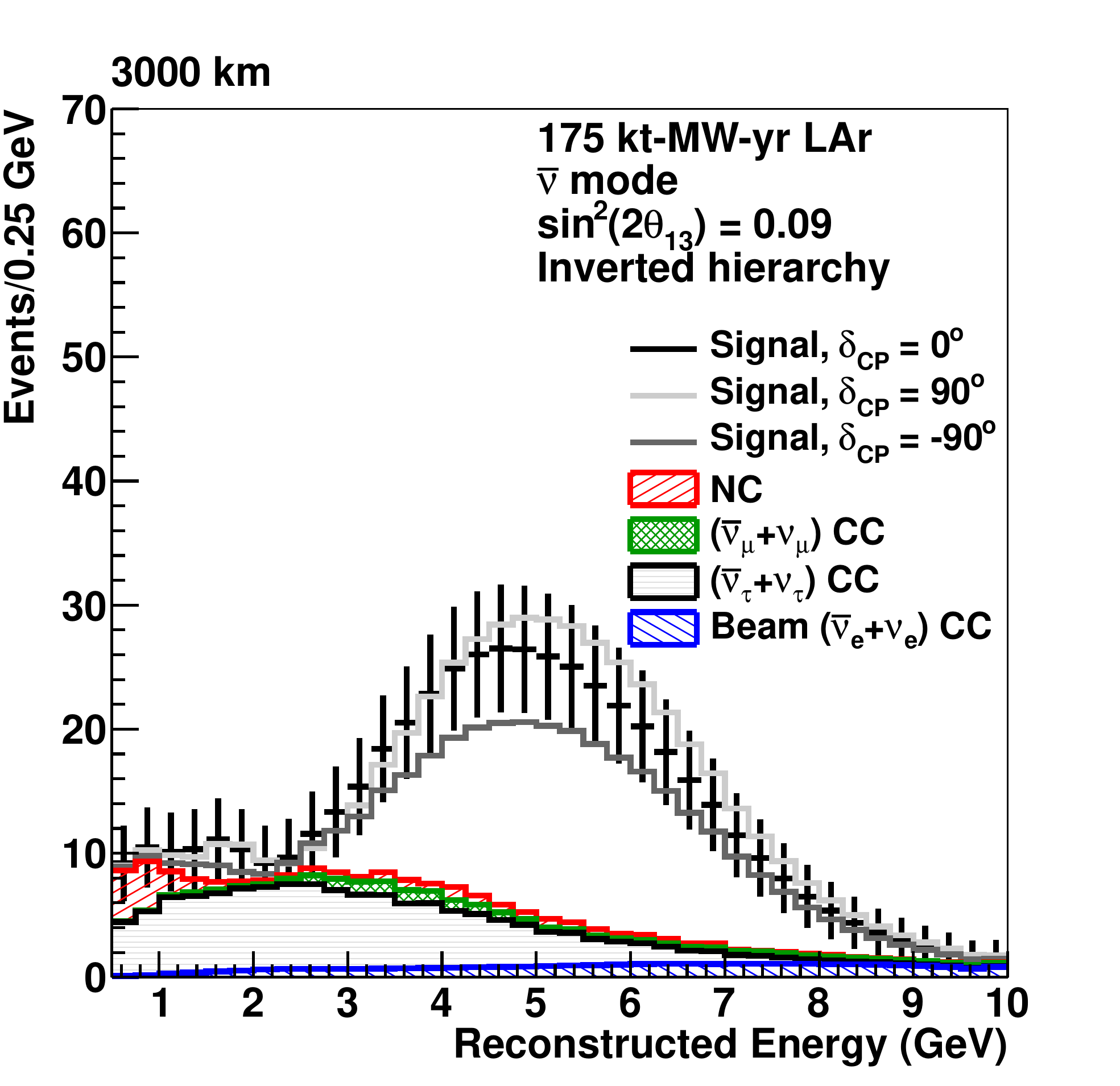}
\caption{Antineutrino spectra for inverted hierarchy: Reconstructed energy distribution of selected $\nu_e$ and $\bar{\nu}_e$ CC-like events assuming a 175~\mbox{kt-MW-yr} exposure in the antineutrino-beam mode at each baseline.  The plots assume inverted mass hierarchy.  The signal contribution is shown for various values of $\delta_{CP}$.}
\label{fig:anuspectra_ih}
\end{figure*}


\begin{table*}[htp]
\caption{Event rates in neutrino mode: The number of signal and background events integrated in reconstructed energy over the first maximum (and second, if possible) for a 175~\mbox{kt-MW-yr} exposure at each baseline in neutrino mode. These rates assume normal hierarchy (inverted hierarchy) and $\delta_{CP}=0$. (Note that the sensitivities presented in this paper are not based solely on rates, but are calculated by fitting the spectra in bins of reconstructed energy.)}
\begin{tabular}{|c|c|ccccc|}
\hline
& Signal &  & & Background & &\\
Baseline~(km)~&~~~~~$\nu_e$~CC~~~~~&~~~Total~~~~~&~~~~$\nu_e$~CC~~~~~&~~~~$\nu_{\mu}$~CC~~~~~&~~~~$\nu_{\tau}$~CC~~~~~~&~~~~~~NC~~~~~\\ \hline
300 & 752 (532) & 276 & 157 & 75 & 1 & 43 \\
500 & 692 (504) & 230 & 139 & 45 & 2 & 44 \\
750 & 738 (470) & 266 & 133 & 60 & 9 & 64 \\
1000 & 1049 (555) & 392 & 144 & 101 & 37 & 110 \\
1300 & 1107 (486) & 422 & 146 & 102 & 95 & 79 \\
1700 & 906 (288) & 330 & 127 & 68 & 92 & 43 \\
2000 & 1030 (267) & 446 & 114 & 69 & 217 & 46 \\
2500 & 854 (130) & 331 & 93 & 49 & 164 & 25 \\
3000 & 844 (77) & 370 & 81 & 38 & 226 & 25 \\ \hline
\end{tabular}
\label{tab:rates_nu}
\end{table*}


\begin{table*}[htp]
\caption{Event rates in antineutrino mode: The number of signal and background events integrated in reconstructed energy over the first maximum (and second, if possible) for a 175~\mbox{kt-MW-yr} exposure at each baseline in antineutrino mode. These rates assume normal hierarchy (inverted hierarchy) and $\delta_{CP}=0$. (Note that the sensitivities presented in this paper are not based solely on rates, but are calculated by fitting the spectra in bins of reconstructed energy.) }
\begin{tabular}{|c|c|ccccc|}
\hline
& Signal & & & Background & &\\
Baseline~(km)~&~($\bar{\nu}_e$+$\nu_e$)~CC~~&~~~Total~~~~&~($\bar{\nu}_e$+$\nu_e$)~CC~&~($\bar{\nu}_{\mu}$+$\nu_{\mu}$)~CC~&~($\bar{\nu}_{\tau}$+$\nu_{\tau}$)~CC~~&~~~~~NC~~~~~\\ \hline
300 & 181 (157) & 96 & 46 & 22 & 1 & 27 \\
500 & 167 (192) & 91 & 49 & 15 & 2 & 25\\
750 & 181 (240) & 111 & 50 & 21 & 6 & 34 \\
1000 & 266 (348) & 186 & 57 & 37 & 30 & 62 \\
1300 & 243 (378) & 190 & 57 & 37 & 53 & 43 \\
1700 & 166 (344) & 161 & 53 & 26 & 57 & 25 \\
2000 & 164 (362) & 201 & 47 & 26 & 104 & 24 \\
2500 & 112 (331) & 156 & 39 & 19 & 84 & 14 \\
3000 & 89 (316) & 158 & 31 & 15 & 100 & 12 \\ \hline
\end{tabular}
\label{tab:rates_anu}
\end{table*}

\begin{figure*}[htp]
\includegraphics[width=0.49\textwidth]{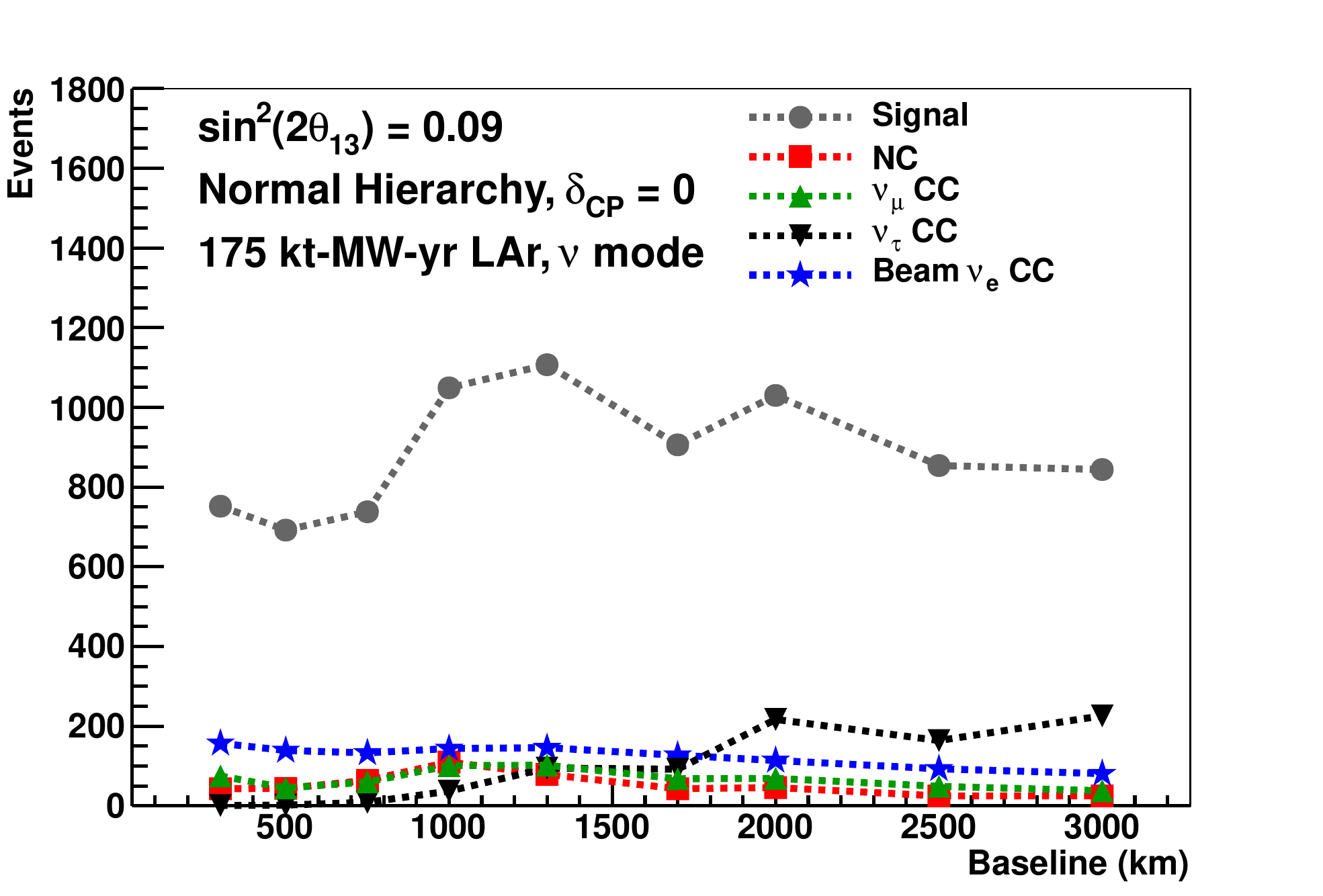}
\includegraphics[width=0.49\textwidth]{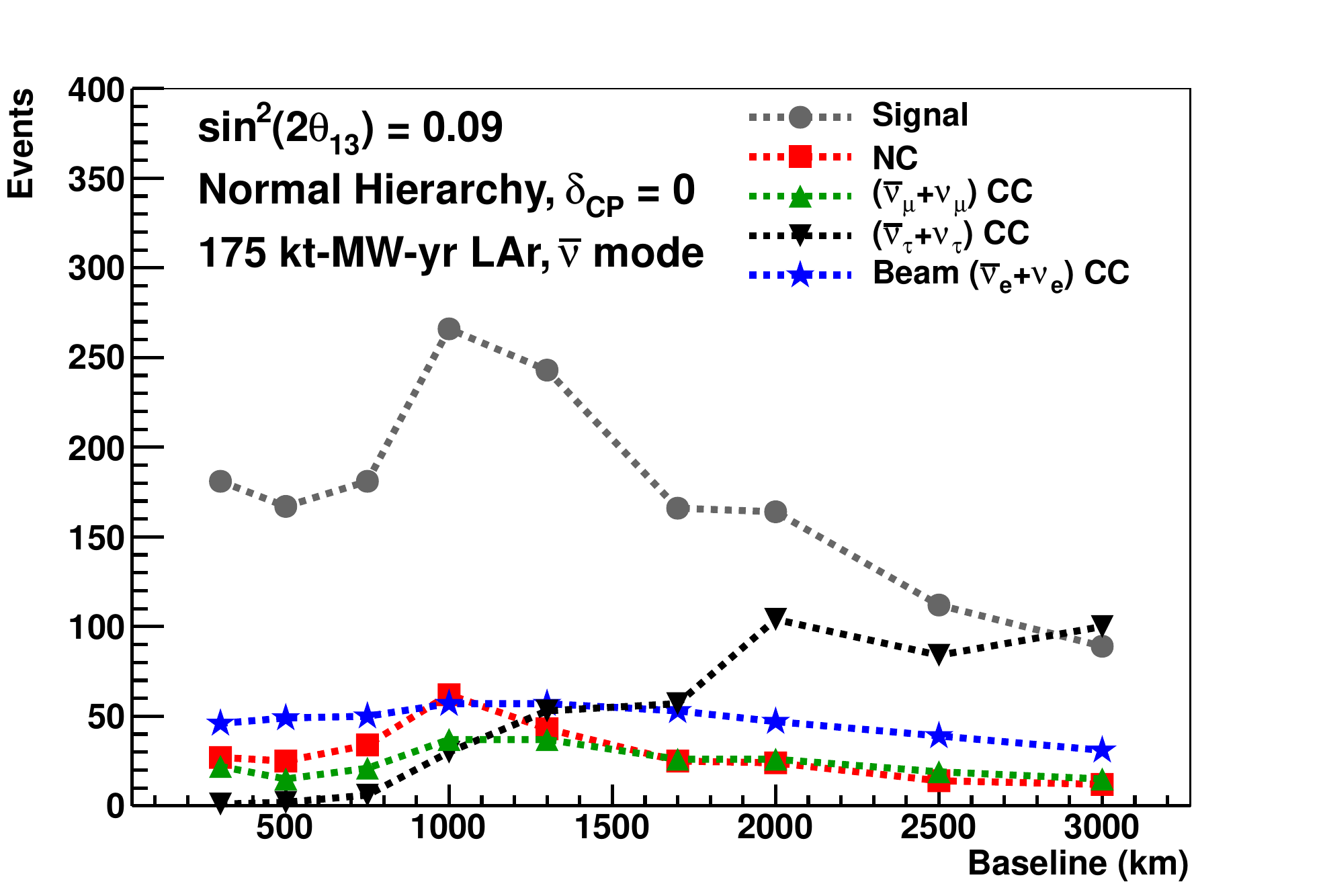}
\caption{Event rates vs baseline: The number of signal and background events integrated in reconstructed energy over the first maximum (and second, if possible) for a 175~\mbox{kt-MW-yr} exposure at each baseline in neutrino (left) and antineutrino (right) modes.  These rates assume normal hierarchy and $\delta_{CP}=0$. (Note that the sensitivities presented in this paper are not based solely on rates, but are calculated by fitting the spectra in bins of reconstructed energy.)}
\label{fig:rates}
\end{figure*}

The expected neutrino and antineutrino spectra at each baseline are shown in Figures \ref{fig:nuspectra_nh}-\ref{fig:anuspectra_ih}.  The expected event rates, assuming normal hierarchy, for neutrino and antineutrino-beam modes are shown in Tables \ref{tab:rates_nu} and \ref{tab:rates_anu}, respectively. Figure \ref{fig:rates} shows the rates as a function of baseline.

The neutrino signal event rates with realistic flux simulation and detector effects are roughly constant across the baselines considered in the study (Figure \ref{fig:rates}).  The antineutrino rates are lower than the neutrino rates due to both lower $\pi^{-}$ production rates and lower antineutrino cross-sections.  The antineutrino signal rate decreases slightly with baseline due to the energy dependence of the $\pi^{-}$ production, which has a steeper drop-off in energy than the $\pi^{+}$ production rate.

\section{Analysis}

To compare the sensitivity at each baseline, we use GLoBES to calculate the significance of the mass hierarchy, CP violation, and octant determination.  Predicted neutrino and antineutrino spectra are generated for appropriate values of $\delta_{CP}$ and the mass hierarchy for the hypothesis being tested, and a $\chi^2$ minimization is performed on the spectra.  The $\chi^2$ minimization takes correlations between the oscillation parameters into account and considers both octant solutions for $\theta_{23}$ in the mass hierarchy and CP violation analysis.  The $\chi^2$ is given by
\begin{align}
 &\chi^{2}(\mathbf{n}^{true},\mathbf{n}^{test},f)\notag\\
 & = 2 \sum_i^{N_{reco}} (n_i^{true} \ln\frac{n_i^{true}}{n_i^{test}(f)} + n_i^{test}(f) - n_i^{true}) + f^2
\end{align}
where $\mathbf{n}$ are the event rate vectors in $N_{reco}$ bins of reconstructed neutrino energy and $f$ is a nuisance parameter to be profiled.  The nuisance parameters include the already-measured oscillation parameters and signal and background normalizations. The oscillation parameters are constrained within their experimental uncertainties given in Section~\ref{sec:assumptions}.  We perform a combined fit to the $\nu_e$ appearance and $\nu_{\mu}$ disappearance spectra.  We assume a 1\% (5\%) uncertainty in the signal normalization for $\nu_{e}$ ($\nu_{\mu}$) and a 5\% (10\%) uncertainty in the background normalization for $\nu_{e}$ ($\nu_{\mu}$) in the fit, where the normalization uncertainties are uncorrelated among the four ($\nu_e$, $\bar{\nu}_e$, $\nu_{\mu}$, $\bar{\nu}_{\mu}$) modes.  Achieving this level of precision will require a well-designed near detector and careful analysis of detector efficiencies and other systematic errors. See \cite{Adams:2013qkq} for a detailed discussion of the expected systematic uncertainties based on current and former experiments. We will not analyze this issue in this paper. 

For the calculations at each baseline, we assume an equal amount of exposure in neutrino-beam mode and antineutrino-beam mode.  The relative amount of neutrino and antineutrino beam time could be optimized for each baseline in future studies.  For example, it is slightly more optimal at longer baselines to collect more neutrino (antineutrino) data assuming normal mass hierarchy (inverted mass hierarchy).

\subsection{Mass Hierarchy}

To calculate the mass hierarchy significance, the $\chi^2$ minimization considers only those solutions that have the opposite hierarchy from that which was used to generate the true spectra.  For example, if the true hierarchy is assumed to be normal, we assume the inverted hierarchy for the observed spectra in the $\chi^2$ calculation.  This allows us to determine the significance at which we can exclude the inverted hierarchy given the true hierarchy is normal.  

To properly interpret the mass hierarchy physics sensitivity, special attention should be paid, as the mass hierarchy determination has only two possible outcomes (normal vs. inverted hierarchy).  Ref.~\cite{Qian:2012zn} carefully examines the statistical nature of this problem and shows the connection between an expected average $\Delta \chi^2$ and probability of mass hierarchy determination.  In particular, an experiment with physics sensitivities determined by $\overline{\Delta \chi^{2}}=$ 9 and 25 (corresponding to 3 and 5$-\sigma$ of average separation between the two hypotheses) would have 93.32\% and 99.38\% probabilities of rejecting the wrong mass hierarchy, with a 6.68\% and 0.62\% probability of incorrect identification, respectively.  

\begin{figure*}[htp]
\includegraphics[width=0.49\textwidth]{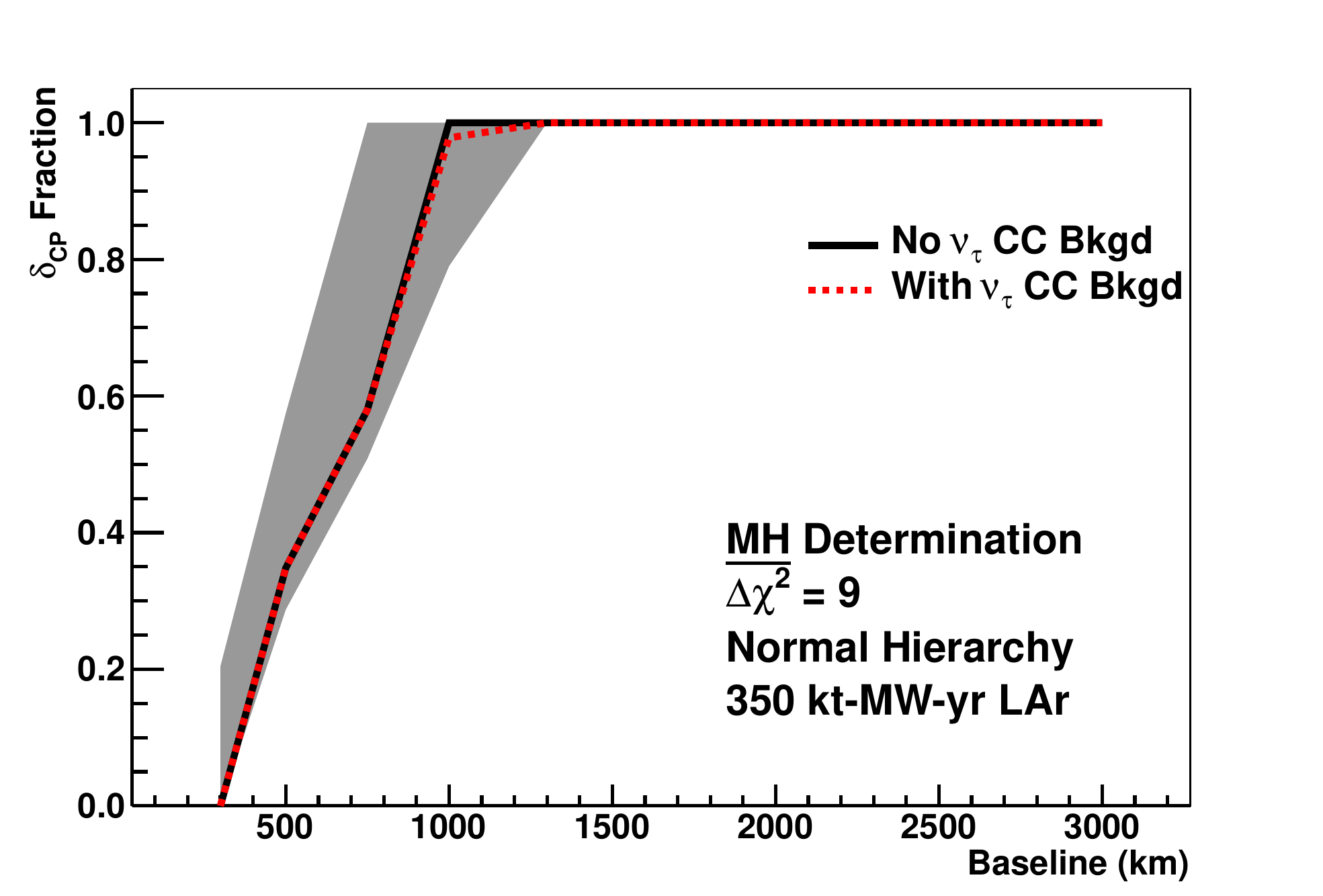}
\includegraphics[width=0.49\textwidth]{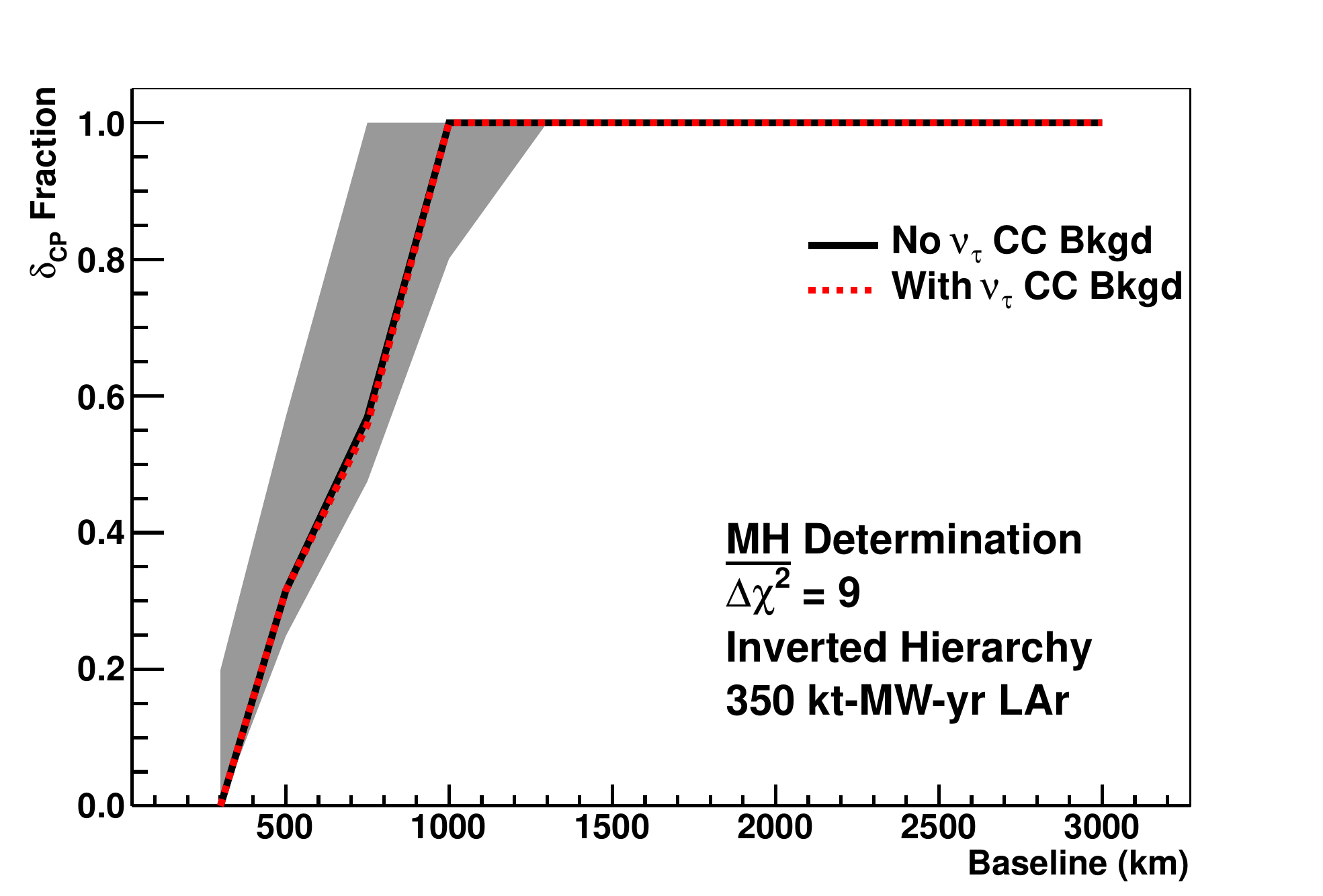}
\caption{The fraction of all possible $\delta_{CP}$ values for which we can determine normal (left) or inverted (right) mass hierarchy with a minimum value of $\overline{\Delta\chi^{2}} = 9$ as a function of baseline.  An expected average value of $\overline{\Delta\chi^{2}} = 9$ corresponds to a 93.32\% probability of determining the correct mass hierarchy according to the analysis in \cite{Qian:2012zn}.  The solid black (red dashed) line shows the result including zero (maximum) $\nu_{\tau}$ CC background.  The shaded band shows the possible range in the fraction due to the uncertainty in the other oscillation parameters and considers both octant solutions for $\theta_{23}$. }
\label{fig:cpfrac_MH_3}
\end{figure*}

\begin{figure*}[htp]
\includegraphics[width=0.49\textwidth]{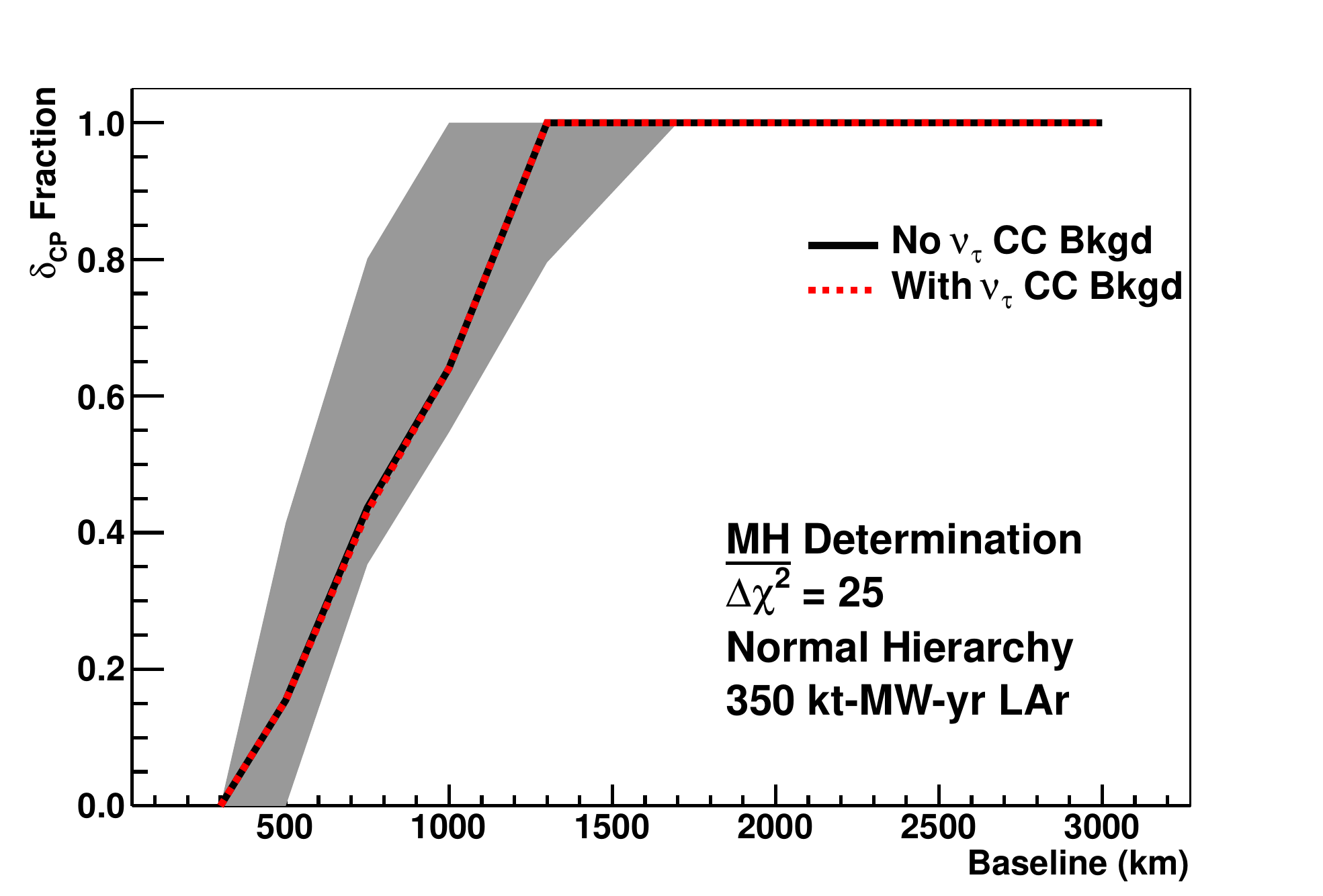}
\includegraphics[width=0.49\textwidth]{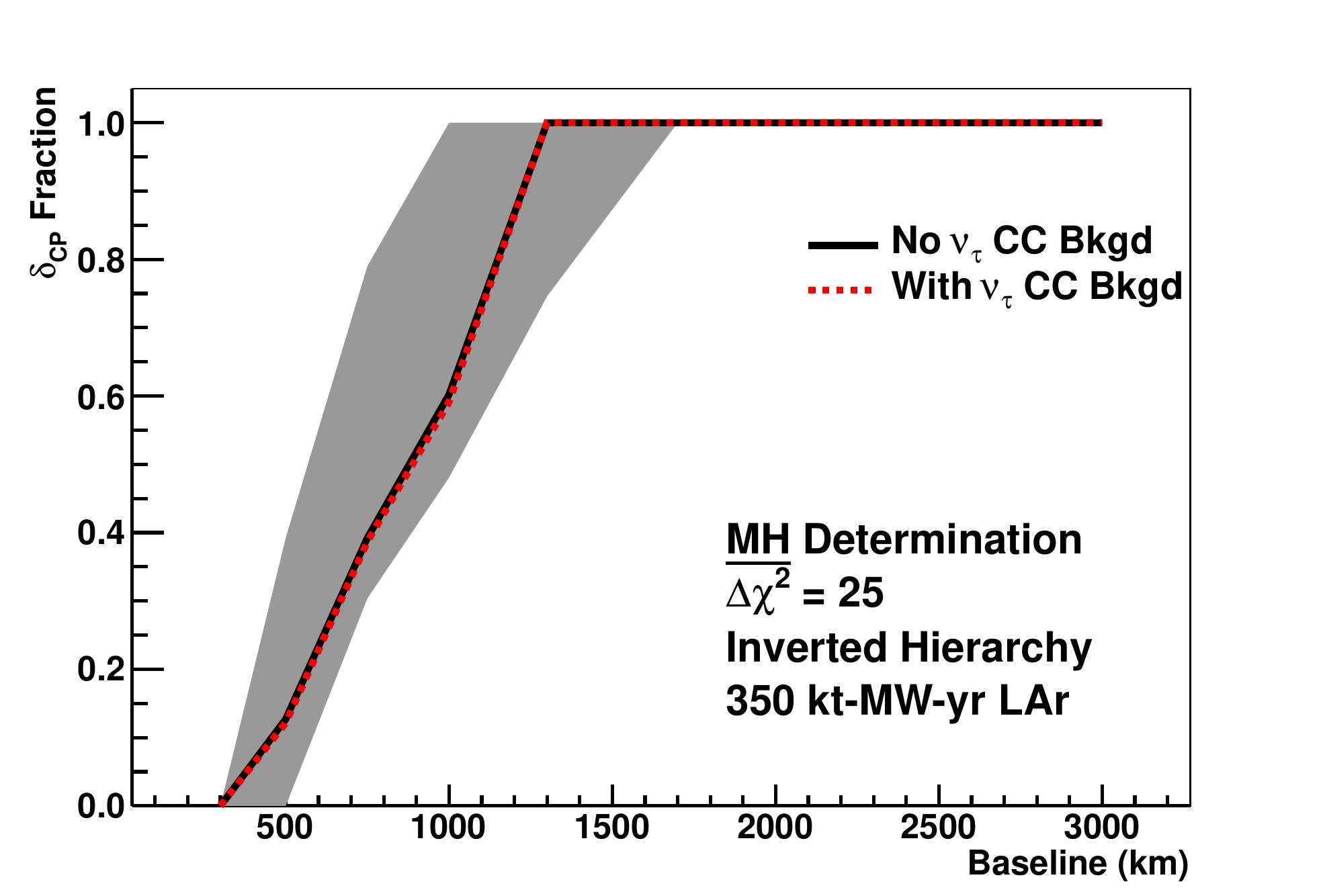}
\caption{The fraction of all possible $\delta_{CP}$ values for which we can determine normal (left) or inverted (right) mass hierarchy with a minimum value of $\overline{\Delta\chi^{2}} = 25$ as a function of baseline.  An expected average value of $\overline{\Delta\chi^{2}} = 25$ corresponds to a 99.38\% probability of determining the correct mass hierarchy according to the analysis in \cite{Qian:2012zn}.  The solid black (red dashed) line shows the result including zero (maximum) $\nu_{\tau}$ CC background.  The shaded band shows the possible range in the fraction due to the uncertainty in the other oscillation parameters and considers both octant solutions for $\theta_{23}$. }
\label{fig:cpfrac_MH_5}
\end{figure*}

Figure \ref{fig:cpfrac_MH_3} (\ref{fig:cpfrac_MH_5}) shows the fraction of all possible true $\delta_{CP}$ values for which we can determine normal or inverted hierarchy with a minimum value of $\overline{\Delta\chi^{2}} = 9$ (25) as a function of baseline.  We find the mass hierarchy can be determined for 100\% of all $\delta_{CP}$ values at a baseline of at least 1000~km (1300~km) with a minimum value of $\overline{\Delta\chi^{2}} = 9$ (25).  The inclusion of the maximum $\nu_{\tau}$ CC background does not have a noticeable effect on the mass hierarchy sensitivity.  The shaded band shows the possible range in the fraction due to the uncertainty in the other oscillation parameters, dominated by the uncertainty in $\theta_{23}$.   Both octant solutions for $\theta_{23}$ are considered by the shaded region.

Existing electron neutrino appearance experiments (\cite{Patterson:2012zs,Abe:2013hdq}) and proposed reactor antineutrino experiments \cite{dayabay2} will seek to constrain the mass hierarchy before a next-generation long-baseline experiment such as LBNE begins taking data. However, unambiguous determination of the mass hierarchy for all possible values of $\delta_{CP}$ is very difficult for existing electron neutrino appearance experiments because of the degeneracy between the matter and CP asymmetries. The electron neutrino appearance and reactor antineutrino disappearance methods are complementary, so it is preferable to make both measurements. Therefore, the mass hierarchy sensitivity is still a relevant consideration in choosing the optimal baseline.


\begin{figure*}[htp]
\includegraphics[width=0.49\textwidth]{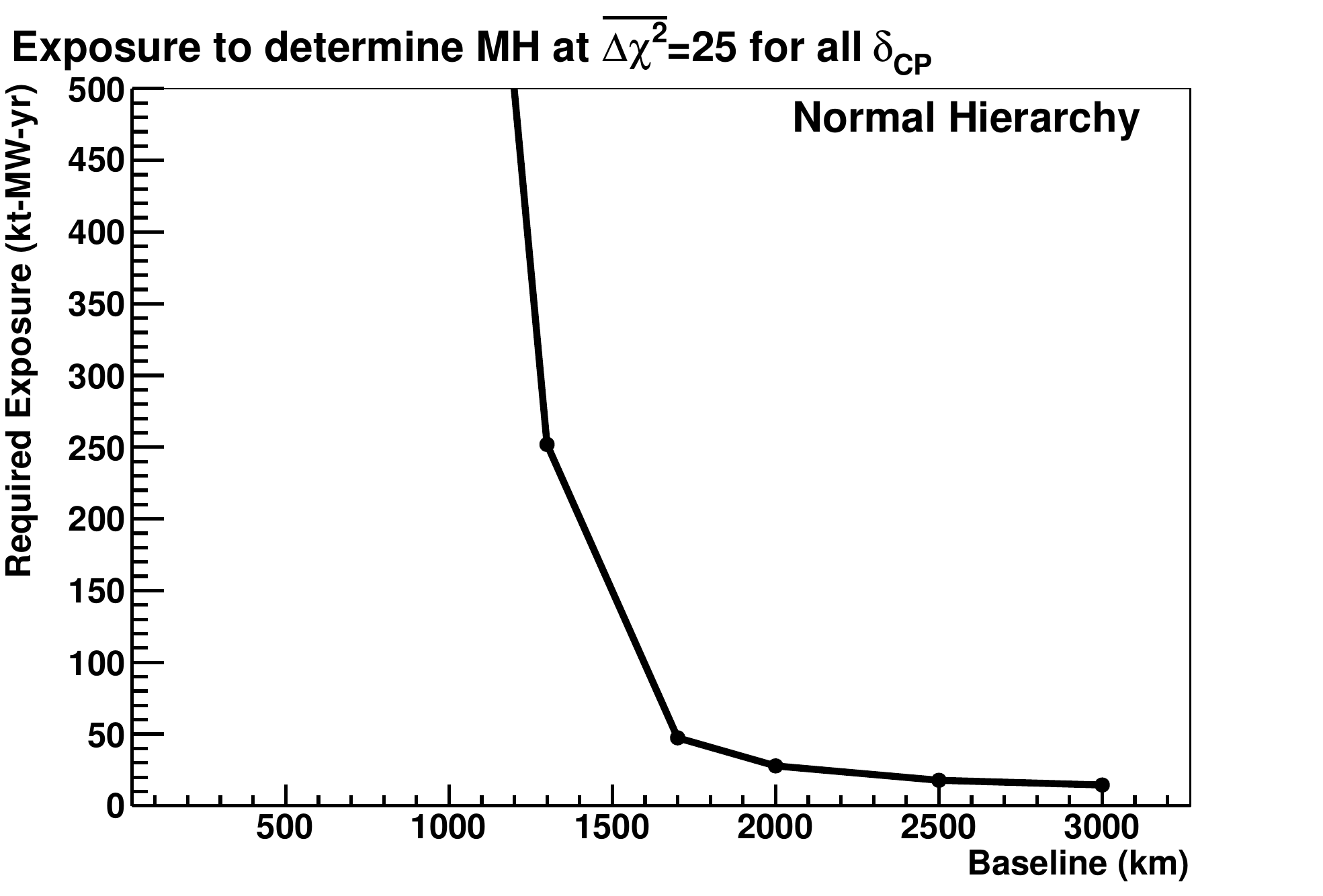}
\includegraphics[width=0.49\textwidth]{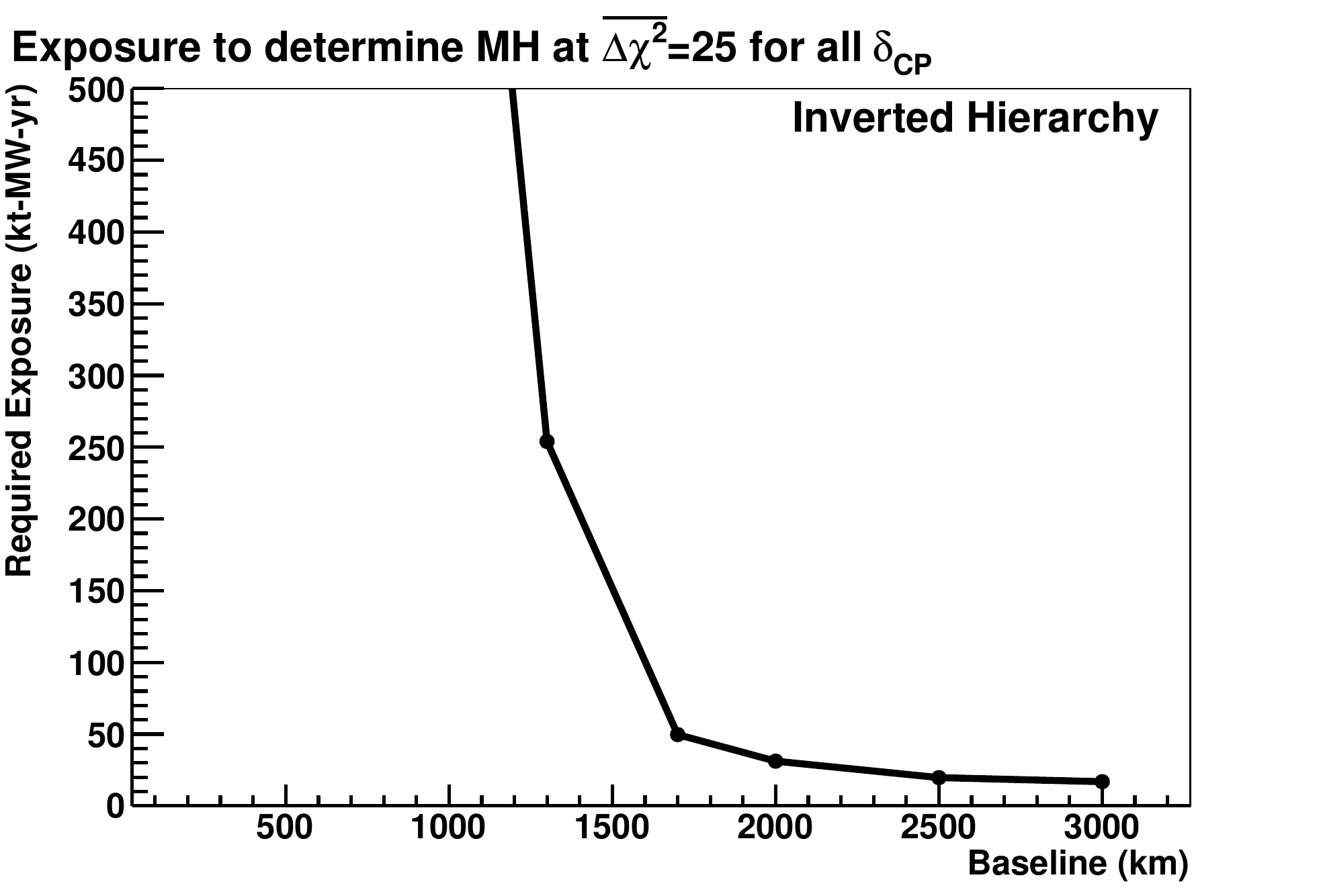}
\caption{The exposure required to determine normal(left) or inverted (right) mass hierarchy with a significance of $\overline{\Delta\chi^{2}} = 25$ for all possible values of $\delta_{CP}$ as a function of baseline. We assume no $\nu_{\tau}$ CC background.}
\label{fig:MH_expandbl}
\end{figure*}

\begin{figure*}[htp]
\includegraphics[width=0.49\textwidth]{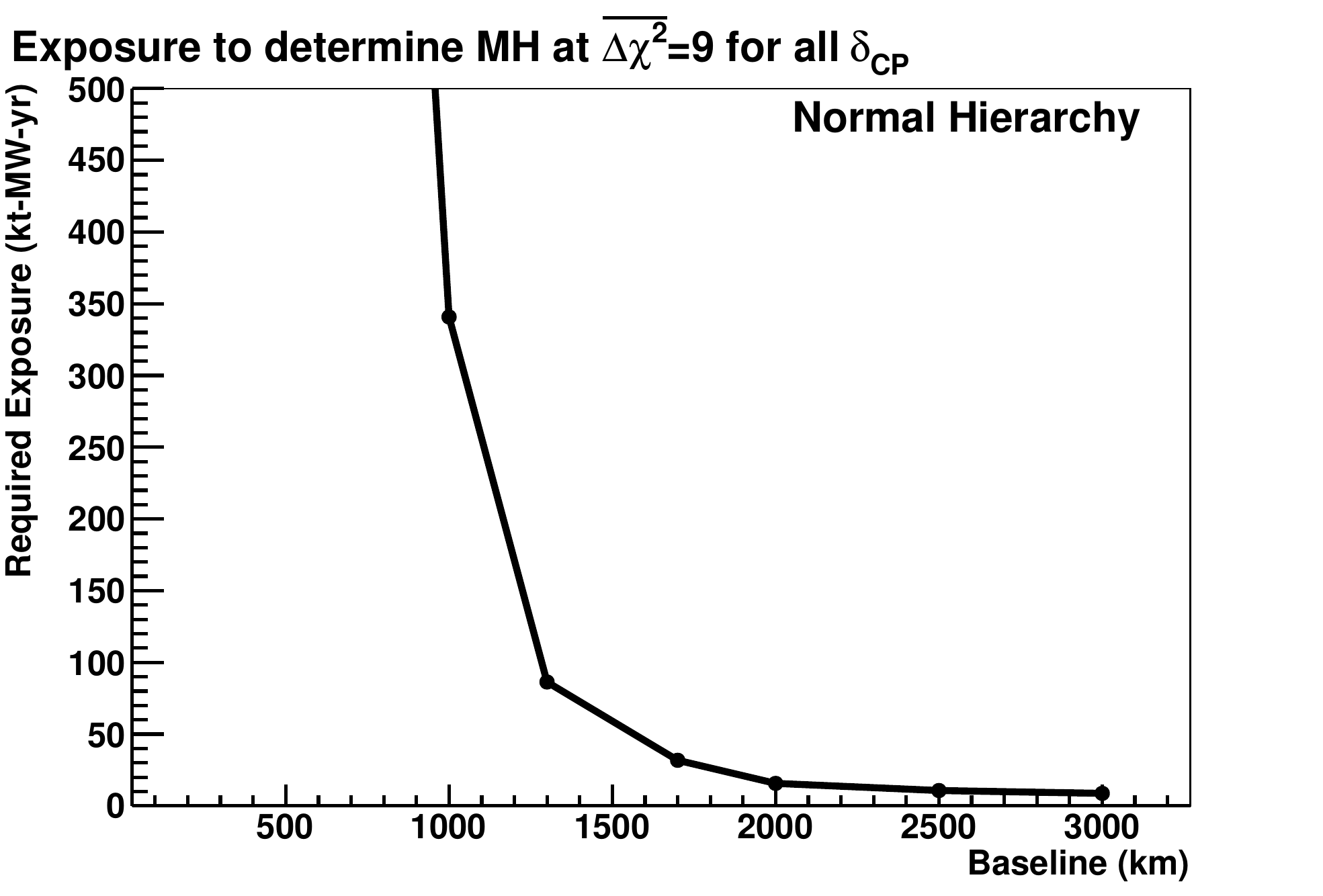}
\includegraphics[width=0.49\textwidth]{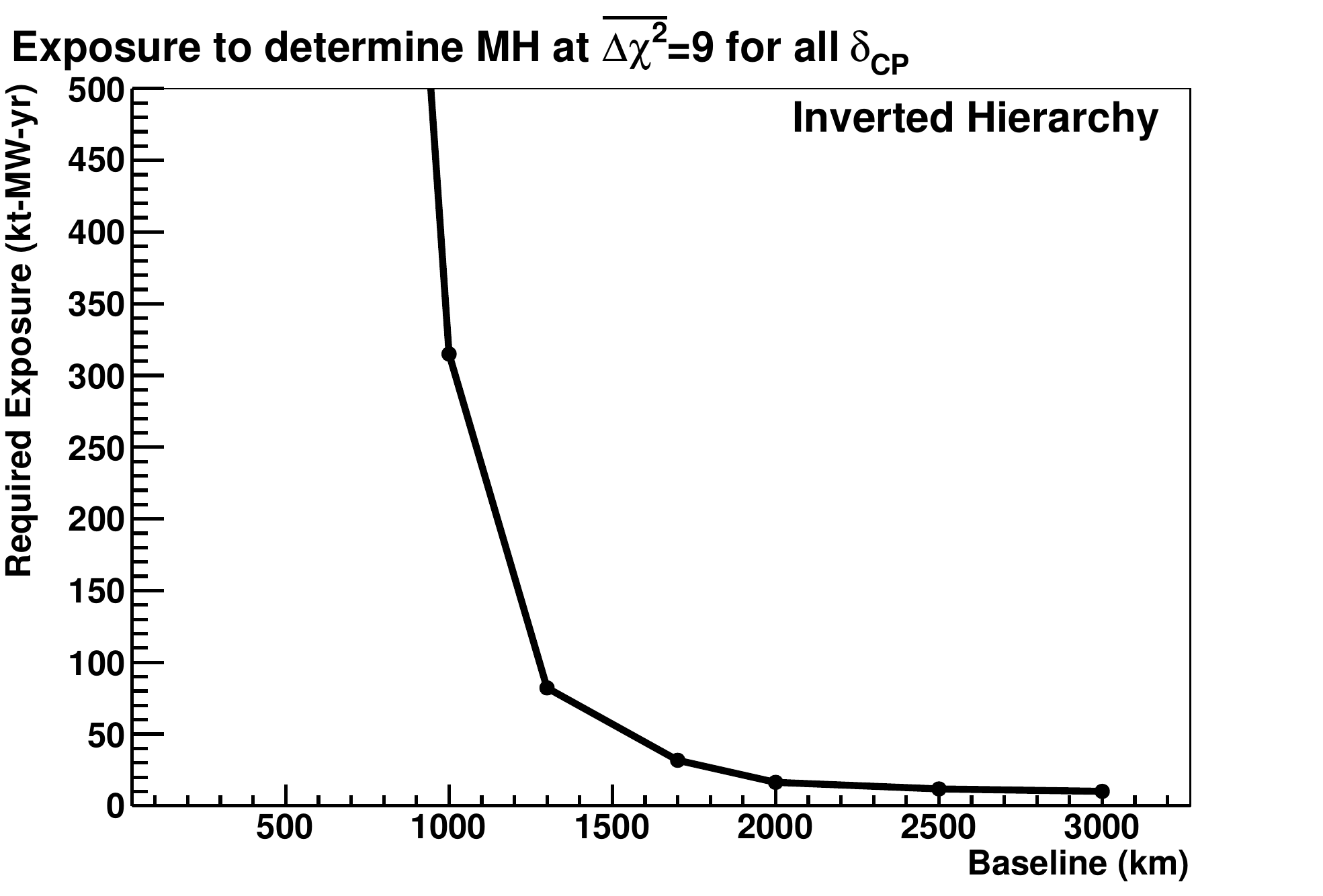}
\caption{The exposure required to determine normal(left) or inverted (right) mass hierarchy with a significance of $\overline{\Delta\chi^{2}} = 9$ for all possible values of $\delta_{CP}$ as a function of baseline. We assume no $\nu_{\tau}$ CC background.}
\label{fig:MH_expandbl_3s}
\end{figure*}

The sensitivity calculations above assume a 350~\mbox{kt-MW-yr} exposure. Figure \ref{fig:MH_expandbl} (\ref{fig:MH_expandbl_3s}) shows the exposure required to determine normal (left) or inverted (right) mass hierarchy with a significance of $\overline{\Delta\chi^{2}} = 25$ ($\overline{\Delta\chi^{2}} = 9$) for all possible values of $\delta_{CP}$ as a function of baseline.


\subsection{CP Violation}
\begin{figure*}[htp]
\includegraphics[width=0.49\textwidth]{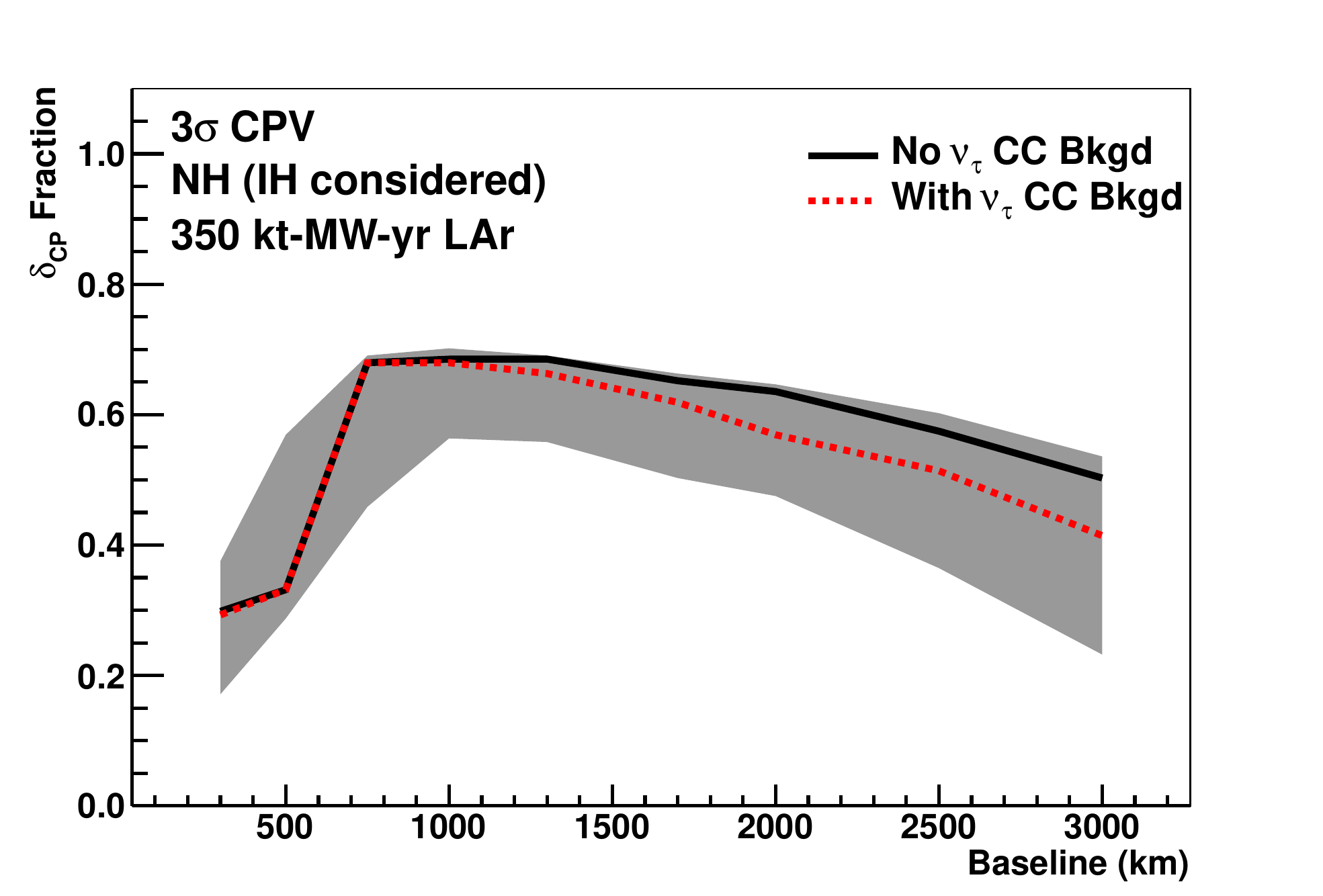}
\includegraphics[width=0.49\textwidth]{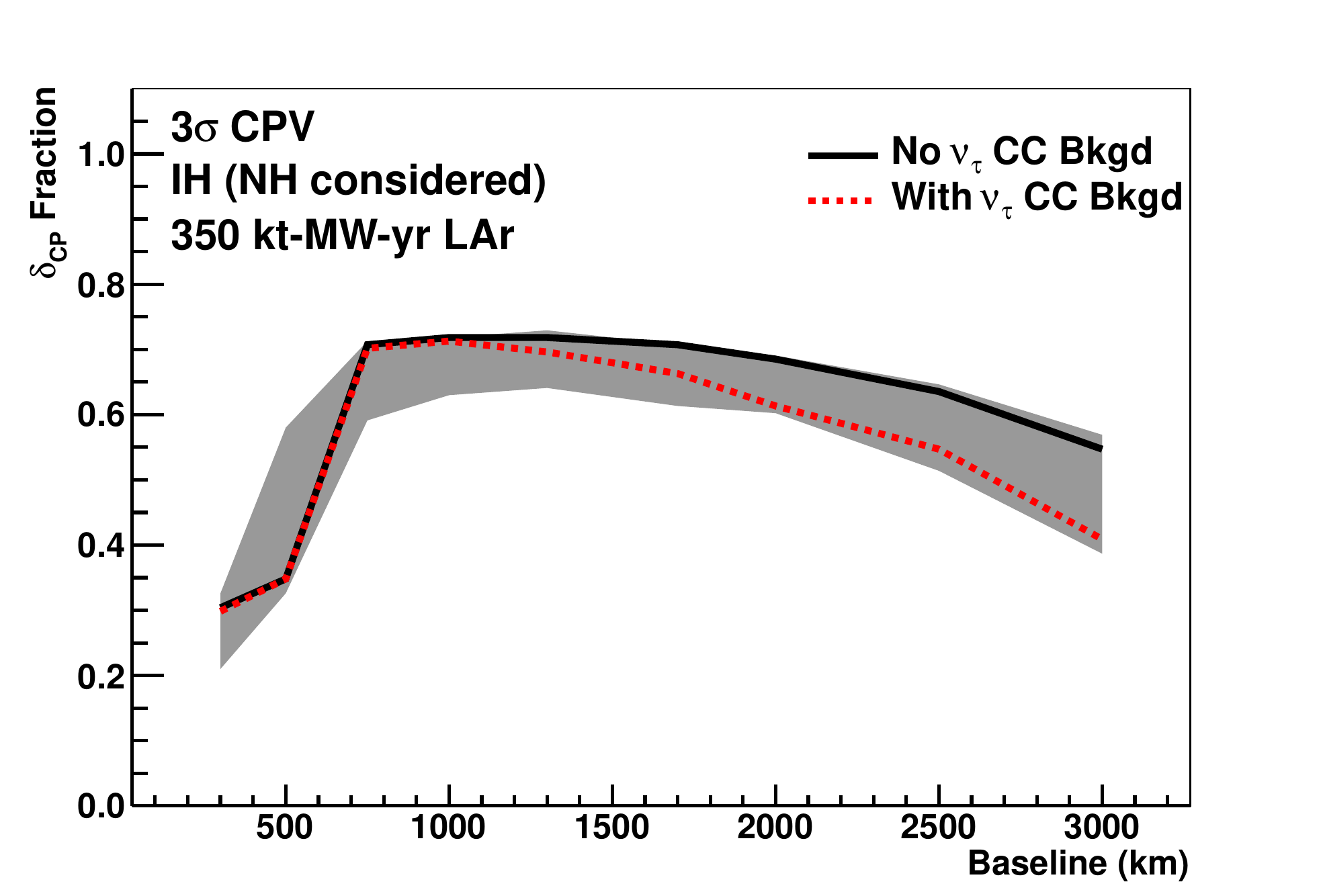}
\caption{The fraction of all possible $\delta_{CP}$ values for which we can observe CP violation with a sensitivity of at least 3$\sigma$ ($\Delta \chi^{2} = 9$) for normal (left) and inverted (right) mass hierarchy as a function of baseline.  The true mass hierarchy is assumed to be unknown. The solid black (red dashed) line shows the result including zero (maximum) $\nu_{\tau}$ CC background.  The shaded band shows the possible range in the fraction due to the uncertainty in the other oscillation parameters and considers both octant solutions for $\theta_{23}$.}
\label{fig:cpfrac_cpv}
\end{figure*}

To determine the sensitivity to CP violation, we calculate the significance of excluding the CP-conserving values of $\delta_{CP} = 0,\pi$.  The significance of the CP violation measurement is defined as $\sigma = \sqrt{\Delta\chi^{2}}$. Figure \ref{fig:cpfrac_cpv} shows the fraction of all possible true $\delta_{CP}$ values for which we can exclude CP-conserving values of $\delta_{CP}$ with a sensitivity of at least 3$\sigma$ ($\Delta \chi^{2} = 9$) as a function of baseline, for both normal and inverted hierarchy.  In these plots we assume that the true hierarchy is unknown by considering both hierarchy solutions in the minimization.  The maximum sensitivity to CP violation is achieved for baselines between 750~km and 1500~km, with the very short baselines having the worst sensitivity.  If the maximum $\nu_{\tau}$ CC background is included, the sensitivity decreases for very long baselines.  The shaded band shows the possible range in the fraction due to the uncertainty in the other oscillation parameters, dominated by the uncertainty in $\
\theta_{23}$.  Both octant solutions for $\theta_{23}$ are considered by the shaded region.

\begin{figure*}[htp]
\includegraphics[width=0.49\textwidth]{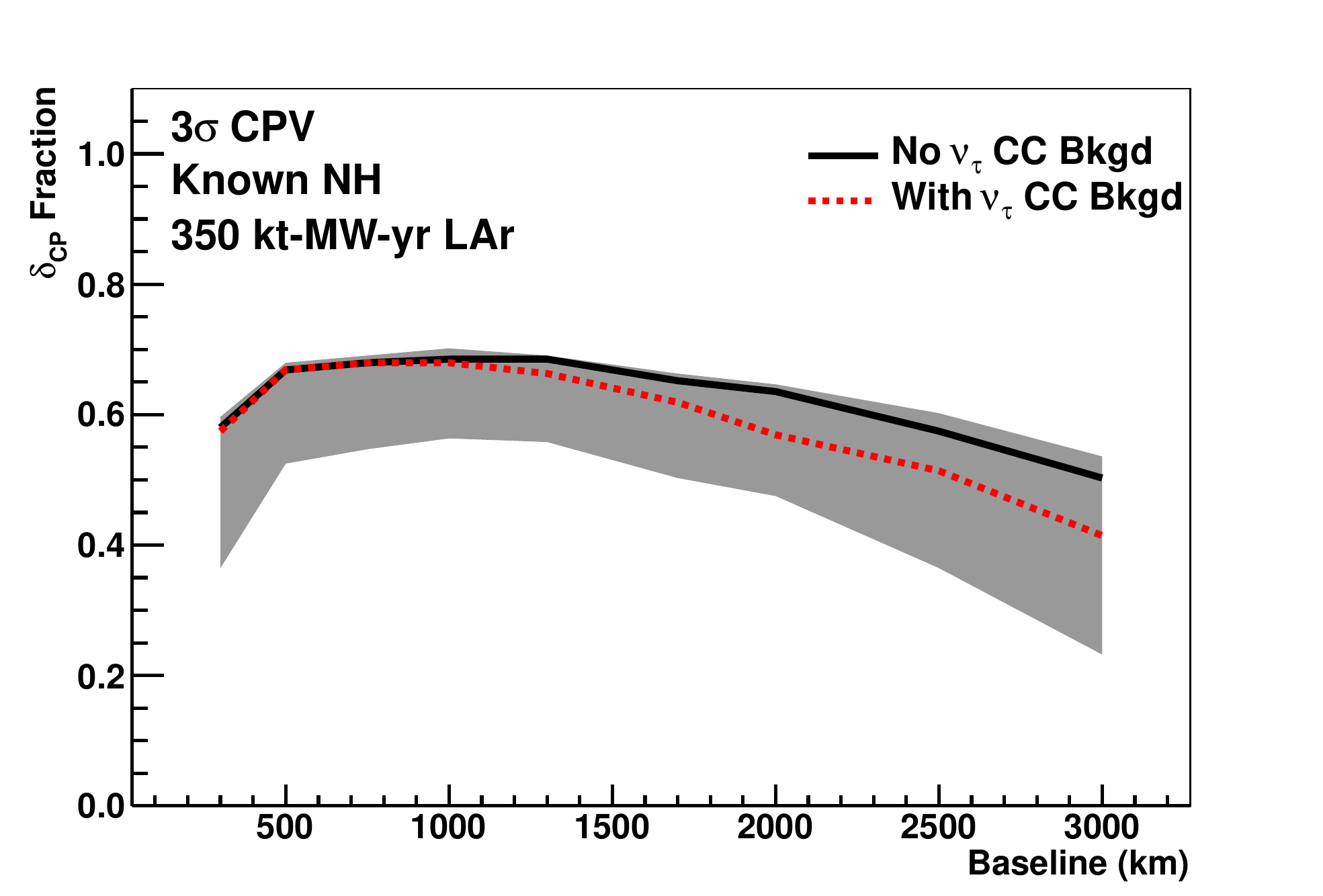}
\includegraphics[width=0.49\textwidth]{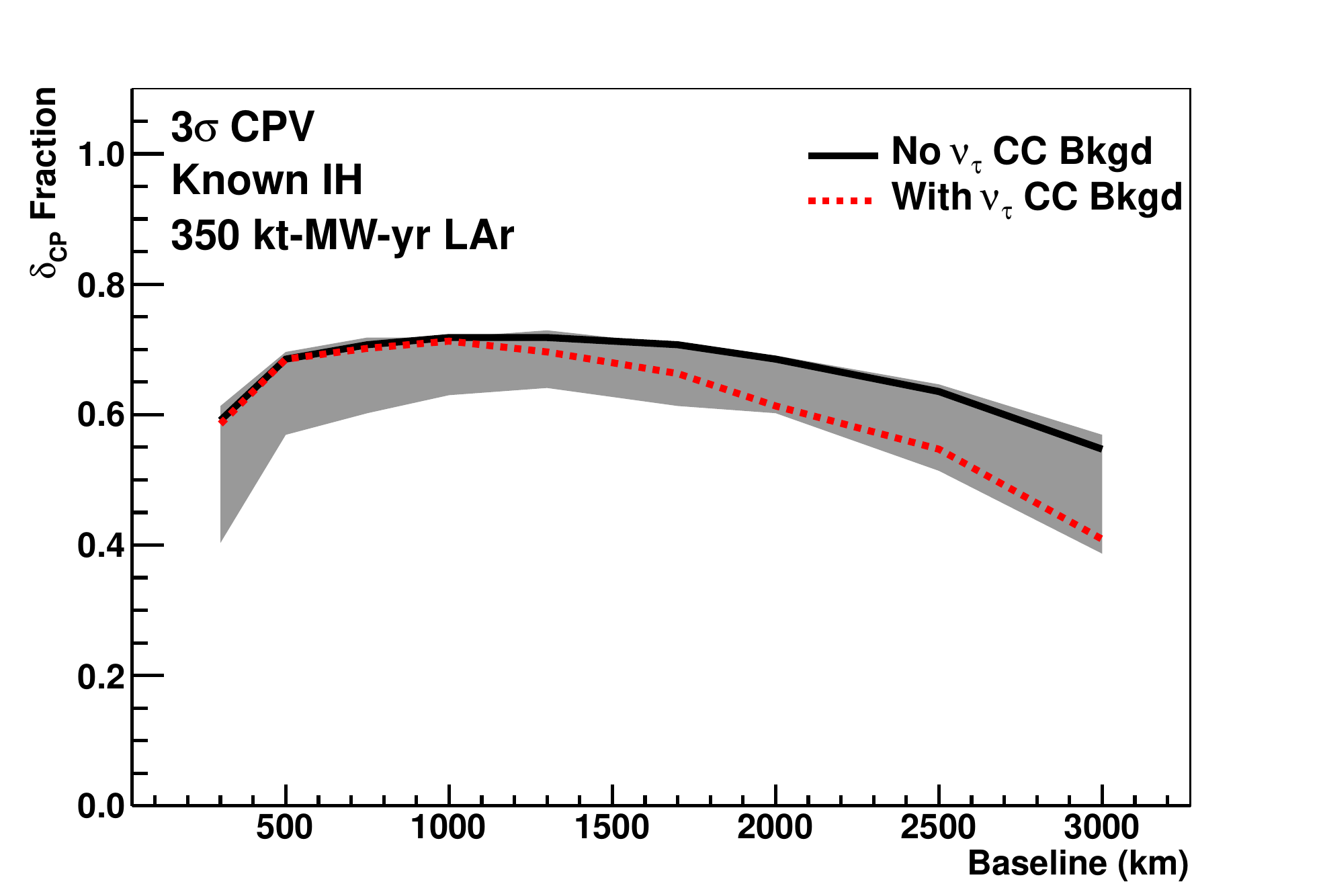}
\caption{The fraction of all possible $\delta_{CP}$ values for which we can observe CP violation with a sensitivity of at least 3$\sigma$ ($\Delta \chi^{2} = 9$) for normal (left) and inverted (right) mass hierarchy as a function of baseline.  The true mass hierarchy is assumed to be perfectly known. The solid black (red dashed) line shows the result including zero (maximum) $\nu_{\tau}$ CC background.  The shaded band shows the possible range in the fraction due to the uncertainty in the other oscillation parameters and considers both octant solutions for $\theta_{23}$.}
\label{fig:cpfrac_cpv_knownmh}
\end{figure*}

\begin{figure*}[htp]
\includegraphics[width=0.32\textwidth]{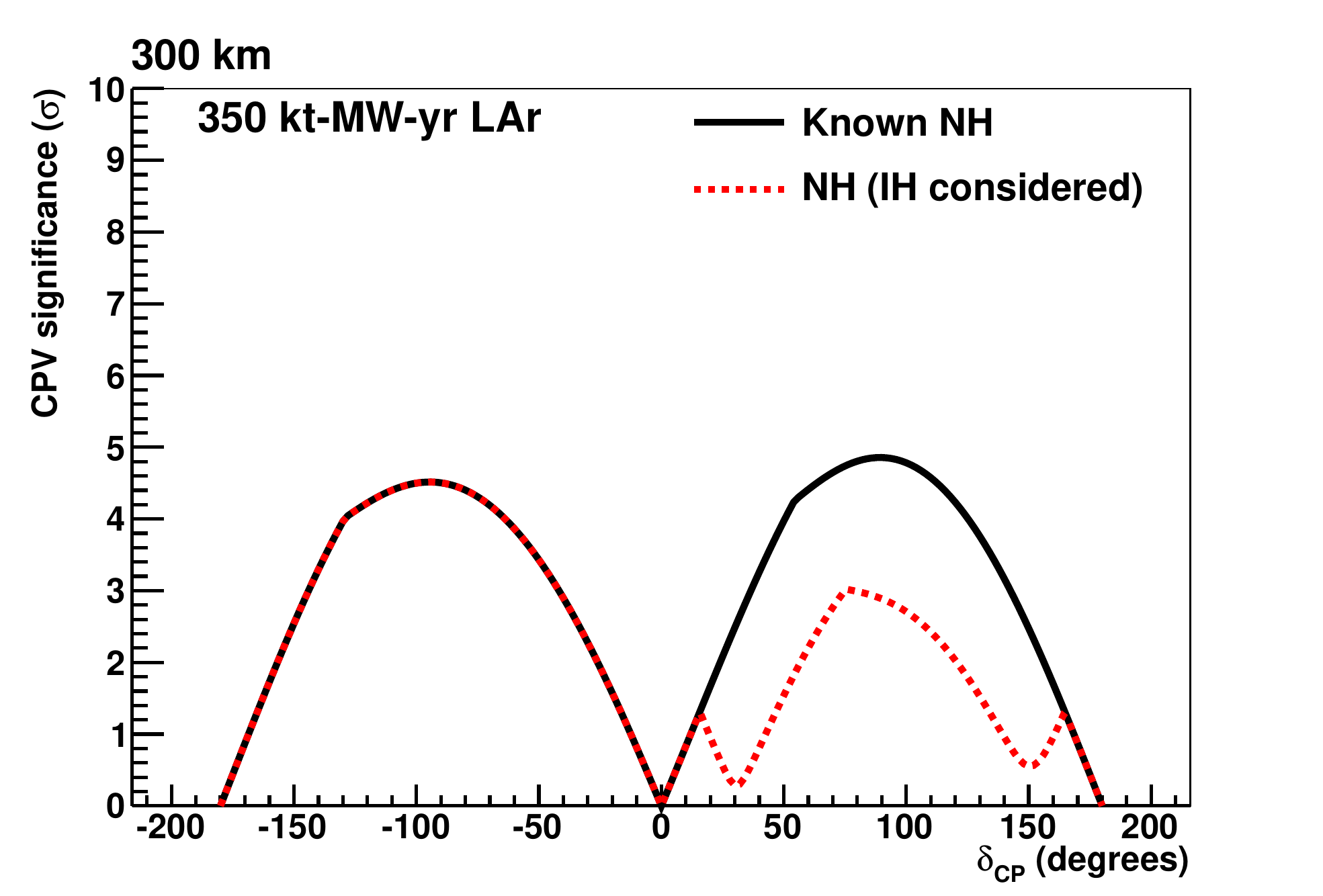}
\includegraphics[width=0.32\textwidth]{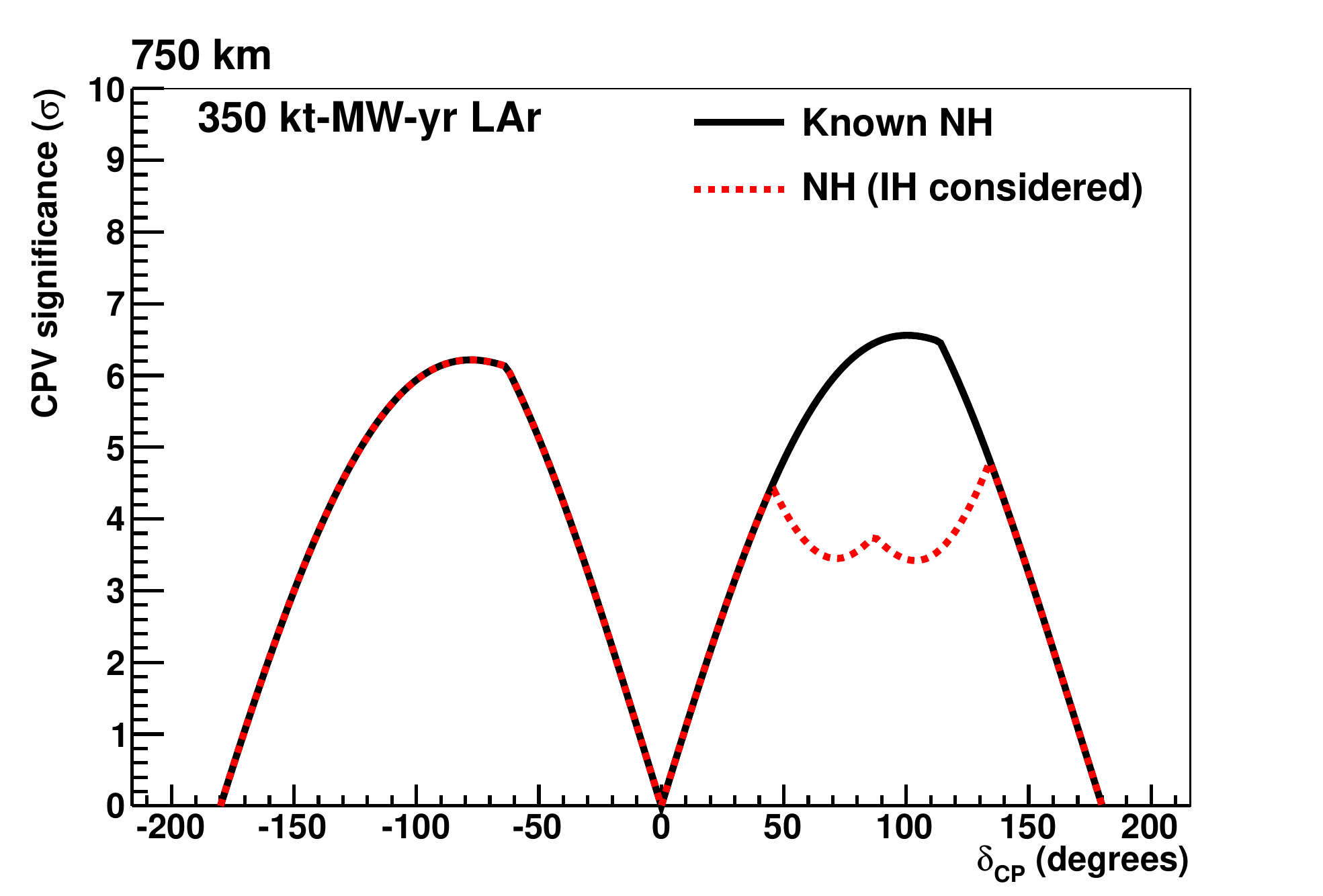}
\includegraphics[width=0.32\textwidth]{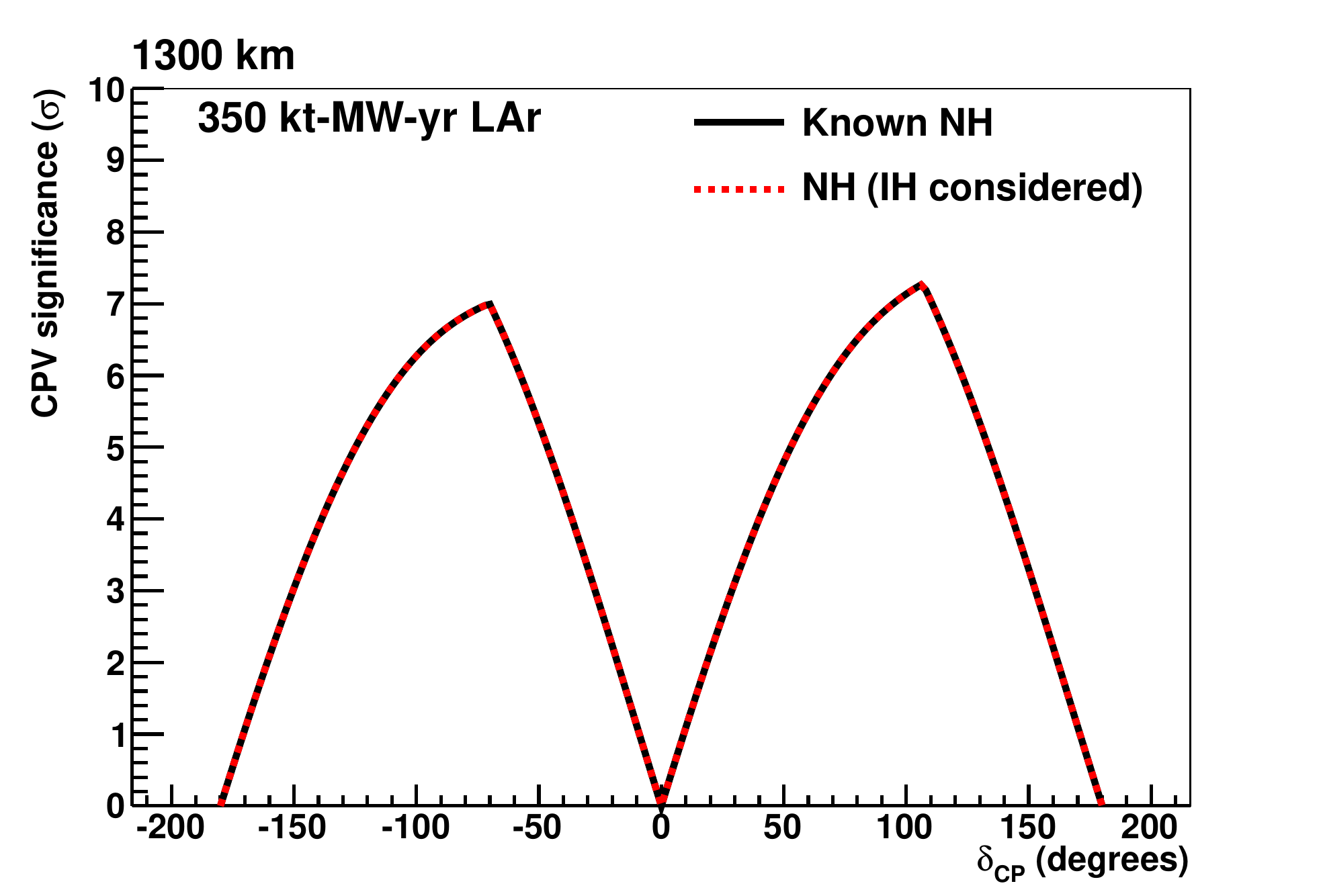}
\caption{The significance ($\sigma = \sqrt{\Delta \chi^{2}}$) at which CP violation can be observed as a function of the true value of $\delta_{CP}$ assuming normal hierarchy for 300~km, 750~km, and 1300~km baselines.  The significance is shown assuming the hierarchy is perfectly known (black-solid) and assuming both hierarchy solutions are considered (red-dashed).  Knowing the hierarchy improves the CP sensitivity at 300~km and 750~km, but has no effect at 1300~km (or baselines $>$ 1300~km).}
\label{fig:cpsig}
\end{figure*}

Figure \ref{fig:cpfrac_cpv_knownmh} shows CP violation sensitivities in plots similar to Figure \ref{fig:cpfrac_cpv}, in which we assume that the true hierarchy is known and consider only those solutions corresponding to the true hierarchy in the minimization.  Knowing the mass hierarchy significantly increases the CP violation sensitivity at shorter baselines.   This effect is illustrated in Figure \ref{fig:cpsig}, which shows the significance as a function of the true value of $\delta_{CP}$ for 300~km, 750~km, and 1300~km baselines.  The shorter baselines do not have the advantage of the large CP asymmetry in the second oscillation maximum and therefore the CP measurement at short baselines suffers from the ambiguity of the matter asymmetry and the CP asymmetry in the first oscillation maximum.  If the hierarchy is known, this ambiguity is removed.  The differences in sensitivity among the baselines are smaller when the mass hierarchy is known, but, as noted in the previous section, an unambiguous  measurement of the mass hierarchy using electron neutrino appearance will remain a high priority.

\begin{figure*}[htp]
\includegraphics[width=0.49\textwidth]{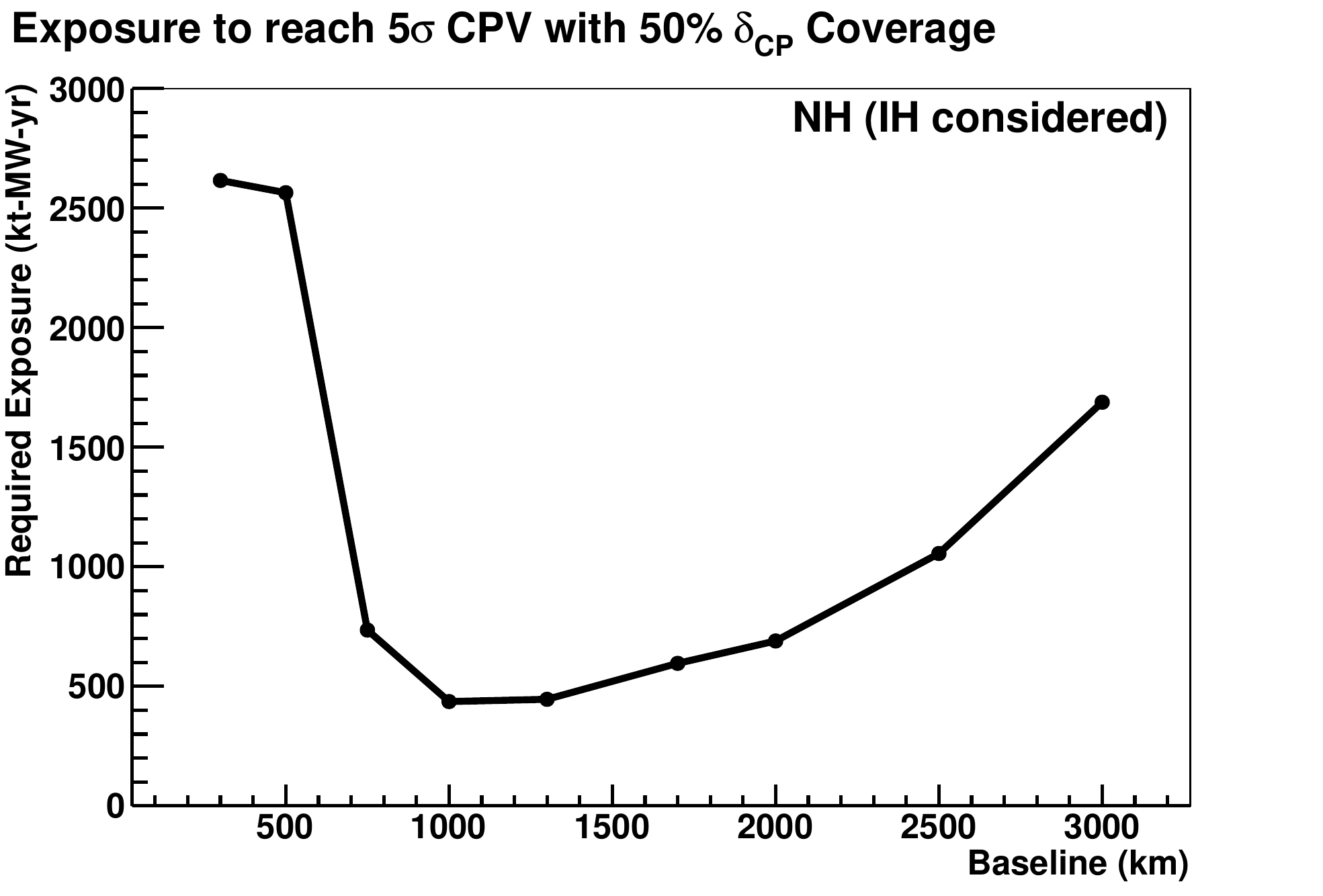}
\includegraphics[width=0.49\textwidth]{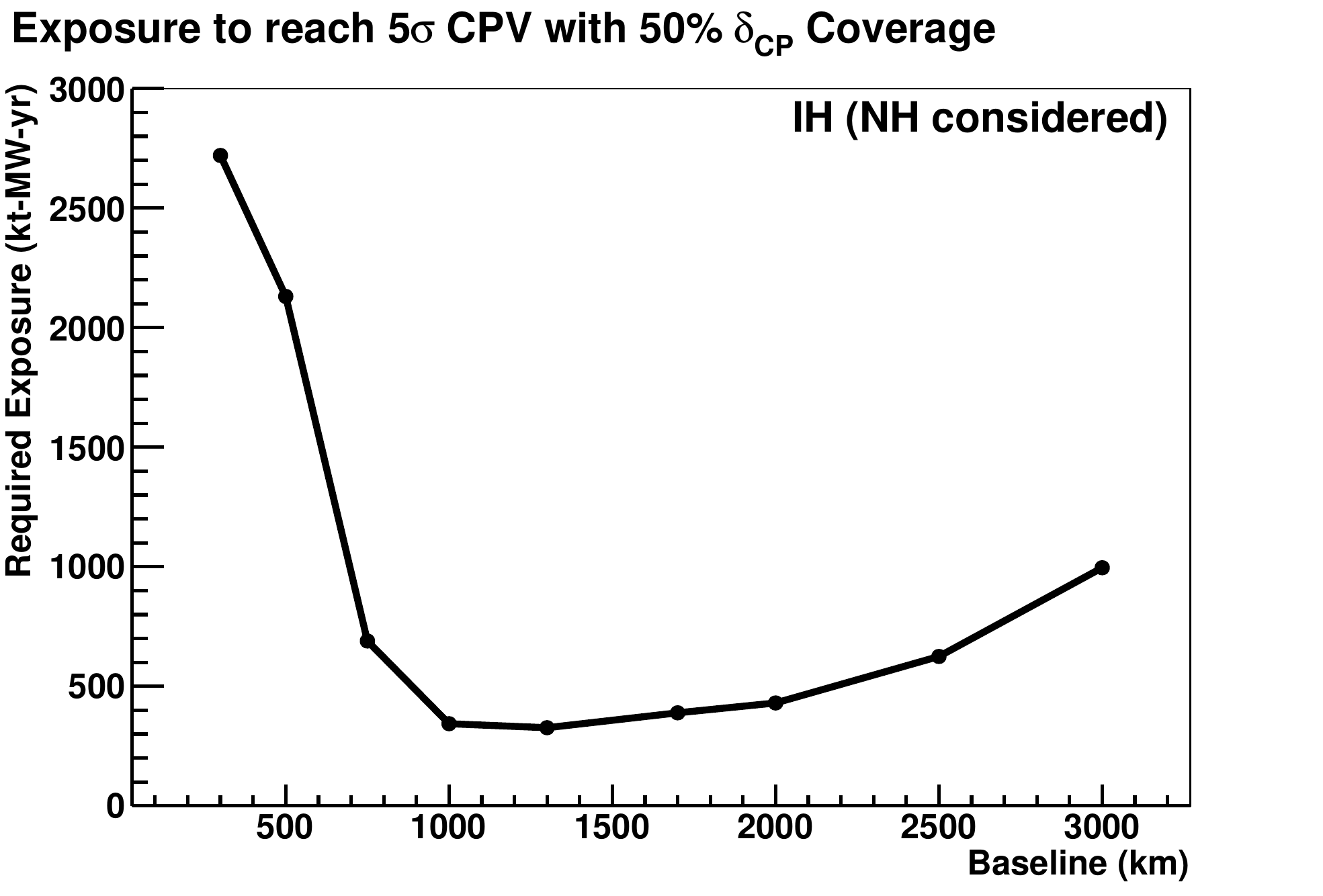}
\caption{The exposure required to observe CP violation with a significance of 5$\sigma$ for 50\% of all possible values of $\delta_{CP}$, assuming normal (left) and inverted (right) mass hierarchy, as a function of baseline.  The true mass hierarchy is assumed to be unknown, and we assume no $\nu_{\tau}$ CC background.}
\label{fig:CP_expandbl}
\end{figure*}

\begin{figure*}[htp]
\includegraphics[width=0.49\textwidth]{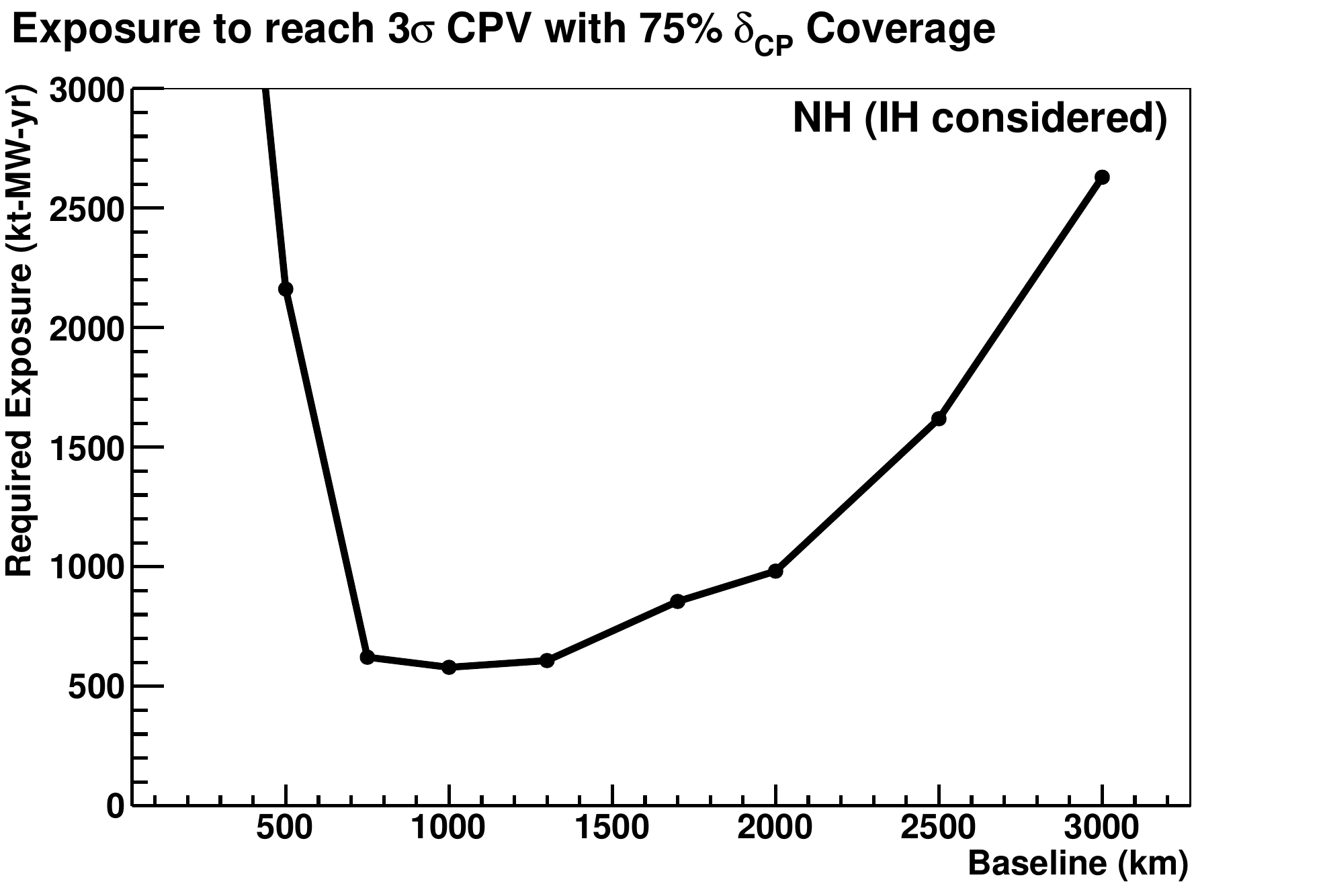}
\includegraphics[width=0.49\textwidth]{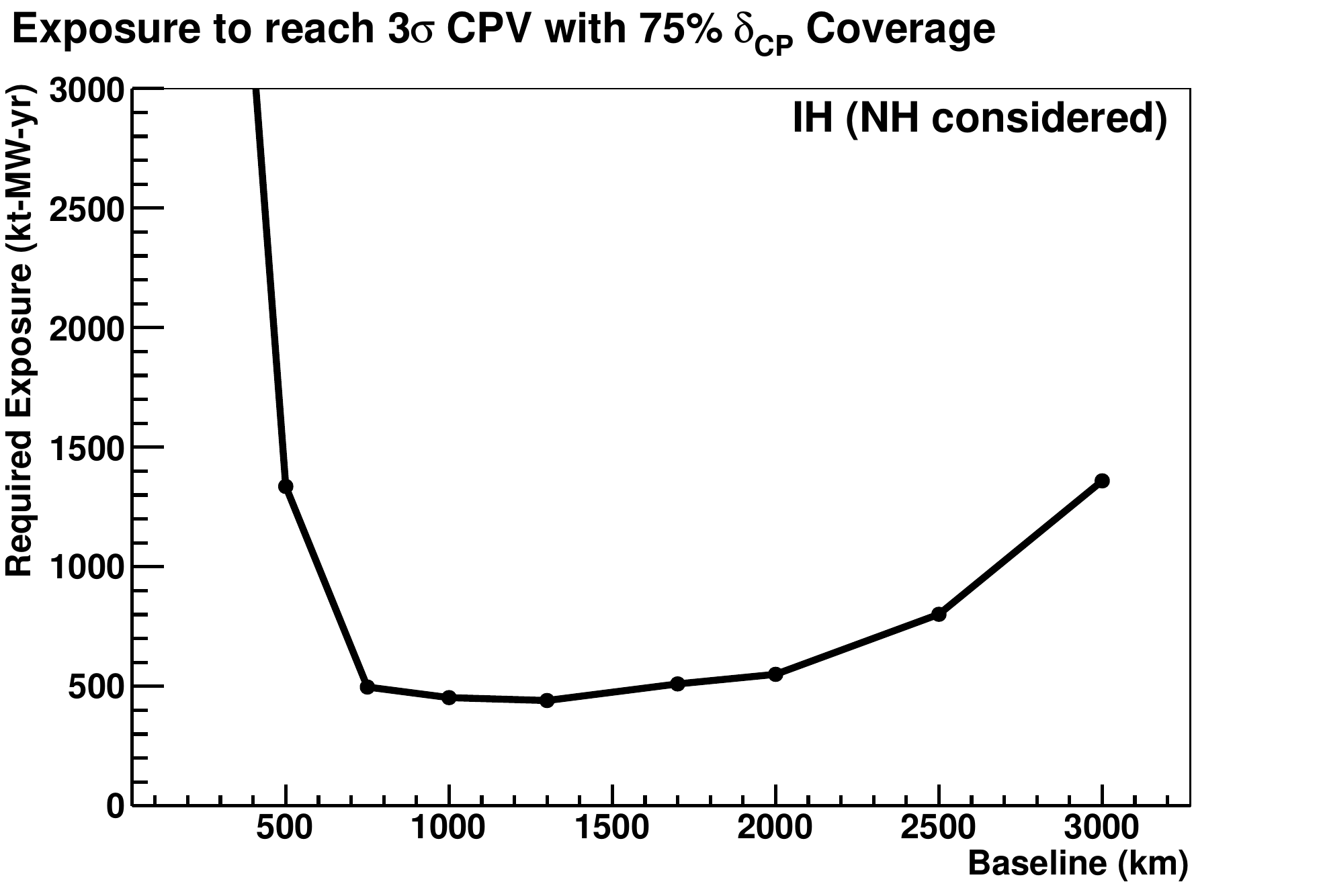}
\caption{The exposure required to observe CP violation with a significance of 3$\sigma$ for 75\% of all possible values of $\delta_{CP}$, assuming normal (left) and inverted (right) mass hierarchy, as a function of baseline.  The true mass hierarchy is assumed to be unknown, and we assume no $\nu_{\tau}$ CC background.}
\label{fig:CP_expandbl_3s}
\end{figure*}

The sensitivity calculations above assume a 350~\mbox{kt-MW-yr} exposure. Figure \ref{fig:CP_expandbl} (\ref{fig:CP_expandbl_3s}) shows the exposure required to observe CP violation with a significance of 5$\sigma$ (3$\sigma$) for 50\% (75\%) of all possible values of $\delta_{CP}$ as a function of baseline.


\begin{figure*}[htp]
\includegraphics[width=0.49\textwidth]{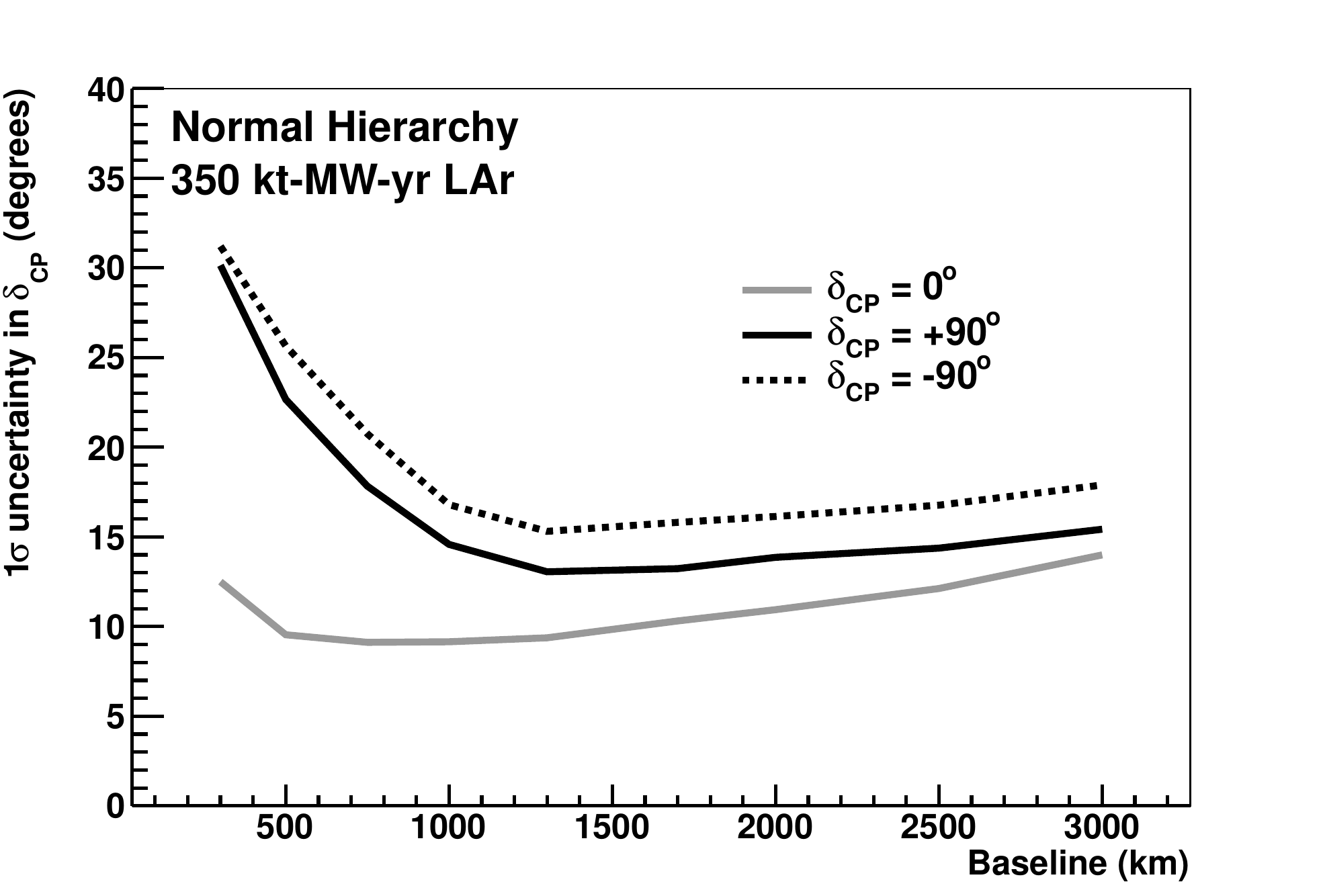}
\includegraphics[width=0.49\textwidth]{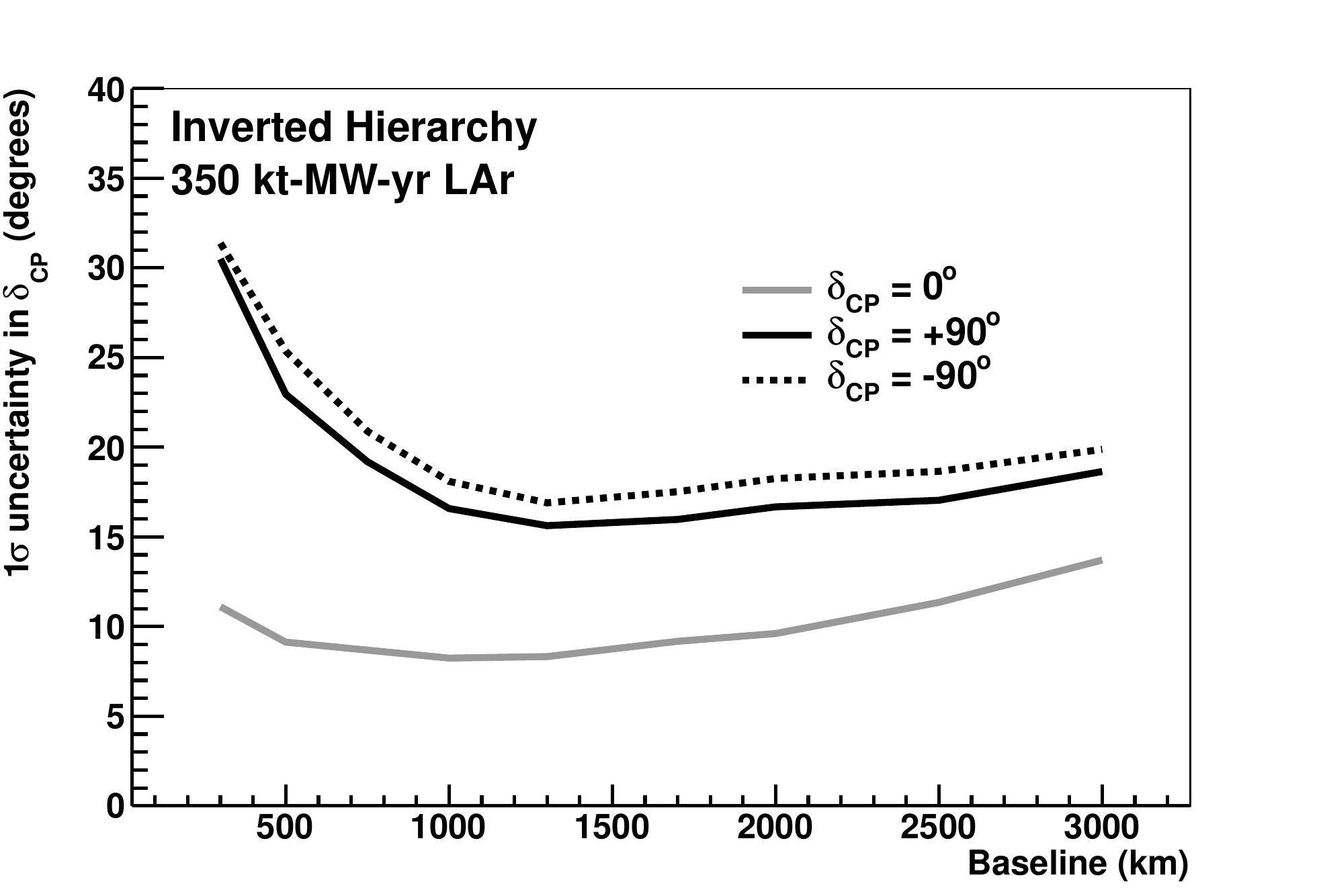}
\caption{The 1$\sigma$ uncertainty in $\delta_{CP}$ as a function of baseline assuming the true value of $\delta_{CP}$ is 0$^{\circ}$ (gray solid), +90$^{\circ}$ (black solid), and -90$^{\circ}$ (black dashed) for normal (left) and inverted (right) mass hierarchy.  The true mass hierarchy is assumed to be perfectly known.}
\label{fig:deltares_pm90}
\end{figure*}

\begin{figure*}[htp]
\includegraphics[width=0.49\textwidth]{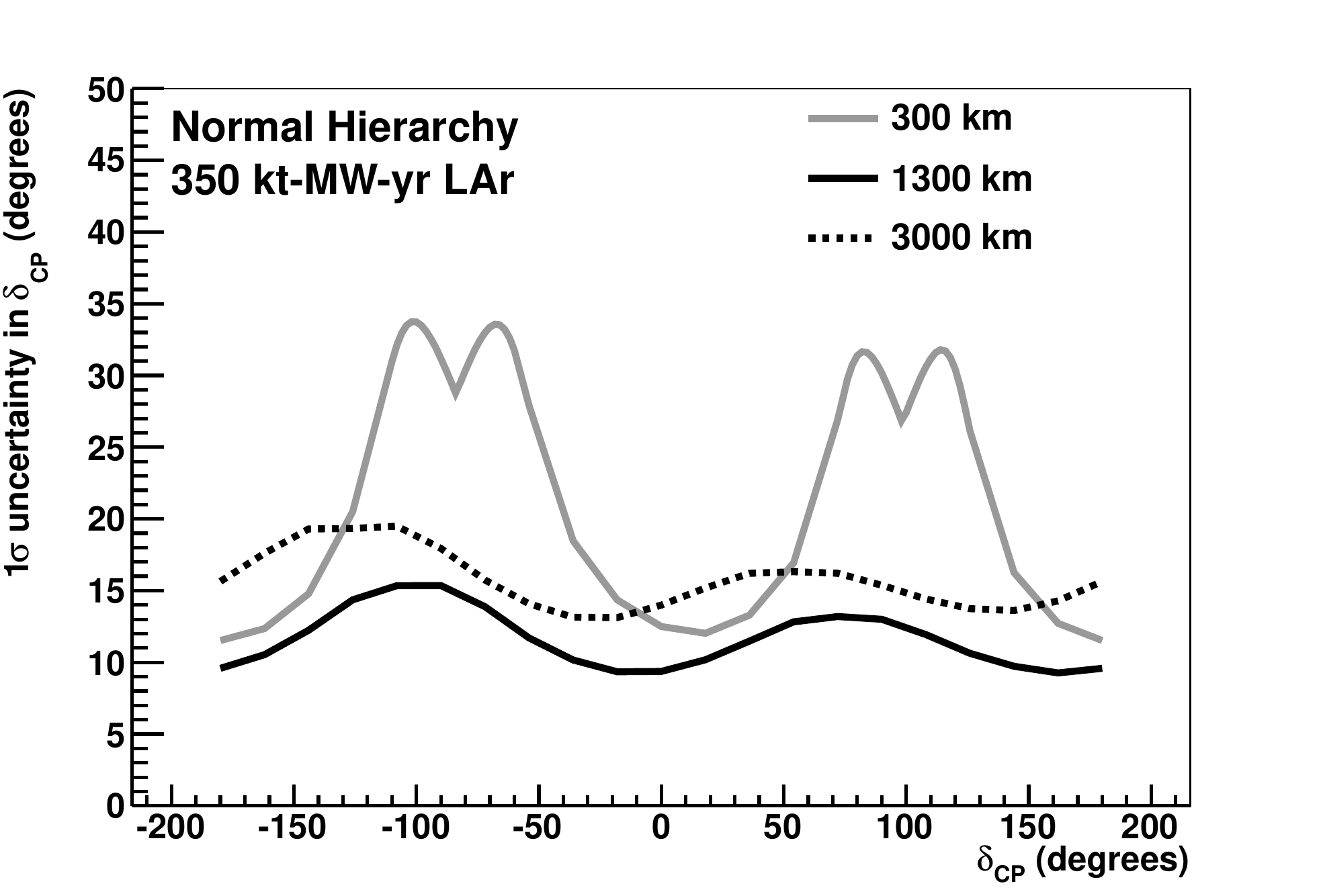}
\caption{The 1$\sigma$ uncertainty in $\delta_{CP}$ as a function of the true value of $\delta_{CP}$, assuming the true hierarchy is known to be normal, for baselines of 300~km, 1300~km, and 3000~km.}
\label{fig:deltaresvsdelta}
\end{figure*}

\begin{figure*}[htp]
\includegraphics[width=0.49\textwidth]{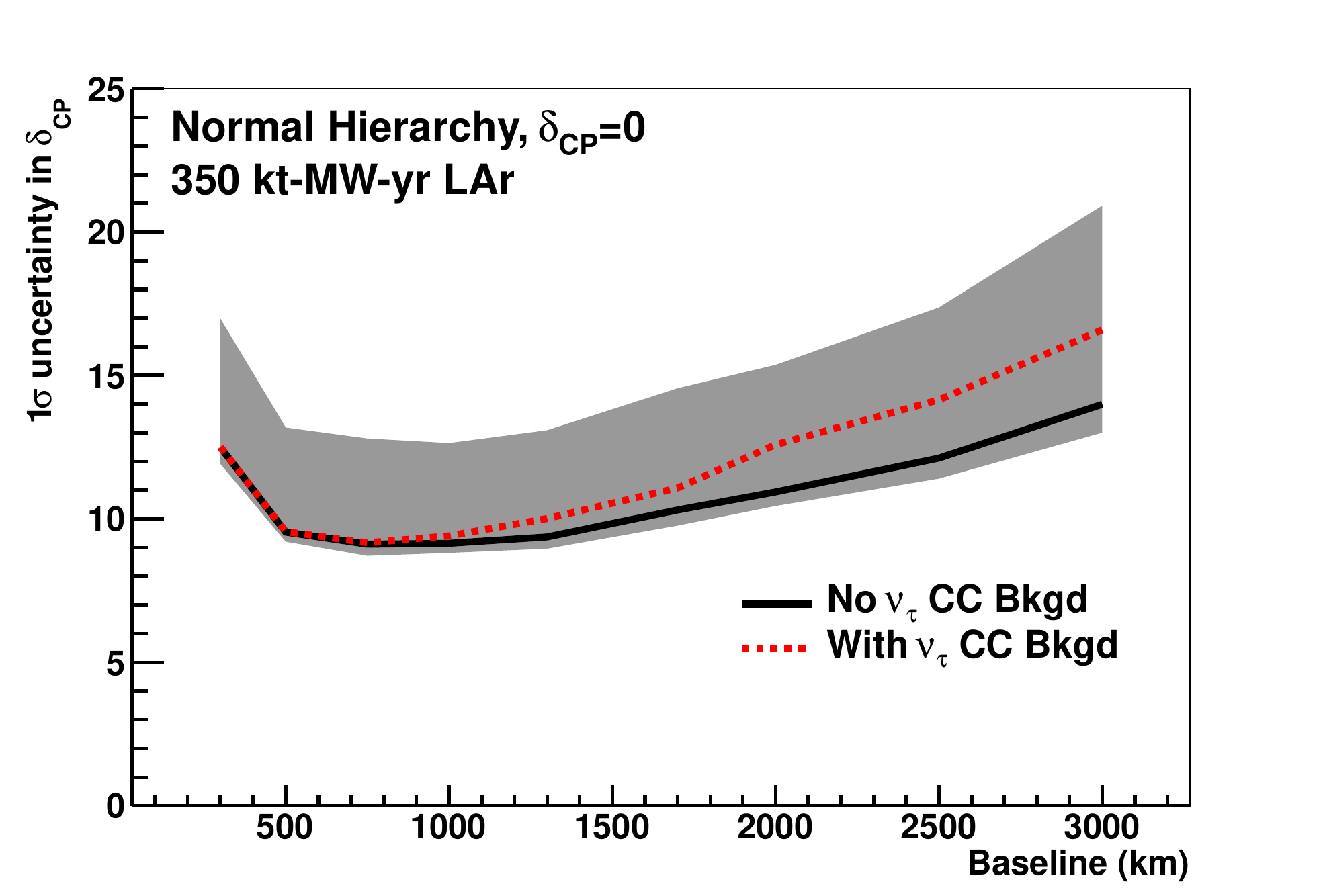}
\includegraphics[width=0.49\textwidth]{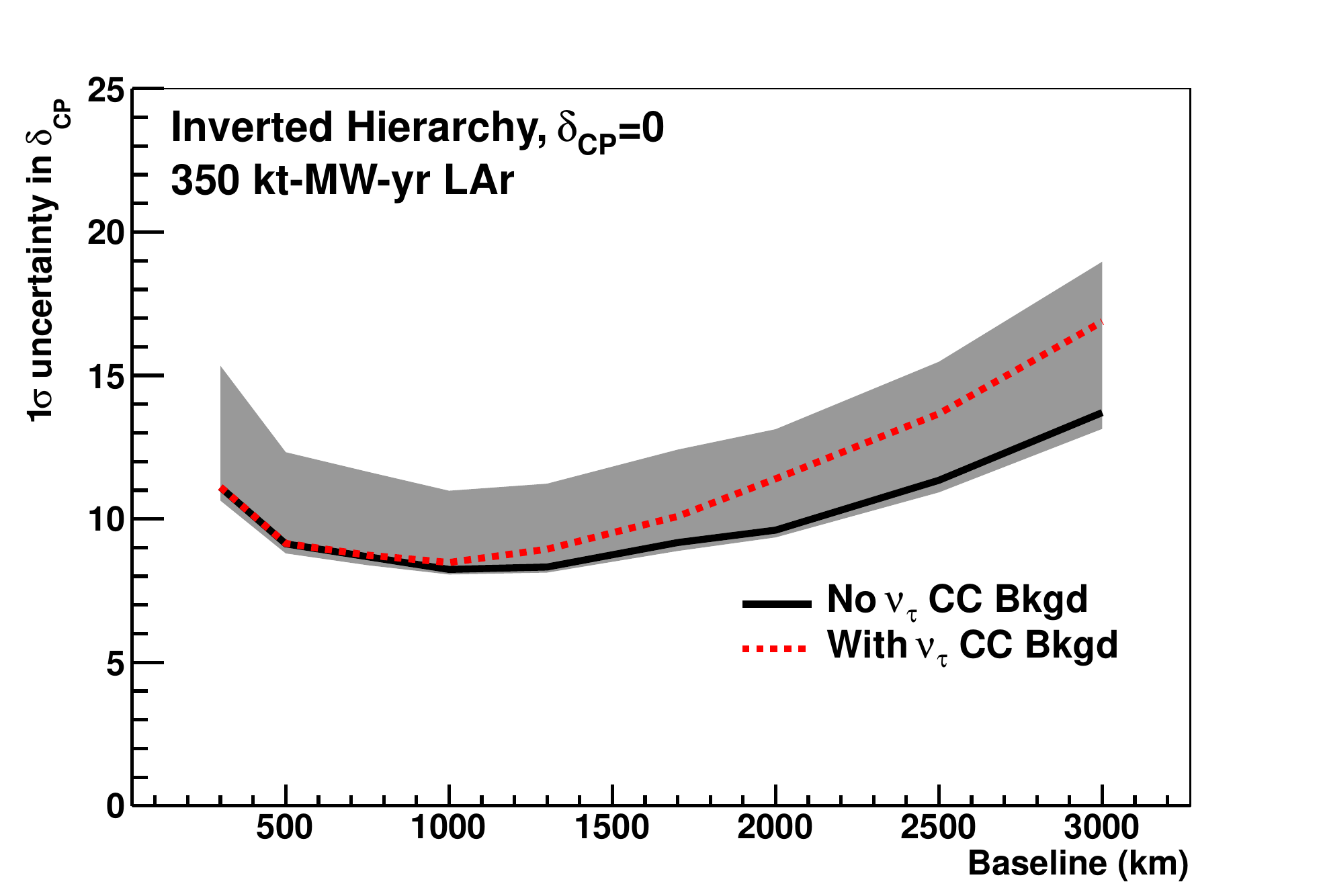}
\caption{The 1$\sigma$ uncertainty in $\delta_{CP}$ as a function of baseline, assuming the true value of $\delta_{CP}$ is 0$^{\circ}$, for normal (left) and inverted (right) mass hierarchy.  The true mass hierarchy is assumed to be perfectly known.  The solid black (red dashed) line shows the result including zero (maximum) $\nu_{\tau}$ CC background.  The shaded band shows the possible range in the resolution due to the uncertainty in the other oscillation parameters and considers both octant solutions for $\theta_{23}$.}
\label{fig:deltares_0}
\end{figure*}

We consider not only the significance of determining CP violation, but also the precision with which the value of $\delta_{CP}$ can be measured.  Figure \ref{fig:deltares_pm90} shows the $\delta_{CP}$ resolution (1$\sigma$ uncertainty, equivalent to $\Delta\chi^2=1$) as a function of baseline for different true values of $\delta_{CP}$.  The dependence of the resolution on the value of $\delta_{CP}$ is shown explicitly in Figure \ref{fig:deltaresvsdelta} for different baselines.  Figure \ref{fig:deltares_0}  shows the resolution for each baseline when $\delta_{CP}=0^{\circ}$, and compares the resolutions obtained when we include the maximum $\nu_{\tau}$ CC background and the range of allowed values for the oscillation parameters.  These plots assume that the mass hierarchy is known.  Even if the mass hierarchy is perfectly known, the resolution is poorest for short baselines, particularly for $\delta_{CP} = \pm90^{\circ}$~\footnote{For the measurement of $\delta_{CP}$, there are physical boundaries: $|\sin(\delta_{CP})| < 1$. Therefore, it may not be accurate to determine the 68\% confidence interval of $\delta_{CP}$ with the simple rule $\Delta\chi^2 = 1$ when the true value of $\delta_{CP}$ approaches $\pm 90$~degrees. The actual resolution of $\delta_{CP}$ at these values may be slightly better than what is shown in this study. For the purposes of this study, however, it is not crucial.}

\subsection{$\theta_{23}$ Octant}

\begin{figure*}[htp]
\includegraphics[width=0.49\textwidth]{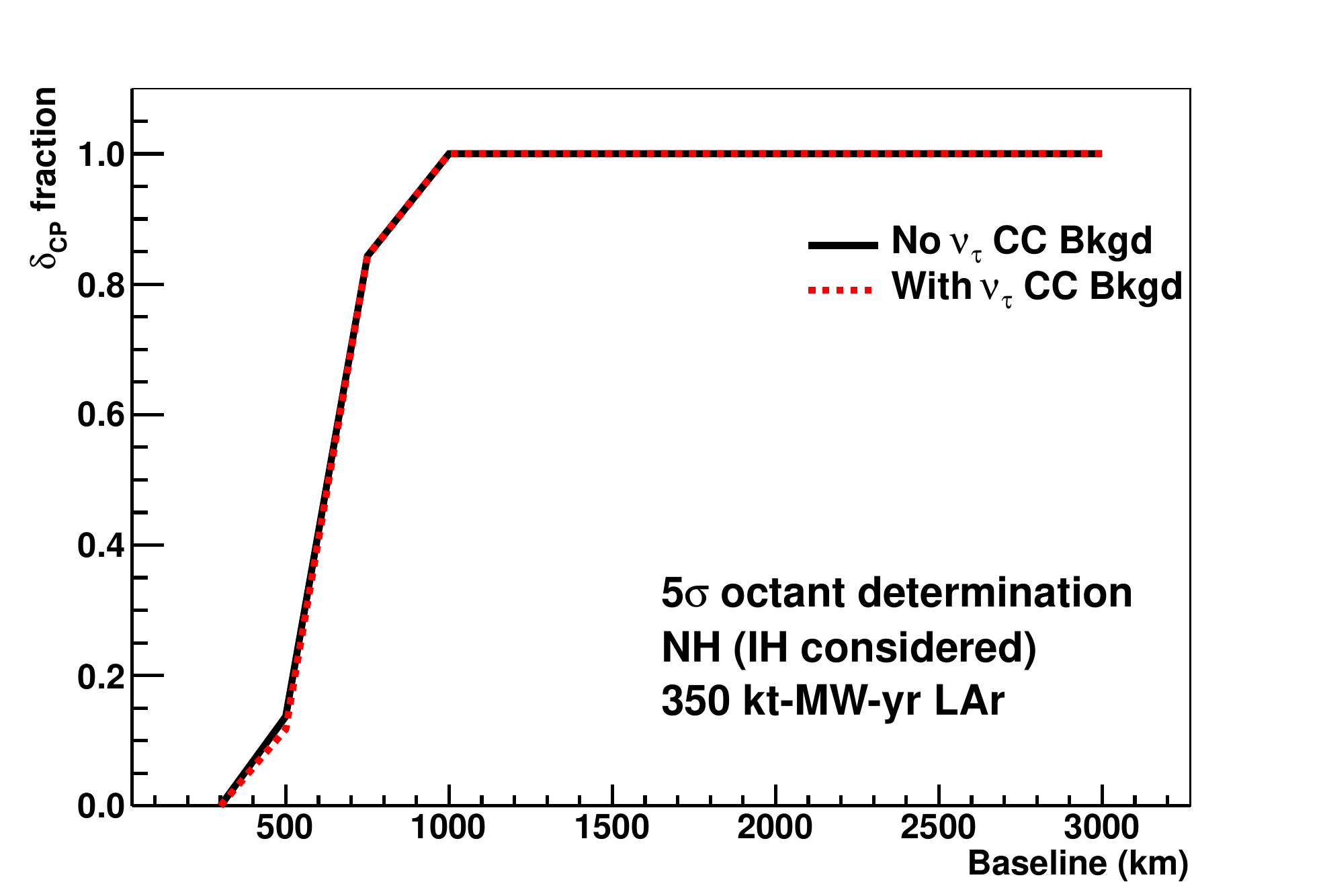}
\includegraphics[width=0.49\textwidth]{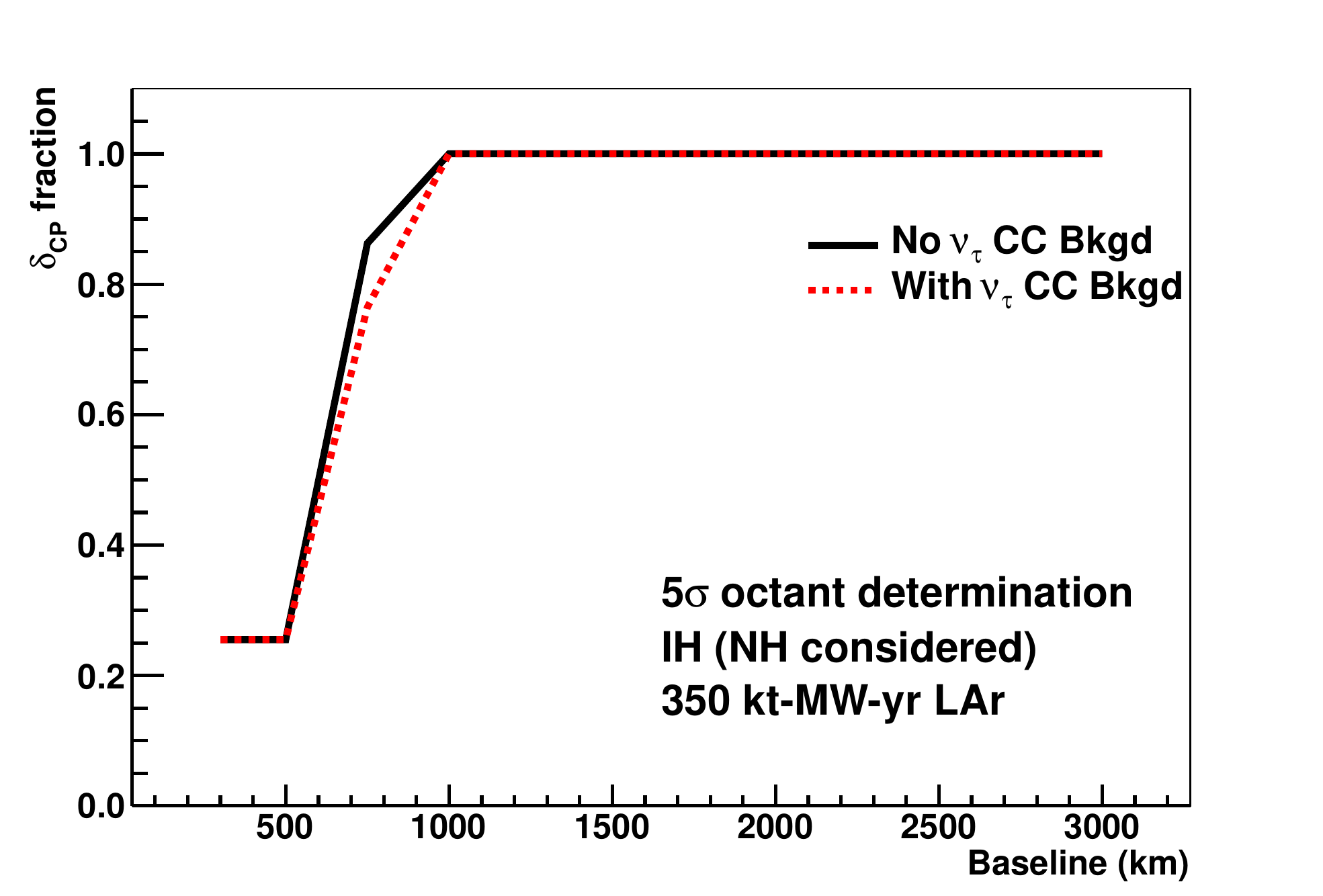}
\caption{The fraction of all possible $\delta_{CP}$ values for which we can determine the $\theta_{23}$ octant with a sensitivity of at least 5$\sigma$ ($\Delta \chi^{2} = 25$) for normal (left) and inverted (right) mass hierarchy as a function of baseline.  The true mass hierarchy is assumed to be unknown. The solid black (red dashed) line shows the result including zero (maximum) $\nu_{\tau}$ CC background.}
\label{fig:cpfrac_octant}
\end{figure*}

To calculate the significance of determining the $\theta_{23}$ octant, the $\chi^2$ minimization only considers solutions that have the opposite octant from that which is used to generate the true spectra. For example, if the true value of $\theta_{23}$ is assumed to be in the first octant, we assume $\theta_{23}$ in the second octant for the observed spectra in the $\chi^2$ calculation.  This allows us to determine the significance at which we can exclude the second octant solution given the true value of $\theta_{23}$ is in the first octant.  The significance of the octant measurement is defined as $\sigma = \sqrt{\Delta\chi^{2}}$.  
Figure \ref{fig:cpfrac_octant} shows the fraction of all possible true $\delta_{CP}$ values for which we can determine the octant with a sensitivity of at least 5$\sigma$ ($\Delta \chi^{2} = 25$) as a function of baseline assuming normal or inverted mass hierarchy, if the true value of $\theta_{23}$ is within the 1-$\sigma$ allowed region \cite{Fogli:2012ua}.  In these plots we assume that the true hierarchy is unknown by considering both hierarchy solutions in the minimization.  We find the octant can be determined at 5$\sigma$ for 100\% of all $\delta_{CP}$ values at a baseline of at least 1000~km.  We find that the sensitivity at the shortest baselines could increase slightly if the true mass hierarchy is known, but the long baselines still have the best sensitivity.

\subsection{Precision Mixing Angle Measurements}

\begin{figure*}[htp]
\includegraphics[width=0.49\textwidth]{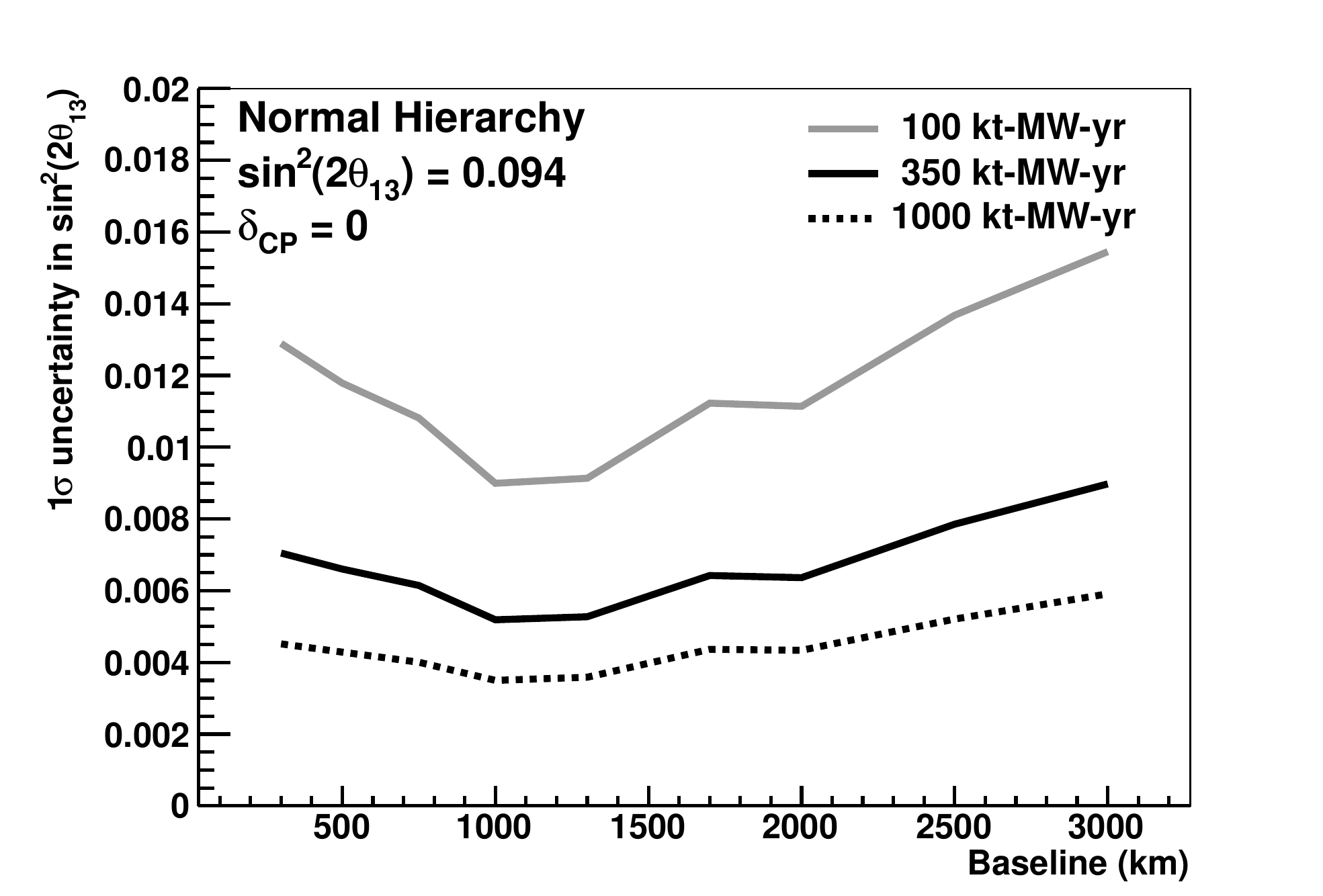}
\includegraphics[width=0.49\textwidth]{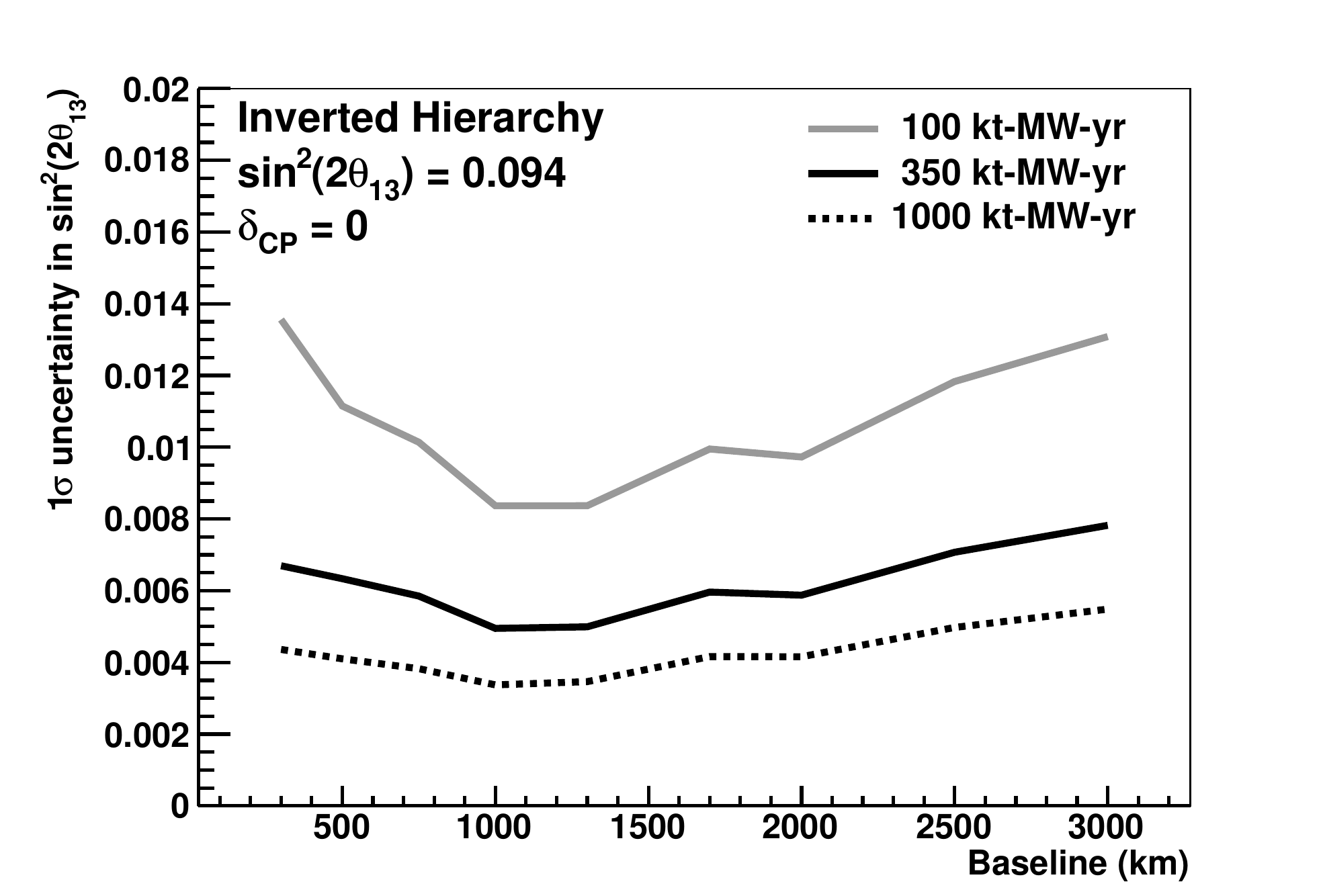}
\caption{The 1$\sigma$ uncertainty in $\sin^2(2\theta_{13})$ as a function of baseline assuming $\sin^2(2\theta_{13}) = 0.094$ and $\delta_{CP}=0$ for normal (left) and inverted (right) mass hierarchy.  The true mass hierarchy is assumed to be perfectly known, and we assume no $\nu_{\tau}$ CC background.}
\label{fig:th13vbaseline}
\end{figure*}

A long-baseline experiment will also make precision measurements of the mixing angles.  Figure \ref{fig:th13vbaseline} shows the resolution of $\sin^2(2\theta_{13})$ as a function of baseline for the nominal exposure of 350~\mbox{kt-MW-yr}.  The true mass hierarchy is assumed to be known, and we assume no $\nu_{\tau}$ CC background for this calculation.  The curves for 100 and 1000~\mbox{kt-MW-yr} exposures are also shown.  The best resolution can be achieved at baselines between 1000 and 1500~km.

\begin{figure*}[htp]
\includegraphics[width=0.49\textwidth]{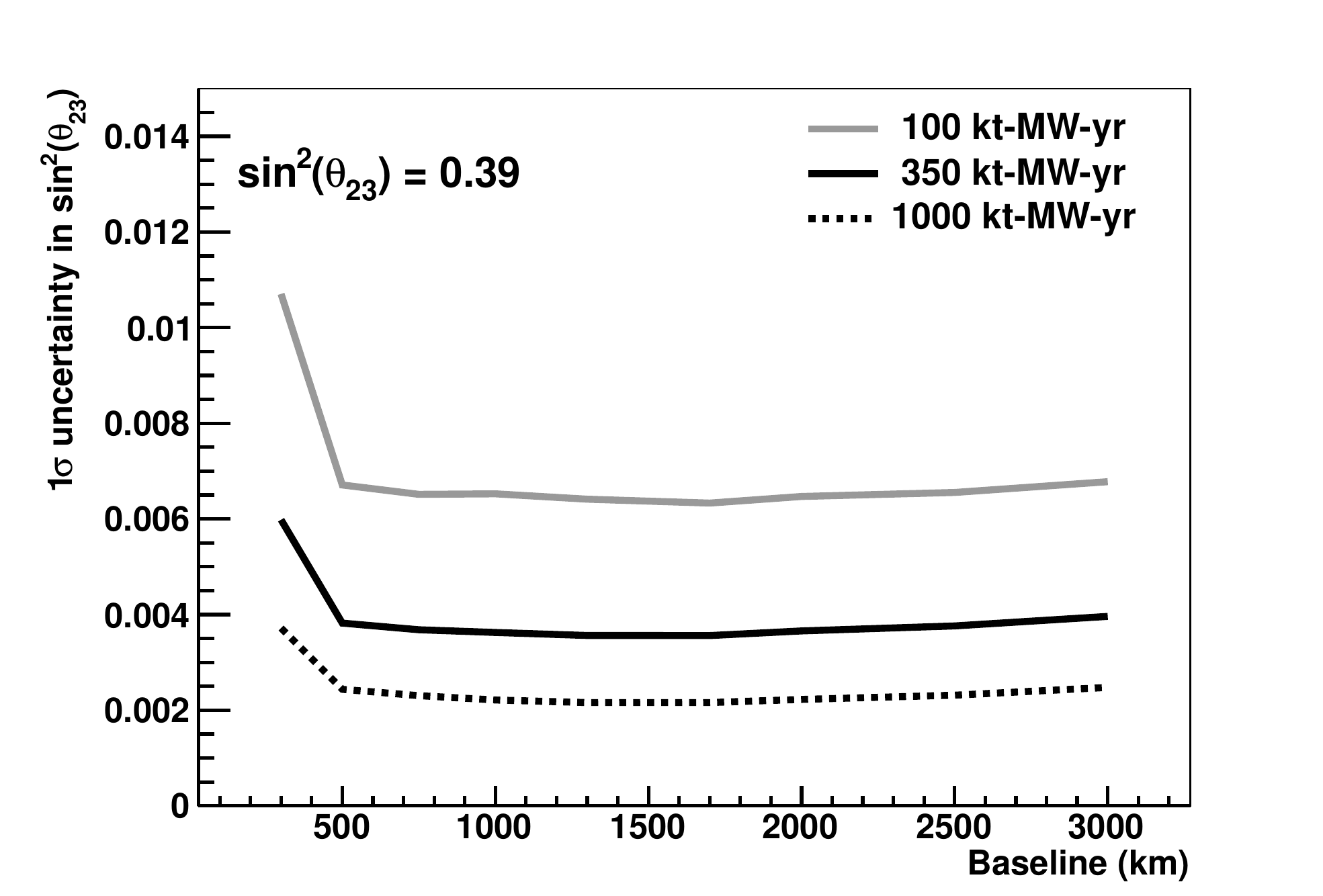}
\caption{The 1$\sigma$ uncertainty in $\sin^2(\theta_{23})$ as a function of baseline assuming $\sin^2(\theta_{23}) = 0.39$.}
\label{fig:th23vbaseline}
\end{figure*}

Figure \ref{fig:th23vbaseline} shows the resolution of $\sin^2\theta_{23}$ as a function of baseline for the nominal exposure of 350~\mbox{kt-MW-yr}.  Normal hierarchy is assumed for this calculation, but the choice of hierarchy has negligible impact on this measurement.  The curves for 100 and 1000~\mbox{kt-MW-yr} exposures are also shown.  The resolution is roughly constant as a function of baseline for baselines 500~km and greater.

\section{Conclusion}

We have studied the sensitivity to the key measurements for an electron neutrino appearance experiment as a function of baseline using a wide-band muon neutrino beam and assuming a nominal exposure of 350~\mbox{kt-MW-yr}. The fluxes are optimized for each baseline considered, assuming achievable beam power and energy from the Fermilab proton complex. We find that a detector at a baseline of at least 1000 km is optimal. In particular, baselines of $\sim$1000-1500~km are optimal to observe CP violation and measure $\delta_{CP}$, the mass hierarchy is resolved for all $\delta_{CP}$ with $\overline{\Delta \chi^{2}}=25$ for baselines greater than 1300~km, and the octant is resolved at 5$\sigma$ for all $\delta_{CP}$ for baselines greater than 1000~km.


\begin{acknowledgments}
We would like to thank Josh Klein, William Louis, Alberto Marchionni, and Michael Mooney for their careful reading and helpful suggestions during the preparation of this manuscript.  This material is based upon work supported by the U.S. Department of Energy, Office of Science, Office of High Energy Physics.
\end{acknowledgments}

\bibliography{lbne-baseline-prd}

\end{document}